\documentclass{article}
\usepackage[authoryear]{natbib}
\usepackage{graphicx} % Required for inserting images
\usepackage[utf8]{inputenc}
\usepackage{amsmath}
\usepackage{amssymb}
\usepackage{gensymb}
\usepackage{hyperref}
\usepackage{natbib}
\usepackage{wrapfig}
\usepackage{xcolor}
\usepackage{booktabs}
\usepackage{longtable}
\usepackage{caption}
\usepackage{float} % For precise float placement
\usepackage{tikz}
\usepackage{lscape}
\usepackage[affil-it]{authblk}
\hypersetup{ 
backref=false, % link backreferences (i.e. citations to their list)
bookmarks=true, % show bookmarks bar?
pdftoolbar=true, % show Acrobats toolbar?
pdfmenubar=true, % show Acrobats menu?
pdffitwindow=false, % window fit to page when opened
pdfstartview={FitH}, % fits the width of the page to the window
pdftitle={}, % title
pdfauthor={}, % author
pdfsubject={}, % subject of the document
pdfkeywords={}{}, % list of keywords
pdfnewwindow=true, % links in new window
colorlinks=true, % false: boxed links; true: colored links
linkcolor=teal, % colour of internal links
citecolor=violet, % colour of links to bibliography
urlcolor=blue,
}

\defcitealias{ipcc2023}{IPCC 2023}
\defcitealias{ipcc2023_1}{IPCC, 2023}
\defcitealias{nasav2}{Beaudoing \& Rodell (2019,2020)}
\defcitealias{nasav21}{Beaudoing \& Rodell (2020)}
\defcitealias{population_data}{WorldPop \& CIESIN (2018)}
\defcitealias{population1}{JRC-EC \& CIESIN (2021)}
\defcitealias{population2}{CIESIN \& CIAT (2005)}

\title{Temperature Sensitivity of Residential Energy Demand on the Global Scale: A Bayesian Partial Pooling Model}

\author[a]{Peer Lasse Hinrichsen\thanks{Corresponding author: hinrichsen@economics.uni-kiel.de}}
\author[a]{Katrin Rehdanz}
\author[b,c,d,e,f]{Richard S.J.Tol}

\affil[a]{Institute for Environmental, Resource and Spatial Economics, Kiel University, Germany}
\affil[b]{Department of Economics, University of Sussex, UK}
\affil[c]{Institute for Environmental Studies and Department of Spatial Economics, Vrije Universiteit, Amsterdam, The Netherlands}
\affil[d]{Tinbergen Institute, Amsterdam, The Netherlands}
\affil[e]{CESifo, Munich, Germany}
\affil[f]{Payne Institute for Public Policy, Colorado School of Mines, Golden, CO, USA}

\date{} % Remove date

\begin{document}

\maketitle
    \begin{minipage}{0.95\textwidth} % Begin a minipage for the abstract
    \begin{abstract}
This paper contributes to the limited literature on the temperature sensitivity of residential energy demand on a global scale. Using a Bayesian Partial Pooling model, we estimate country-specific intercepts and slopes, focusing on non-linear temperature response functions. The results, based on data for up to 126 countries spanning from 1978 to 2023, indicate a higher demand for residential electricity and natural gas at temperatures below -5\celsius{} and a higher demand for electricity at temperatures above 30\celsius. For temperatures above 23.5\celsius, the relationship between power demand and temperature steepens. Demand in developed countries is more sensitive to high temperatures than in less developed countries, possibly due to an inability to meet cooling demands in the latter.
\end{abstract}
    \end{minipage}

    \vspace{1cm} % Space between keywords and abstract
 
 \begin{minipage}{0.95\textwidth} % Begin a minipage covering the full text width
    \end{minipage}

\newpage
\section{Introduction}
The past century has seen a substantial increase in global temperatures and future scenarios suggest further warming  (\citetalias{ipcc2023}) with significant implications for the energy sector. The energy sector, which is responsible for a substantial part of global greenhouse gas emissions, plays a dual role in both influencing and being influenced by climate change (\citetalias{ipcc2023}).
\paragraph{}
Our paper clarifies how temperature influences residential energy demand, emphasizing the sensitivity of energy use to temperature variations. Prior research has typically focused on the micro-level, considering socio-economic and geographic factors (see e.g. \citealt{tran2023}, for a recent review). This micro-perspective is advantageous for assessing particular policy measures and region-specific problems (e.g., identifying energy-poor households or analysing distributional effects of specific policies) and for understanding how energy demand responds to temperature changes in specific regions or countries. Macro-level studies, on the other hand, use data from multiple countries to gain insights into diverse energy uses, technologies, economic circumstances, and climates. This broader perspective is essential for future scenarios of climate change and the related economic consequences. We herefollow this approach.
\paragraph{}
Limited data availability is a major challenge for macro-level analyses. For example, \cite{decian2013} include data from 31 countries. Other studies cover more countries but have significant data gaps (e.g., \citealt{damm2017, ECM_enerdata, liddle2021}). To address this issue, our study employs a Bayesian Partial Pooling Model to make better use of the available data. This method shares information across countries and provides estimates for countries with limited data without relying solely on sparse information. It also allows for a more detailed analysis of the temperature and residential energy demand relationship by examining the distribution of key parameters. In addition, slopes vary between countries, enabling a detailed analysis of country-specific deviations from the global mean.
\paragraph{}
%We begin our analysis with the application of the often-used LSDVC Estimator.
Our dataset covers data for 126 countries for the period 1978 to 2023 and, like previous studies, includes temperature, income, and price data as predictors \citep{ECM_enerdata,liddle2021}. In contrast to previous studies, we abandon the heating and cooling degree specification in favor of multiple temperature intervals, to better capture the non-linear relationship between temperature and residential energy demand. However, using multiple prior structures, we also assess the often theorized V- or Hockey-stick-shaped temperature response function of residential energy demand (\citealt{response_functions}).
%The results of the LSDVC specification feed into our Bayesian model, which uses uninformative priors for various effects and variances.
\paragraph{}
The paper is organized as follows: Section 2 reviews the relevant literature. Section 3 outlines our methodology, including the determinants of energy use and temperature intervals. Section 4 describes the data. Section 5 presents our findings, including the effects of heating and cooling and additional analyses. Section 6 discusses the results and concludes.

\section{Prior research}
The majority of contributions to date have addressed the topic of temperature sensitivity of residential energy demand either at a micro level (see e.g., \citealt{tran2023}, for a recent review) or using country or regional time-series data (e.g. \citealt{assadorian2008}). These studies concentrate on specific countries without aiming at large-scale representativeness. Their methods vary widely, as do the results. In addition, the role of temperature is often not a central focus of the analysis.
\paragraph{}
Studies using multi-country panel data are less common. Most of these studies focus on industrialized countries such as G7, OECD or EU countries (e.g., \citealt{Bigano2006,bessec2008,Eskeland2010,Sihvola2010,cialani2018,castanorosa2021,emenekewe}). Studies that also cover non-OECD countries are few (\citealt{lescaroux2011,decian2013,ECM_enerdata,liddle2021}) and vary widely in their coverage of countries. \cite{lescaroux2011}, for example, covers 101 countries and three regional aggregates for the period 1960-2006. \cite{decian2013} include 26 OECD and five non-OECD countries for the period 1978-2000. In terms of the type of energy, most of the studies focus on electricity demand (e.g., \citealt{bessec2008,damm2017,emenekewe}) and only few consider multiple energy types (e.g, \citealt{Bigano2006, chen2016, ECM_enerdata}). While there are a few studies that utilize daily electricity loads instead of annual data (e.g., \citealt{damm2017,wenz2017}), these studies cannot differentiate between sectors, but analyse countries' total electricity demand. None of the latter studies cover non-OECD countries. \cite{damm2017} covers 26 OECD countries for the period 2006-2013 while \cite{wenz2017} includes 35 OECD countries for the period 2006-2012.
\paragraph{}
These studies differ not only in terms of country coverage, coverage of energy type and time period, but also in terms of their econometric approach (e.g., error-correction model, multivariate regression model) and how they specify the relationship between temperature and energy demand. The temperature response function reflects how a household’s energy demand changes with temperature  (\citealt{response_functions}). Approaches applied so far include, e.g., average annual temperature, average seasonal temperature, and degree days. Another approach is to account for non-linearities in the response to temperature by clustering the sample into groups (e.g., temperate or tropical, depending on the baseline temperature level of the country; see \citealt{ECM_enerdata}).
\paragraph{}
Most multi-country panel studies use heating degree days (HDDs) and cooling degree days (CDDs). The degree-days approach defines a temperature range or comfort zone (e.g., 17\celsius–22\celsius), in which neither heating nor cooling is required. Cooling or heating therefore is only required when outdoor temperatures fall outside the comfort zone. For example, \cite{Eskeland2010} use the concept of HDD and CDD to analyse residential electricity demand in 31 European countries. However, the degree-day approach has been subject to criticism due to its a priori and sometimes arbitrary choice of threshold values (\citealt{bessec2008}).
\paragraph{}
To address this issue, we follow the micro-level literature on residential energy demand (e.g., \citealt{aufhammer2011}) and use temperature bins, i.e. exposure to different temperature ranges, to model annual energy demand. For each temperature bin, a separate coefficient is estimated. In this way, without imposing a specific functional form, the shape of the response functions can be identified from the data. However, this approach is quite data-intensive which may be one reason why it has not been used more extensively in macro-level analyses.

\section{Modeling Determinants of Residential Energy Demand}
To account for heterogeneous temperature levels within countries and years, we constructed a measure of regionalized temperature exposure. Ignoring the geographical distribution of the population within a country would lead to an inaccurate measure of a country's temperature exposure. Our measure takes account of the fact that the population is very unevenly distributed in countries such as Canada and Russia. Using gridded temperature and population data, we constructed a temperature exposure index, which measures the average fraction of people living in each country in each year who are exposed to a given temperature interval. 
\paragraph{}
This approach is formalized as follows. Let $T_{i,j,h,t}$ be the $h$th three-hourly mean temperature of grid-cell $j$ in country $i$ in year $t$ and denote $p_{j,t}$ as the corresponding population count of grid-cell $j$ for that year.\footnote{Note that most climate impact studies us \emph{fixed} population weights, typically the first or final year of observation. Our approach more accurately reflects the climate change experienced by the people in a country rather than the warming experienced by the atmosphere \citep{Tol2017}.} Furthermore, denote $b(.)$ as the function that assigns a temperature record to the corresponding bin $k=1,...,K$. Let $J_i$ denote the set of all grid-cells in country $i$ and $I_{b(T_{i,j,h,t})}(k)$ be the function that indicates that the temperature record $T_{i,j,h,t}$ is assigned to bin $k$.
\paragraph{}
If we now define the population weight $w_{i,j,t}=\frac{p_{j,t}}{\sum_{j \in J_i}p_{j,t}}$ we can write the fraction of people living in the country $i$ who were exposed to temperatures in the range of bin $k$ in the time slot $h$ in the year $t$ as 
\begin{align*}
    f_{i,h,t}^k=\sum_{j \in J_i}w_{i,j,t}I_{b(T_{i,j,h,t})}(k) .
\end{align*}
Averaging this index for each year
\begin{align*}
    F_{i,t}^k=\frac{1}{H_t} \sum_{h=1}^{H_t} f_{i,h,t}^k 
\end{align*}
yields our indicator variable for each bin $k$ which we use to estimate the impact of temperature changes on residential energy demand, where $H_t$ is the total count of three-hour time-periods in a given year \footnote{For our analysis we focus on the temperatures between 6 am to 9 pm since these are the hours in which residents actively control indoor heating and cooling.}.
\paragraph{}
For our estimation, we chose nine different bin configurations of distinct granularity. The temperature bins are defined in the range of -5\celsius{} to 30\celsius, with two additional outer bins to capture everything below and above this range. We choose for these specifications the bin widths 1\celsius{} to 5\celsius{} in 0.5\celsius{} increments to capture non-linear effects while still remaining somewhat parsimonious. For a geographic visualization of the average index values for the 3.5\celsius{} bin width specification, see Figure \ref{fig:geo_vis}. 

%Previous studies suggest that the effect of temperature on residential energy demand is nonlinear and it is expected that for high temperatures when cooling demand increases, a positive effect on energy consumption can be observed see \cite{response_functions,review}. In addition, energy demand is elevated for lower temperatures as well, reflecting heating demand. This effect is also observable for electricity consumption. Conventional electric heating has a high prevalence in e.g Scandinavian countries and countries with warmer climates, which lack conventional heating equipment and instead use electric acclimatization devices to raise indoor temperatures during colder months.

\paragraph{}
The advantages of using Bayesian Hierarchical Models, which allow for partial pooling and borrowing across individuals, have made them increasingly popular for conducting meta-analyses and aggregating results from multiple studies, especially in medicine and psychology, but also in economics (compare e.g. \citealt{meta_depression,psychology_multilevel,meta_economics}). Others have started to apply this model to single empirical studies, such as \cite{Electricity_bayes}.
\paragraph{}
As we expect that temperature responses differ between countries, we build on these advancements and apply a Partial Pooling Model, which not only allows to estimate individual intercepts, but also individual temperature responses for each country. In addition, a population-wide intercept and population-wide temperature responses are estimated. The model is specified as follows:
\begin{align*}
\ln(y_{i,t}) &\sim \text{Normal}(\mu_{i,t}, \sigma_e) \\
\mu_{i,t} &= \nu \mu_{i,t-1}+ \alpha + \alpha_{i} + \sum_{k=1}^K [(\beta_k +\beta_{i,k}) F_{i,t}^k]+\boldsymbol{\gamma} \boldsymbol{X}_{i,t} \\
\begin{bmatrix}
\alpha_{i} \\
\beta_{i,1}\\
\vdots\\
\beta_{i,K}
\end{bmatrix}
&\sim \text{MVNormal}
\left(
\boldsymbol{0}_{1 \times (K+1)} 
, \boldsymbol{\Sigma}
\right)
\\
\boldsymbol{\Sigma} &= \textbf{SRS}
\\
\alpha &\sim \text{Normal}(0, 1) \\
\beta_k &\sim \text{Normal}(0, 1) \quad \forall k \in 1,\dots, K \\
\gamma_l &\sim \text{Normal}(0, 1)  \quad \forall l \in 1,\dots,L  \\
\nu &\sim \text{Normal}(0, 1)\\
\sigma_e &\sim \text{t}_{3}(0,1)^{+}\\
\textbf{S} &= \text{diag}(\sigma_{\alpha_i},\sigma_{\beta_{i,1}},\dots,\sigma_{\beta_{i,K}})
\\
\sigma_{a_i} &\sim \text{Normal}(0, 1)^{+} \quad \forall a_i \\
\sigma_{\beta_{i,k}} &\sim \text{Normal}(0, 1)^{+} \quad \forall \beta_{i,k}\\
\textbf{R} &\sim \text{LKJ}(2)
\end{align*}
The dependent variable, $ \ln(y_{i,t}) $, represents the natural logarithm of per capita residential energy demand in country $i \in N$ in year $t \in T$. The likelihood for each individual observation is modeled to be normally distributed, characterized by a mean ($ \mu_{i,t} $) and a standard deviation ($ \sigma_e $). 
\paragraph{}
The mean $ \mu_{i,t} $ integrates various factors that reflect both the overarching effects of the population and the dynamics of the individual groups. By convention (compare, for example, \citealt{ECM_enerdata,liddle2021}) we model variations in residential energy demand by variations in income, energy prices, and temperature. Energy is a normal good: its demand increases with income and decreases with its own price. Households are assumed to take one period to adapt to price changes, so the price variable is lagged by one period.
\paragraph{}
The term $ \alpha $ represents the baseline intercept for the entire population, $ \beta $ and $ \gamma $ quantify the overall population impacts of our primary predictors. The temperature indices for the bin $k \in K$ in country $i$ in year $t$, denoted as $F_{i,t}^k$, are crucial for assessing the influence of temperature variations on residential energy demand. Therefore, the estimates of $\beta_k$ and $\beta_{i,k}$ are the main focus of this study. Additionally, $X_{i,t} $ accounts for $L$ other covariates, mainly per capita GDP and energy prices, and their broad population effects. $\gamma_l, \beta_k, \nu $ and $ \alpha $ have normal priors, which are weakly informative, based on the expectation that most estimates tend to be close to zero, rationalizing the central positioning of these priors. Additionally, a one-period lag of the dependent variable is added to capture some of the persistent but time-varying explanatory power that is not accounted for, and to distinguish between temporary and sustained changes (\citealt{koyck1954}).
\paragraph{}
In addition to temperature, income, and prices, numerous factors influence the demand for household energy. These factors are often unobserved due to their ideosyncratic nature or limited data availability. This is particularly evident when examining energy demand across different countries. To overcome this challenge, we utilize a panel data set that allows us to account for time-invariant, country-specific, unobserved factors that affect energy demand. These country-specific variations are captured through $ \alpha_{i} $ and $ \beta_{i,k}$, representing country-specific intercepts and slopes linked to temperature-related predictors. We hypothesize a correlation between these country-specific parameters, modeling their prior to follow a multivariate normal distribution with zero mean and covariance matrix $ \Sigma $.
\paragraph{}
Group-level standard deviations are modeled using weakly informative half-normal distributions. $\sigma_e$ follows a half t prior with 3 degrees of freedom, as per the BRMS package, with parameterization based on the data \citep{prior}. Lastly, the correlation structure between $ \alpha_{i} $ and $ \beta_{i,k} $ is modeled by $\boldsymbol{R}$ which is defined as a $(N+1) \times (N+1)$ matrix with ones on the diagonal and correlation coefficients on the off diagonals. $\boldsymbol{R}$ is assigned an $ LKJ(2) $ prior \citep{lkj}. This weakly-informative prior puts a high weight on no correlations so that any posterior correlations will not be due to prior assumptions \citep{prior}.
\paragraph{}
With this structure for our model priors, we remain agnostic, while allowing for some regularization of key parameters. Despite our mostly weakly informative priors, this set-up can make use of the "borrowing strength" property \citep{rethinking}. Intuitively speaking, we use information on the general relationship between temperature and energy demand to inform the individual country estimates. This approach allows us to use data more efficiently and obtain meaningful estimates even for countries with relatively low data coverage. For countries with very sparse data, the population mean dominates and the estimates are shrunk towards this mean \citep{rethinking}.
\paragraph{}
The Bayesian model is estimated with the R-package BRMS which serves as an interface to the probabilistic programming language STAN while still providing intuitive lmer syntax. In our case, the No-U-Turn Sampler (NUTS) is used to obtain the draws from the posterior distribution \citep{prior}. 
\newpage
\section{Data}
Data on residential electricity, natural gas, light fuel oil demand, the corresponding energy prices, and real per capita GDP as a proxy of income are retrieved from ENERDATA for the period 1978 to 2023. Data electricity demand data were retrieved for 126 countries, for a total of 3,261 observations. For natural gas demand, data for 58 countries was retrieved, summing to a total of 1523 observations. For light fuel oil demand, data for 45 countries is available, summing to a total of 1170 observations. A complete list of countries and their respective period counts, can be found in Figure \ref{fig:country_year} to \ref{fig:country_year_oil}. Summary statistics are presented in Table \ref{tab:summary}.
\paragraph{}
We use three-hourly average temperature values taken from the high-resolution gridded dataset of NASA Earthdata \citep{nasav2} available on a 0.25-degree grid. We transform the gridded temperature averages to the temperature exposure indices at the country level according to the procedure described in Section 3. Temperature data are available for most countries and for all years of interest. All temperature variables are expressed in degrees Celsius. These exposure indicators are regionalized using population data obtained from \citetalias{population_data,population1,population2}.
% Title                                                              Unit   
% -------------------------------------------------------------------------
% Natural gas final consumption in residential sector                ktoe   
% Electricity final consumption in residential sector                ktoe   
% Diesel final consumption in residential sector                     ktoe   
% Price of electricity residential (USD2015/toe incl. taxes)         $15/toe
% Price of natural gas residential (USD2015/toe incl. taxes)         $15/toe
% Price of light fuel oil (LFO) residential (USD2015/toe incl. taxes) $15/toe
% GDP per capita (constant local currency)                           kUS$15 
% Population                                                         k     

%ktoe were convertet to toe
\begin{landscape}
\begin{table}[!htpb]
\begin{tabular}{@{\extracolsep{3pt}}lccccc} 
\\[-1.8ex]\hline 
\hline \\[-1.8ex] 
Variable & N & Mean & St. Dev. & Min & Max \\
\hline \\[-1.8ex] 
Electr.Demand in toe/ k capita & 7,041 & 68.006 & 102.956 & 0.140 & 738.120 \\
Nat. Gas Demand in toe/ k capita & 2,817 & 111.223 & 139.494 & 0.001 & 806.645 \\
Light Oil Demand in toe/ k capita & 2,164 & 71.944 & 134.501 & 0.002 & 1,104.070 \\
GDP/capita in k 2015 USD & 7,674 & 10.753 & 16.309 & 0.001 & 112.418 \\
Electr. Price in 2015 USD/toe & 3,412 & 1,643.744 & 931.545 & 40.821 & 9,285.304 \\
Nat. Gas Price in 2015 USD/toe & 1,680 & 664.977 & 426.117 & 3.926 & 3,073.777 \\
Light Fuel Oil Price in 2015 USD/toe & 1,752 & 859.755 & 394.035 & 37.006 & 2,384.482 \\
$< -5\celsius$ & 7,674       & 0.017 & 0.048 & 0.000 & 0.423 \\
$[-5, -1.5)\celsius$ & 7,674 & 0.016 & 0.028 & 0.000 & 0.144 \\
$[-1.5, 2)\celsius$ & 7,674  & 0.028 & 0.044 & 0.000 & 0.244 \\
$[2, 5.5)\celsius$ & 7,674   & 0.040 & 0.053 & 0.000 & 0.314 \\
$[5.5, 9)\celsius$ & 7,674   & 0.049 & 0.058 & 0.000 & 0.287 \\
$[9, 12.5)\celsius$ & 7,674  & 0.060 & 0.062 & 0.000 & 0.336 \\
$[12.5, 16)\celsius$ & 7,674 & 0.075 & 0.063 & 0.000 & 0.274 \\
$[16, 19.5)\celsius$ & 7,674 & 0.099 & 0.066 & 0.000 & 0.425 \\
$[19.5, 23)\celsius$ & 7,674 & 0.143 & 0.095 & 0.000 & 0.712 \\
$[23, 26.5)\celsius$ & 7,674 & 0.182 & 0.135 & 0.000 & 0.889 \\
$[26.5, 30)\celsius$ & 7,674 & 0.142 & 0.117 & 0.000 & 0.659 \\
$> 30\celsius$ & 7,674       & 0.141 & 0.165 & 0.000 & 0.719 \\
\hline \\[-1.8ex] 
\end{tabular} 
\caption{Summary Statistics of economic and climate variables.
\scriptsize{Data from ENERDATA; \citetalias{nasav2}, \citetalias{population_data}, \citetalias{population1}, \citetalias{population2}.}}
\label{tab:summary}
\end{table}
\end{landscape}
%From Table \ref{tab:summary} it can read, that the mean value for the temperature index is highest for moderate temperatures in the range of 23\celsius{} to 26.5\celsius. In general it can be observed that mean exposure to colder temperatures is lower. Also, the summary statistics suggest that the distribution of the temperature index has longer tails for relatively warm temperatures. The highest maximum value for a temperature index is 0.889 for 23\celsius{} to 26.5\celsius. These observations reflect the tendency for humans to settle in areas with relatively warm climates like central Europe and north America or even hot climates like South East-Asia. The same picture emerges when looking at the geographical visualization for the temperature index in Figure \ref{fig:geo_vis}.
%\paragraph{}
%Furthermore, it is interesting to note that it can be read from the data that during the time frame of this study about 60\% of the world population was on average exposed to temperatures above 19.5\celsius. About 28\% where on average exposed to temperatures exceeding 26.5\celsius{} possibly requiring cooling. Almost the same percentage was exposed to temperatures below 16\celsius. This directly reflects the settlement choices of humans. The large portion living in relatively high temperature regions can likely be explained by the large populations in countries with tropic or sub tropic climates. In addition, even in milder climates, cities tend to be heat islands, further increasing the proportion of people exposed to hot temperatures \cite{Deilami2018}.
\section{Results}
This section presents the results of our Bayesian Hierarchical Model used to understand the relationship between temperature and residential energy demand. The model formalizes the assumption that the response functions to temperature is neither completely independent across countries nor exactly the same. This leads to partial pooling by making use of the information sharing property of hierarchical Bayes models. In this way, more accurate estimates of the heating and cooling effects can be obtained compared to, e.g., panel fixed effects analyses. To ensure the validity and reliability of the results, we first assess the model's diagnostics.
\subsection{Model Diagnostics}
\begin{figure}[H]
    \centering
    % First minipage for the first graphic
    \begin{minipage}{0.48\textwidth}
        \centering
        \includegraphics[width=\textwidth]{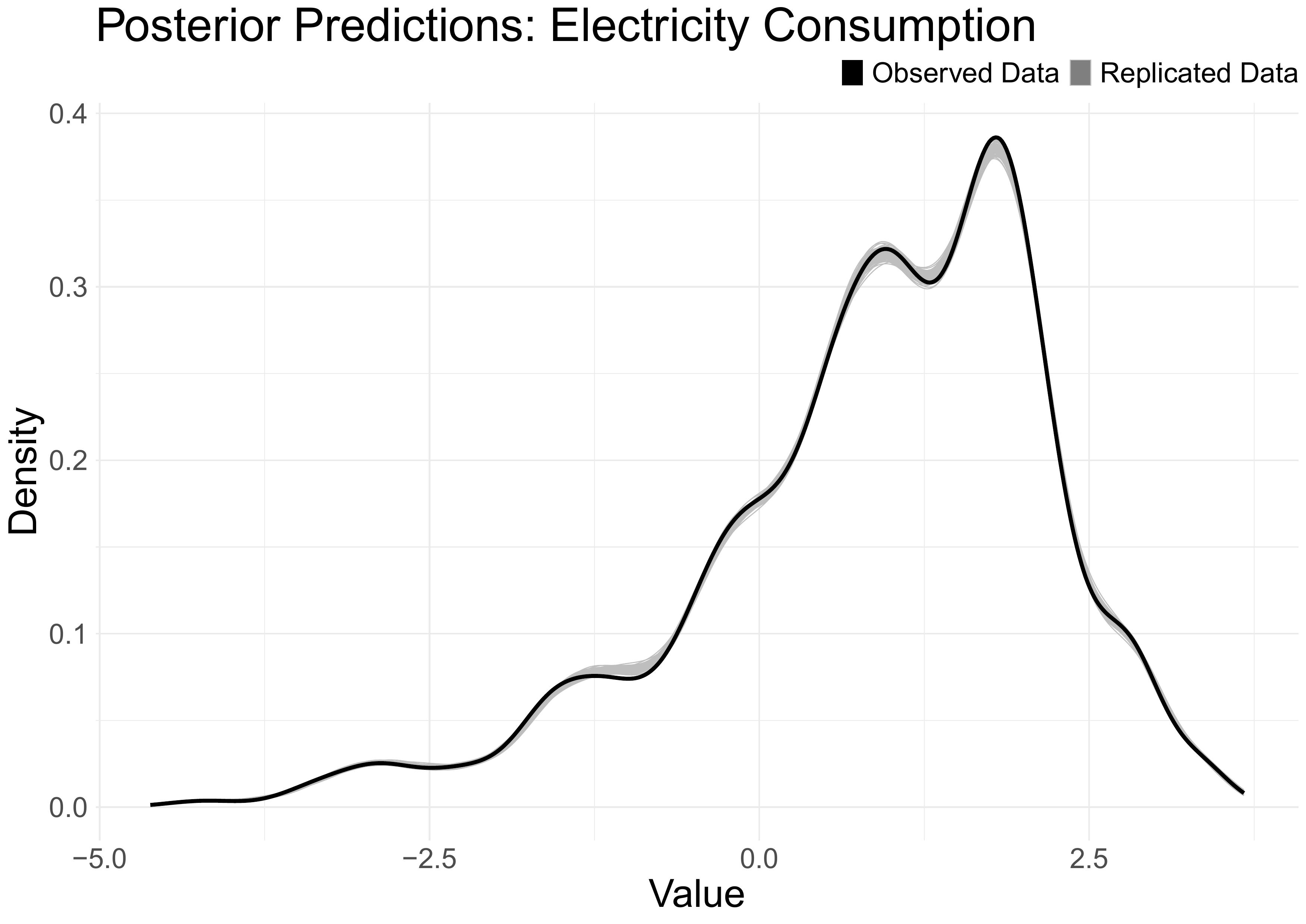}
        \caption{}
        \label{fig:postpcheck_new}
    \end{minipage}
    % Second minipage for the second graphic
    \begin{minipage}{0.48\textwidth}
        \centering
        \includegraphics[width=\textwidth]{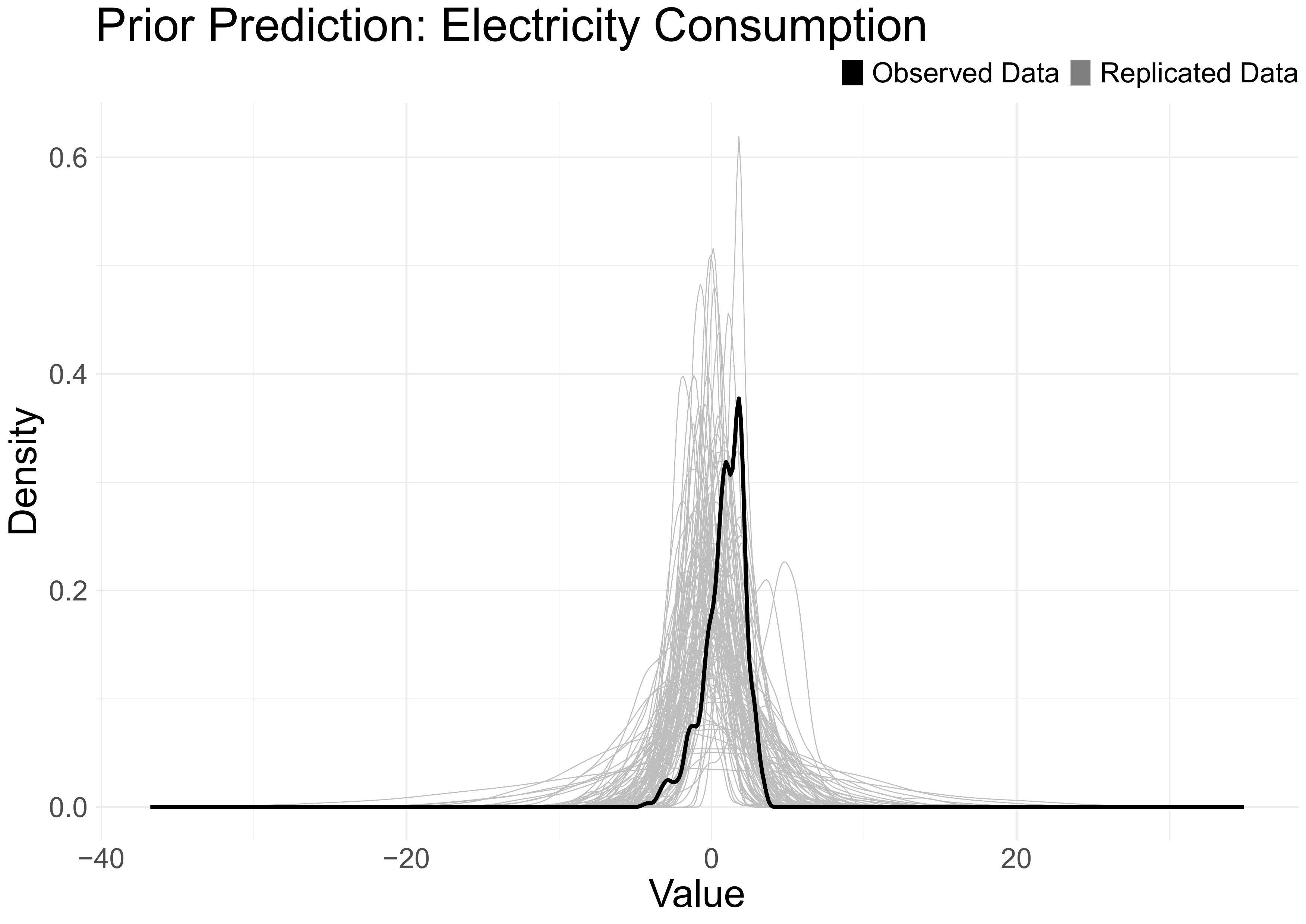}
        \caption{}
        \label{fig:priorpcheck_new}
    \end{minipage}
    \caption*{\scriptsize{Estimated using Bayesian Partial Pooling Model (NUTS-sampler). Data from ENERDATA; \citetalias{nasav2}, \citetalias{population_data}, \citetalias{population1}, \citetalias{population2}.}}
\end{figure}

To check model fit, we use the joint posterior distribution derived from our prior model and the available data, to generate predictions and subsequently evaluate them against the actual data points. A comparison of the empirical distribution of our dependent variable with the corresponding distribution of the predicted values shows that our model adequately captures all the essential characteristics exhibited by the empirical distribution of the dependent variable, indicating a good model fit. For a visual representation of the posterior predictive checks see Figure \ref{fig:postpcheck_new} for electricity demand.
\paragraph{}
Figure \ref{fig:priorpcheck_new} shows prior predictive simulations for electricity. By simulating the values of the dependent variable using only the prior structure without the likelihood, and then comparing them with the actual data, it can be seen that the priors cover the range of plausible values for our dependent variable, without giving much weight to implausibly low or high values. A similar analysis for natural gas and light fuel oil can be found in Figure \ref{fig:predictive_checks}.

\begin{table}[H]
\centering
\begin{tabular}{lrrcc}
\hline
\textbf{Model}            & \textbf{ELPD} & \textbf{s.e.} & \textbf{Random Intercepts} & \textbf{Random Slopes} \\
\hline
Model (1)       & 0.0                 & 0.0               & $\checkmark$               & $\checkmark$           \\
Model (2)          & -46.5               & 11.6              & $\checkmark$               &                        \\
Model (3)              & -85.1               & 24.0              &                            &                        \\
\hline
\end{tabular}
\caption{Model comparison using LOO-CV ELPD differences, standard errors, and model characteristics.}
\label{tab:model_comparison_features}
\end{table}
To evaluate whether the addition of random intercepts and slopes improves the quality of the model, we compare the predictive performance in Table \ref{tab:model_comparison_features}. It shows the results of the comparison of out-of-sample prediction performance using leave-one-out cross-validation for the three models, measured as the expected log-predictive density (ELPD). The first model (model 1) includes both individual intercepts and individual slope estimates for the temperature response. The second model (model 2) only includes individual intercepts, and the third model (model 3) does not estimate any individual-level effects. The difference in expected log-predictive density indicates that the model with both individual intercepts and slopes performs best. Both models 2 and 3 significantly deviate negatively from model 1, indicating a poorer predictive performance of these models.
\paragraph{}
The $\hat{R}$ indicator is used to assess the convergence of the sampling algorithm. Values close to one indicate good convergence (\citealt{Gelman1992}). The $\hat{R}$ values as well as the parameter estimates are shown in Table \ref{tab:maintable}

\subsection{Bayesian Estimates of Global Temperature Response}
\begin{figure}
    \centering
    \textbf{Electricity Demand}
    \\
    \includegraphics[width=0.63\linewidth]{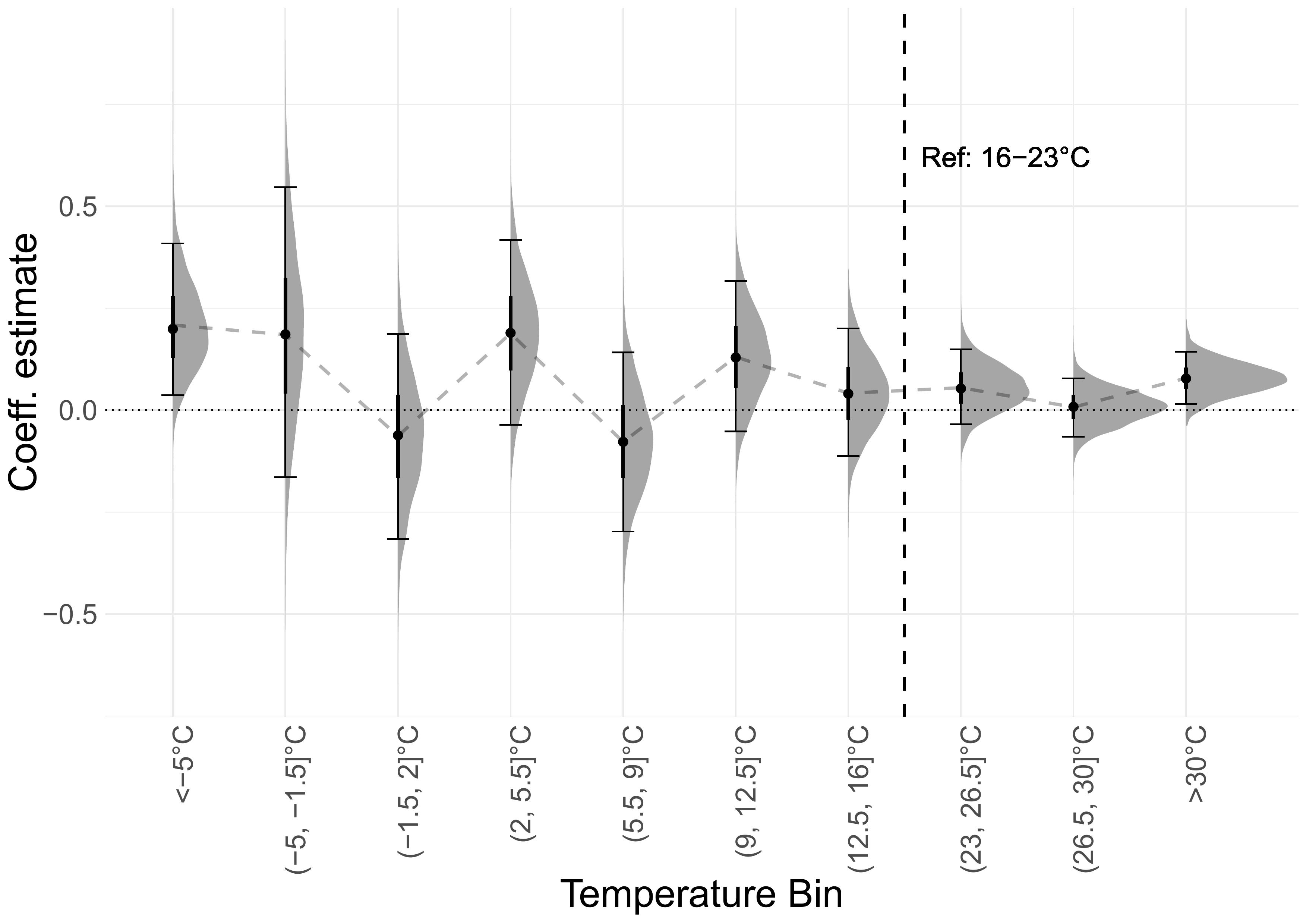}
    \vspace{0.1cm} % Space between images
    
    \textbf{Natural Gas Demand}
    \\
    \includegraphics[width=0.63\linewidth]{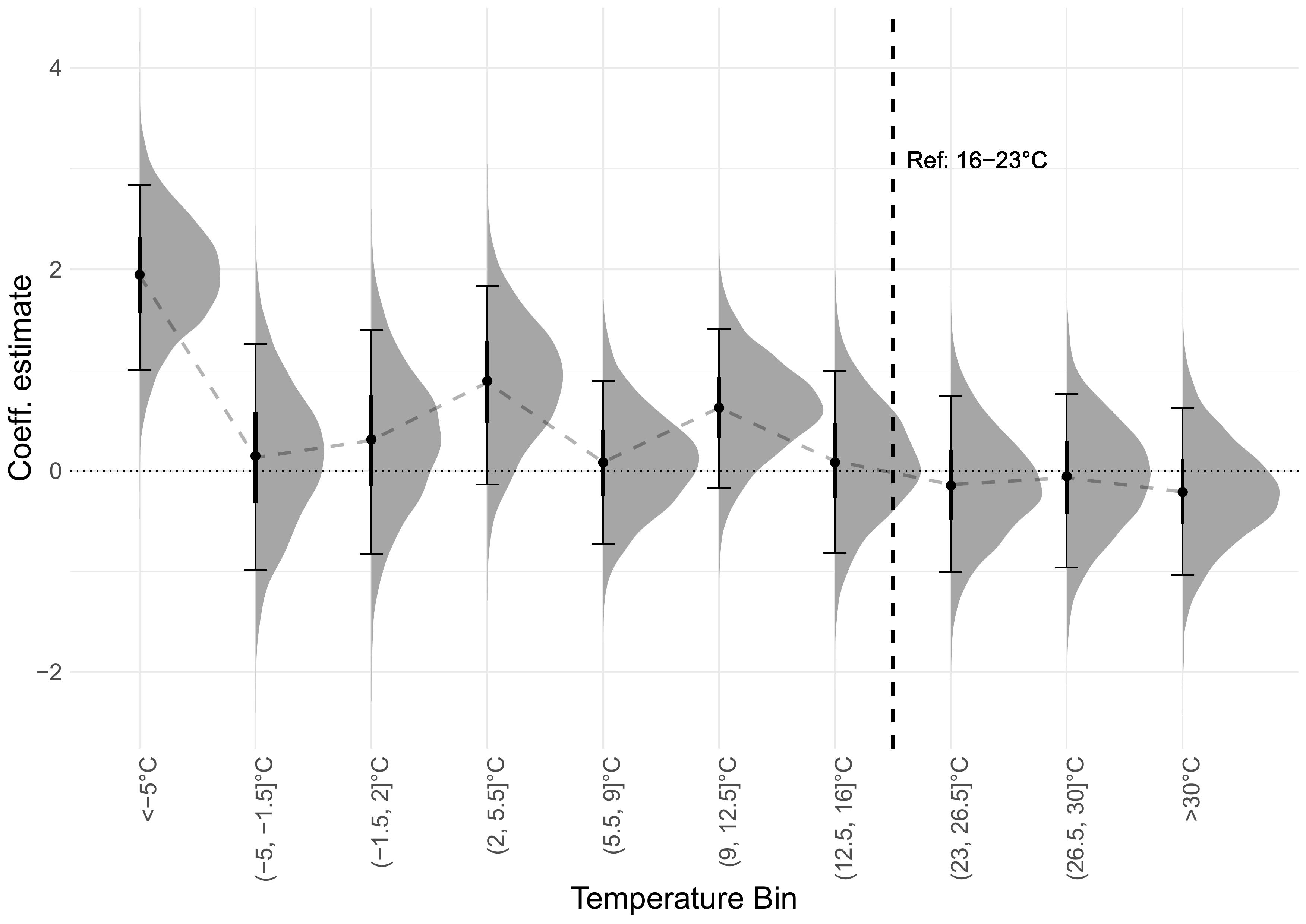}
    \vspace{0.1cm} % Space between images
    
    \textbf{Light Fuel Oil Demand}
    \\
    \includegraphics[width=0.63\linewidth]{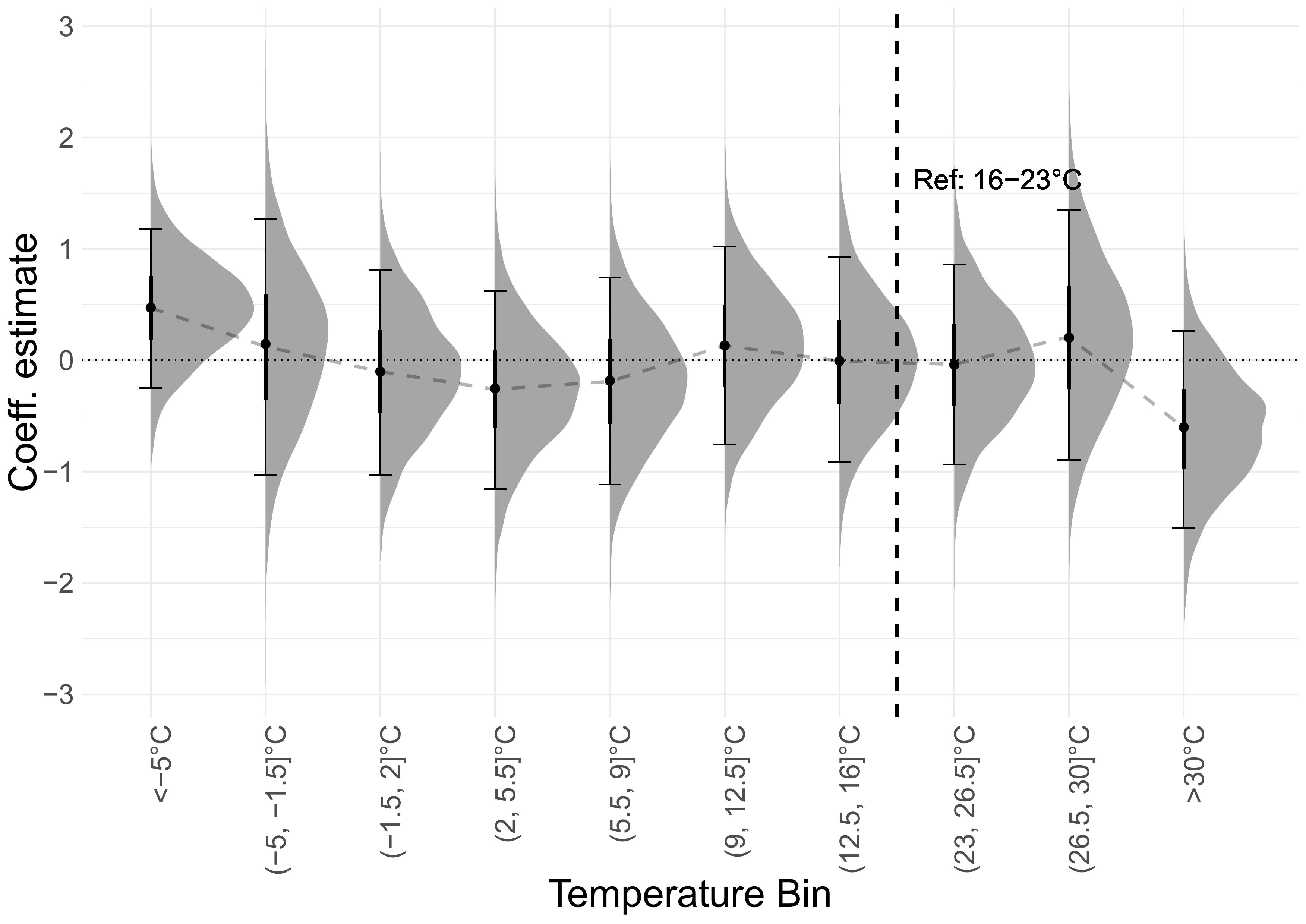}
    
    \caption{Estimated impact of a shift of temperature exposure of the population for ten different temperature bins (\celsius{}), relative to the 16\celsius{} to 23\celsius{} bin, for log residential electricity, natural gas and light fuel oil demand, using a 3.5\celsius{} bin width and including 90\% and 50\% credible intervals.
    \scriptsize{Estimated using Bayesian Partial Pooling Model (NUTS-sampler). Data from ENERDATA; \citetalias{nasav2}, \citetalias{population_data}, \citetalias{population1}, \citetalias{population2}.}}
    \label{fig:temperature-effects}
\end{figure}

The results for the global temperature effects based on the specification detailed in Section 3 are shown in Figure \ref{fig:temperature-effects}.
Note that for our main specification we choose a bin width of 3.5\celsius{} which marks the midpoint used in previous studies (\cite{aufhammer2011,ECM_enerdata,replicated_study,wenz2017}).

\paragraph{}
Since the Bayesian framework estimates the distribution of parameters of interest rather than a single-points, the coefficient plots capture multiple characteristics of the estimates. The center line of the plots visualizes the posterior mean, while the inner and outer bounds represent the 50\% credibility interval (CI) and the 90\% interval, respectively.   
\paragraph{}
The posterior densities in Figure \ref{fig:temperature-effects} show that only the outer temperature bins can be confidently distinguished from zero. For natural gas a rather strong heating effect for temperatures below -5\celsius{} is present. The same effect, but somewhat smaller, is visible for residential electricity demand. For light fuel oil the effect is absent. Only residential electricity demand seems to respond to high temperatures, indicating a small cooling effect that increases residential electricity demand when the temperature rises above 30\celsius.
\paragraph{}
Looking at Table \ref{tab:maintable}, we can evaluate point estimates, such as the posterior mean for the effect of the coldest temperature bin. Starting with the posterior mean for residential electricity demand ($0.21$) and given the definition of our temperature index, the results indicate that a 10 percentage point increase in the proportion of people exposed to the coldest temperature bin is associated with an expected increase in electricity demand of $2.1\%$. The estimated cooling effect for temperatures above 30\celsius{} is less than half this, with an expected increase in residential electricity demand of $0.8\%$.
The posterior mean for the auto-regressive parameter $\nu$ can be used to calculate the long run impact of temperature changes \citep{koyck1954}. For electricity, the parameter estimate is $0.96$ (see Table \ref{tab:maintable}), highlighting the typically strong persistence of electricity demand. Even a relatively small short-term change in temperature can lead to significant changes in the long run.
Assuming constant adaptation patterns, a 10 percentage point increase in the population in the coldest temperature bin would result in a long run increase in residential electricity demand of around $50\%$. For the hottest temperature bin, demand would increase by $20\%$ respectively. 
\paragraph{}
For natural gas, a short-term increase in the proportion of people exposed to temperatures below -5\celsius{} by 10 percentage points is associated with an expected increase in residential natural gas demand of 19.4\% (compare Table \ref{tab:maintable}). Assuming again constant adaptation patterns, this would accumulate in the long run to an increase of 114.1\%. The high uncertainty around the estimates for the temperature response of light fuel oil makes it impossible to draw any conclusions about a heating or cooling effect. 
\paragraph{}
Table \ref{tab:maintable} also provides information on the estimated income and own-price elasticities. Electricity demand exhibits an income elasticity of $0.03$ and a price elasticity of $-0.01$. The long run income and own-price elasticities are $0.75$ and $-0.25$, respectively. For natural gas demand, the short-term income elasticity is estimated at $0.17$, increasing to $1.00$ in the long run. The corresponding price elasticities are $-0.05$ in the short term and $-0.29$ in the long run. Regarding demand for light fuel oil, the short-term income elasticity is $0.01$, while in the long run, it rises to $0.34$. The price elasticity is $-0.04$ in the short term \citep[cf.][]{Meta_price_income} and significantly more elastic at $-1.34$ in the long run. 
\paragraph{}
Another interesting result (see lower half of Table \ref{tab:maintable}), is that, after controlling for the covariates, for electricity the only temperature effects that seem to differ with a high probability between countries are those for the lowest and the highest temperature bins and for the range of 23\celsius{} to 26.5\celsius. For these, the 2.5\% quantiles of the posterior distribution for the group-level standard deviation are $0.03$ to $0.07$. This indicates that when evaluating differences in terms of the temperature response of residential electricity demand across countries, the focus should be on the outer bins as this is where the heterogeneities become apparent. For natural gas and light fuel oil temperature effects appear to vary for any temperature range across countries. Posterior group-level standard deviations are large for natural gas.

\subsection{Robustness Checks}
To test the robustness of our results, we examine the effect of different prior specifications, the evolution of the parameter estimates over time, and alternative bin widths. In addition, we estimate a fixed-effects panel model that is more in line with the existing literature (\citealt{ECM_enerdata,emenekewe,Eskeland2010,replicated_study}) to compare our results.
\paragraph{}
The results for residential electricity demand are robust to different prior specifications including multiple specifications which impose priors on the global as well as on the country level with varying degrees of tightness. Changing the shape parameter of the LKJ prior and thus regulating the amount of correlation between parameters does not influence results either (Figure \ref{fig:prior_sensitivity}). The \emph{Strong V-shape} prior is of particular interest, as it formalizes a common hypothesis on the temperature response function of residential energy demand. Here, the prior probabilities for each temperature bin are affected in such a way that a relatively strong V-shape temperature response is suggested a priori\footnote{Moving away from the reference bin, the prior mean was increased by 0.5 from zero in both directions}. 
Looking at the graph, it can be seen that even this strong V-shaped prior does not alter the results in any meaningful way. The only prior choice that pulls the estimates for the heating and cooling effect closer to zero is the one that imposes extremely small standard deviations for the country level effects. This leads to very strong shrinkage.
\paragraph{}
The same priors were tested for natural gas and light fuel oil demand; however, since for these fuel types a cooling effect is not expected, the V-Shaped prior was replaced by a hockey-stick prior. As can be seen in Figures \ref{fig:prior_sensitivity_gas} and \ref{fig:prior_sensitivity_oil}, the results for these two types of energy are more sensitive to prior specifications. This is likely due to less available data and stronger heterogeneity between countries. It can be seen that the hockey-stick prior as well as very wide priors lead to much higher estimates for the heating effect for both energy types.
\paragraph{}
To examine the evolution of the parameter estimates over time, we use a rolling window analysis, as shown in \ref{fig:window_electricity}. With a window size of 15 years, it can be observed that for electricity demand the heating effect dominates in the early periods and the cooling effect appears only after the year 2004 and increases in the most recent periods. This is likely a result of technological progress and subsequent electrification, as well as the fact that earlier years only include developed countries with mostly temperate or cold climates.
The heating effect for natural gas, depicted in Figure \ref{fig:window_gas}, is also visible for moderate temperatures, in line with the literature. It appears from the year 2000 onward. The results for light fuel oil demand remain inconclusive; see Figure \ref{fig:window_oil}. No effects were found.
\paragraph{}
Turning to the results for alternative bin widths (Figure \ref{fig:alternative_coeff_bayes_ele} for residential electricity and Figure \ref{fig:alternative_coeff_bayes_gas} for natural gas demand\footnote{For natural gas the 1\celsius{} bin width specification did not converge, so no estimates for the posterior densities were obtained.}). The quantitative results are somewhat sensitive to the bin width specification, and estimates tend to be unstable for small bin widths. However, the qualitative interpretation is the same across all bin width specifications, indicating a heating effect for residential electricity and gas demand as well as a cooling effect for electricity demand. For light fuel oil demand, no bin width specification yields interpretable results (Figure \ref{fig:alternative_coeff_bayes_oil}).
\paragraph{}
To investigate how the results based on the Bayesian Partial Pooling Model relate to results based on the common panel fixed-effects approach, we reproduced the study of \cite{replicated_study} which analyses residential electricity demand in the USA. We chose this study because its framework is straightforward and similar to our specification which eases comparison. Figure \ref{fig:replication_study1} shows that using standard techniques such as fixed effects yields very different results for the temperature response of residential electricity demand. The fixed effects estimation suggests a heating effect already at temperatures below 10\celsius{}, which increases with lower temperatures. All coefficients for the temperature bins below the reference bin are statistically significant. A detailed presentation of the results and a comparison with the partial pooling approach can be found in Appendix \ref{appendixD}.

\begin{landscape}
\begin{table}[htbp!]
\centering
\setlength{\tabcolsep}{3pt}
\begin{tabular}{lrrrrrr|rrrrrr|rrrrrr}
\toprule
\textbf{Parameter} & \multicolumn{6}{c}{\textbf{Electricity}} & \multicolumn{6}{c}{\textbf{Natural Gas}} & \multicolumn{6}{c}{\textbf{Light Fuel Oil}} \\
\cmidrule(lr){2-7} \cmidrule(lr){8-13} \cmidrule(lr){14-19}
 & $\hat{R}$ & Mean & SD & 2.5\% & Median & 97.5\% & $\hat{R}$ & Mean & SD & 2.5\% & Median & 97.5\% & $\hat{R}$ & Mean & SD & 2.5\% & Median & 97.5\% \\
\midrule
$\alpha$ & 1.00 & 0.04 & 0.00 & 0.03 & 0.04 & 0.04 & 1.02 & 2.51 & 0.22 & 2.07 & 2.52 & 2.92 & 1.00 & -0.08 & 0.05 & -0.18 & -0.08 & 0.02 \\ 
$\nu$ & 1.00 & 0.96 & 0.00 & 0.95 & 0.96 & 0.96 & 1.00 & 0.83 & 0.01 & 0.81 & 0.83 & 0.85  & 1.00 & 0.97 & 0.01 & 0.95 & 0.97 & 0.99 \\ 
$\beta_{below \, -5\celsius}$ & 1.00 & 0.21 & 0.12 & 0.01 & 0.20 & 0.46 & 1.00 & 1.94 & 0.56 & 0.80 & 1.95 & 3.00 & 1.00 & 0.47 & 0.44 & -0.39 & 0.47 & 1.33 \\ 
$\beta_{-5\celsius \,to \,-1.5\celsius}$ & 1.00 & 0.19 & 0.21 & -0.23 & 0.19 & 0.61 & 1.00 & 0.13 & 0.68 & -1.20 & 0.15 & 1.44  & 1.00 & 0.12 & 0.70 & -1.26 & 0.15 & 1.47 \\ 
$\beta_{-1.5\celsius \,to\, 2\celsius}$ & 1.00 & -0.06 & 0.15 & -0.36 & -0.06 & 0.24 & 1.00 & 0.30 & 0.67 & -1.05 & 0.31 & 1.62 & 1.00 & -0.10 & 0.55 & -1.19 & -0.10 & 0.96 \\ 
$\beta_{2\celsius \,to\, 5.5\celsius}$ & 1.00 & 0.19 & 0.14 & -0.09 & 0.19 & 0.46 & 1.00 & 0.88 & 0.60 & -0.31 & 0.89 & 2.01 & 1.00 & -0.26 & 0.54 & -1.34 & -0.25 & 0.80 \\ 
$\beta_{5.5\celsius\, to\, 9\celsius}$ & 1.00 & -0.08 & 0.13 & -0.34 & -0.08 & 0.17 & 1.00 & 0.08 & 0.49 & -0.85 & 0.08 & 1.06 & 1.00 & -0.19 & 0.57 & -1.28 & -0.18 & 0.93 \\ 
$\beta_{9\celsius \,to\, 12.5\celsius}$ & 1.00 & 0.13 & 0.11 & -0.08 & 0.13 & 0.35 & 1.00 & 0.62 & 0.47 & -0.32 & 0.62 & 1.53 & 1.00 & 0.13 & 0.54 & -0.92 & 0.13 & 1.18 \\ 
$\beta_{12.5\celsius \,to\, 16\celsius}$ & 1.00 & 0.04 & 0.09 & -0.14 & 0.04 & 0.23 & 1.00 & 0.10 & 0.55 & -0.98 & 0.08 & 1.17 & 1.00 & -0.01 & 0.56 & -1.08 & -0.01 & 1.14 \\ 
$\beta_{23\celsius \,to\, 26.5\celsius}$ & 1.00 & 0.06 & 0.06 & -0.06 & 0.05 & 0.17 & 1.00 & -0.14 & 0.53 & -1.17 & -0.15 & 0.90 & 1.00 & -0.04 & 0.54 & -1.10 & -0.04 & 1.01 \\ 
$\beta_{26.5\celsius \,to\, 30\celsius}$ & 1.00 & 0.01 & 0.04 & -0.08 & 0.01 & 0.09 & 1.00 & -0.07 & 0.53 & -1.12 & -0.06 & 0.91 & 1.00 & 0.21 & 0.68 & -1.09 & 0.20 & 1.54 \\ 
$\beta_{above\, 30\celsius}$ & 1.00 & 0.08 & 0.04 & 0.00 & 0.08 & 0.16 & 1.00 & -0.21 & 0.50 & -1.20 & -0.21 & 0.80& 1.00 & -0.61 & 0.54 & -1.67 & -0.60 & 0.46 \\ 
$\gamma_{log(GDP)}$ & 1.00 & 0.03 & 0.00 & 0.02 & 0.03 & 0.04 & 1.00 & 0.17 & 0.03 & 0.11 & 0.17 & 0.23& 1.00 & 0.01 & 0.03 & -0.04 & 0.01 & 0.07 \\ 
$\gamma_{log(Price_{t-1})}$ & 1.00 & -0.01 & 0.00 & -0.02 & -0.01 & -0.01 & 1.00 & -0.05 & 0.02 & -0.09 & -0.05 & -0.02 & 1.00 & -0.04 & 0.03 & -0.10 & -0.04 & 0.01 \\ 
\midrule
\multicolumn{18}{c}{\textbf{Group-Level Standard Deviation Estimates}} \\
\midrule
\textbf{Parameter} & \multicolumn{6}{c}{\textbf{Electricity}} & \multicolumn{6}{c}{\textbf{Natural Gas}} & \multicolumn{6}{c}{\textbf{Light Fuel Oil}} \\
\cmidrule(lr){2-7} \cmidrule(lr){8-13} \cmidrule(lr){14-19}
 & $\hat{R}$ & Mean & SD & 2.5\% & Median & 97.5\% & $\hat{R}$ & Mean & SD & 2.5\% & Median & 97.5\% & $\hat{R}$ & Mean & SD & 2.5\% & Median & 97.5\% \\
\midrule
$sd_{Intercept}$ & 1.00 & 0.01 & 0.00 & 0.01 & 0.02 & 0.02 & 1.00 & 1.57 & 0.15 & 1.30 & 1.56 & 1.89  & 1.00 & 0.04 & 0.03 & 0.00 & 0.03 & 0.11 \\ 
$sd_{< -5^\circ\text{C}}$ & 1.00 & 0.20 & 0.10 & 0.03 & 0.19 & 0.41 &1.01 & 1.53 & 0.68 & 0.20 & 1.55 & 2.87 & 1.00 & 0.64 & 0.40 & 0.04 & 0.61 & 1.50 \\ 
$sd_{-5^\circ\text{C} \text{ to } -1.5^\circ\text{C}}$ & 1.00 & 0.12 & 0.09 & 0.00 & 0.10 & 0.35 & 1.01 & 1.84 & 0.96 & 0.15 & 1.83 & 3.78 & 1.00 & 0.69 & 0.49 & 0.03 & 0.60 & 1.83 \\ 
$sd_{-1.5^\circ\text{C} \text{ to } 2^\circ\text{C}}$ & 1.00 & 0.07 & 0.06 & 0.00 & 0.06 & 0.21 & 1.00 & 3.49 & 0.74 & 2.09 & 3.47 & 4.97 & 1.00 & 0.66 & 0.42 & 0.04 & 0.61 & 1.56 \\ 
$sd_{2^\circ\text{C} \text{ to } 5.5^\circ\text{C}}$ & 1.00 & 0.08 & 0.06 & 0.00 & 0.07 & 0.22 & 1.00 & 1.91 & 0.64 & 0.57 & 1.92 & 3.17 & 1.00 & 0.46 & 0.33 & 0.01 & 0.40 & 1.25 \\ 
$sd_{5.5^\circ\text{C} \text{ to } 9^\circ\text{C}}$ & 1.00 & 0.07 & 0.05 & 0.00 & 0.05 & 0.19 & 1.00 & 0.46 & 0.35 & 0.02 & 0.39 & 1.28& 1.00 & 0.44 & 0.33 & 0.02 & 0.38 & 1.23 \\ 
$sd_{9^\circ\text{C} \text{ to } 12.5^\circ\text{C}}$ & 1.00 & 0.07 & 0.05 & 0.00 & 0.06 & 0.20 & 1.00 & 0.83 & 0.50 & 0.05 & 0.80 & 1.86 & 1.00 & 0.44 & 0.33 & 0.02 & 0.38 & 1.18 \\ 
$sd_{12.5^\circ\text{C} \text{ to } 16^\circ\text{C}}$ & 1.00 & 0.09 & 0.07 & 0.00 & 0.08 & 0.25 & 1.00 & 2.20 & 0.42 & 1.39 & 2.19 & 3.04 & 1.00 & 0.44 & 0.33 & 0.02 & 0.37 & 1.20 \\ 
$sd_{23^\circ\text{C} \text{ to } 26.5^\circ\text{C}}$ & 1.00 & 0.16 & 0.05 & 0.07 & 0.15 & 0.26 & 1.00 & 0.73 & 0.56 & 0.02 & 0.61 & 2.04 & 1.00 & 0.32 & 0.25 & 0.01 & 0.27 & 0.90 \\ 
$sd_{26.5^\circ\text{C} \text{ to } 30^\circ\text{C}}$ & 1.00 & 0.06 & 0.04 & 0.00 & 0.05 & 0.15 & 1.00 & 0.66 & 0.51 & 0.03 & 0.55 & 1.90 & 1.00 & 0.41 & 0.30 & 0.02 & 0.35 & 1.13 \\ 
$sd_{> 30^\circ\text{C}}$ & 1.00 & 0.11 & 0.03 & 0.05 & 0.11 & 0.18 & 1.00 & 1.48 & 0.81 & 0.10 & 1.45 & 3.07 & 1.00 & 0.36 & 0.27 & 0.01 & 0.31 & 1.03 \\ 
\bottomrule
\end{tabular}
\caption{Selected statistics for the estimated posterior densities for population level parameter and group level standard deviations.
\scriptsize{Estimated using Bayesian Partial Pooling Model (NUTS-sampler). Data from ENERDATA; \citetalias{nasav2}, \citetalias{population_data}, \citetalias{population1}, \citetalias{population2}.}}
\label{tab:maintable}
\end{table}
\end{landscape}

\subsection{Individual Country Responses}
\begin{figure}[!htbp]
    \centering
    \includegraphics[width=\textwidth]{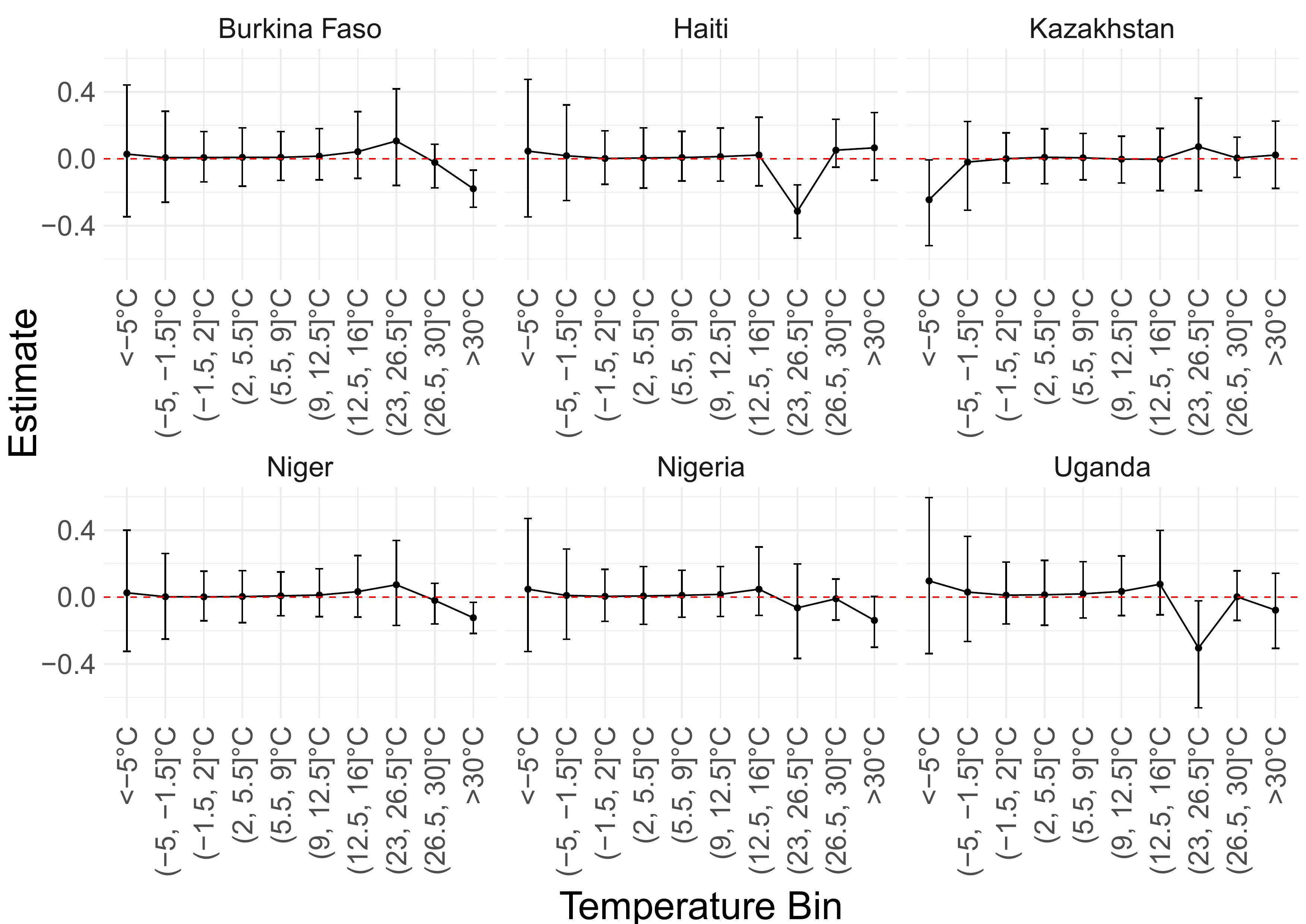}
    \caption{Estimated country specific deviation from the population level estimate for the effect of a shift of temperature exposure of the population to ten different temperature bins (\celsius{}) on log residential electricity demand, relative to the 16\celsius{} to 23\celsius{} bin. Using a 3.5\celsius{} bin width and including 90\% credible intervals}
\scriptsize{Estimated using Bayesian Partial Pooling Model (NUTS-sampler).Data from ENERDATA;Data from ENERDATA;\citetalias{nasav2,population_data,population1,population2}.}
    \label{fig:country_slopes}
\end{figure}

The hierarchical Bayesian method allows us to estimate individual intercepts and slope parameters for each country. The estimated coefficients for individual parameters on the temperature sensitivity of residential energy demand can be understood as deviations from the corresponding global parameter.
\paragraph{}
 Plotting the posterior means of the individual estimates from the regression with log-transformed residential electricity as the outcome variable against each other reveals a positive correlation for higher temperatures, as shown in Figure \ref{fig:intercept_slope_ele}. Although there is a lot of uncertainty in the individual estimates and further research is needed to draw firmer conclusions, the results suggest that a higher baseline demand for electricity is associated with a stronger temperature sensitivity at high temperatures. Figure \ref{fig:intercept_slope_ele} makes clear that this correlation is mainly driven by countries with high temperatures. Countries with a relatively low level of development tend to be on the lower left. These are countries mostly from the African continent, such as Uganda, Cameroon and Nigeria. The upper right of this Figure is dominated by countries from Southeast Asia, such as Cambodia and Vietnam. The same analysis shows no clear pattern for residential gas or oil demand (Figures \ref{fig:intercept_slope_gas} and \ref{fig:intercept_slope_oil}).
\paragraph{}
For illustrative purposes, Figure \ref{fig:country_slopes} shows a selection of countries with deviations from the global average in terms of the temperature response of electricity demand.\footnote{Results for the remaining countries for all energy types are presented in Figures \ref{fig:country_slopes_electricity1} to \ref{fig:country_slopes_oil3}.} Examining the individual estimates for the cooling effect first, especially for less developed countries with very hot temperatures such as Burkina Faso, Niger and Nigeria, the estimate for the cooling effect in the highest temperature bin deviates negatively from the global estimate. This can be interpreted as further evidence for the disproportionally large effect of climate change on less developed countries as it suggests that electric cooling is less common in these countries. The impact on energy use is small, the concomitant impacts on health and productivity are large. In terms of heating effects, Kazakhstan seems to have a lower heating effect for very cold temperatures compared to the average country. This is likely due to the abundance of fossil fuels, so no additional electric heating is needed in Kazakhstan. 
\paragraph{}
Two intriguing cases are Haiti and Uganda. Both show a negative deviation from the global average for some inner temperature bins. For Haiti, this can probably be explained by data anomalies caused by the devastating earthquake in 2010. This earthquake destroyed many homes and infrastructure, followed by an immense inflow of aid from other countries, which distorted electricity demand. For Uganda data are only available from 2001 to 2012. During this preiod (2005 and 2006) Uganda faced low water levels in Lake Victoria and a severe energy crisis.
\paragraph{}
Figure \ref{fig:country_slopes_gas} illustrates deviations from the global temperature effect for residential natural gas demand. We observe a strong deviation from the global temperature coefficients for Germany for the temperature ranges -1.5 to 5.5\celsius{} and 12.5 to 16\celsius. This indicates that there might be a heating effect in Germany at these temperatures, whereas there is none for the global mean. The estimates also indicate the possibility of a heating effect for Turkey for temperatures in the range of -5\celsius{} to 2\celsius. The coefficients for Peru show a negative deviation from the global mean for temperatures between -1.5\celsius{} to 5.5\celsius{} and for 12.5\celsius{} to 16.5\celsius.

%HAB GRAD NE IDEE: 
%\paragraph{} 
%Possibly the absence of detectable temperature sensitivty to cold but not extreme cold temperatures could be due to the interplay of technical heating capacity, thermostats and behavioral inertia. So when it get´s colder people are turning on their heating equipment as they always do during that time of the year. However, as temperature drop even lower, they not necessarily change the setting of their thermostat. Now usually this would still lead to an increase in energy demand since the heating equipment will need to heat more to keep the room temperature constant. But, maybe the heating equipment is already close to it´s maximum heating capacity given the thermostat setting and the home insulation. Then the heating equipment can not react as strongly. The only way to react to changing temperatures is then to change the duration of heating. However, during moderate cold temperatures people might be willing to accept slightly colder indoor temperatures and only as temperatures get really cold they resort to measures like extending the heating phase, like e.g not turning down heating for the night or when they are away.  
%\section{Discussion}
%\begin{itemize}
%    \item Argument 1 
%    \item Argument 2
%\end{itemize}
\section{Discussion and Conclusion}
Using the convenient properties of Bayesian Hierarchical Models, this study provides robust evidence on the long run relationship between temperature and residential energy demand, offering important insights for both energy and climate policy.
\paragraph{}
Our findings, based on data for 126 countries that span from 1978 to 2023, show pronounced asymmetries in the response of residential energy demand to extreme temperatures. For example, a 10 percentage point increase in the population exposed to temperatures below -5\celsius{} is associated with a 2\% increase in residential electricity demand and a substantial 19.4\% increase in residential natural gas demand. These effects are even more pronounced in the long run, with residential electricity demand increasing by 50\% and natural gas demand doubling (114.1\%). Exposure to extreme heat (temperatures above 30\celsius{}) is associated with a smaller increase in electricity demand. A similar 10 percentage point increase in the population exposed to such high temperatures results in a 0.8\% increase in residential electricity demand, which accumulates to a 20\% increase in the long run. This suggests that cooling demand is so far less responsive to extremely hot temperatures compared to heating demand in extremely cold conditions. For light fuel oil, no effects were observed. A general downward trend in the usage of heating oil might cause this. It is also possible that consumers usually plan their oil demand and buy in bulk such that the current temperature does not affect current demand, but rather the demand for the next period.
\paragraph{}
In contrast to studies using the common panel fixed-effects approach, our results indicate that there are no significant heating or cooling effects for moderate temperature variations. Only extreme temperatures show measurable effects on residential energy demand. Furthermore our results show that most countries do not deviate strongly from this global pattern, emphasizing the need for policymakers to focus on these extremes when designing energy resilience strategies.
\paragraph{}
At the level of individual countries, our analysis uncovers a positive correlation between countries' individual estimates of intercepts and their temperature responsiveness to hot climates. Poorer countries in hot climates are more vulnerable to rising temperatures, highlighting the disproportionate burden of climate change on already economically disadvantaged countries.
\paragraph{}
Given these estimates, and taking into account increasing electrification and the geographical distribution of the world's population, with the majority living in warm and hot climates, we expect the increasing demand for cooling to outweigh any potential reductions in energy demand for heating. In 2022, almost 2.5 billion people lived in regions with an annual average temperature below 18\celsius, while 4.2 billion lived in regions with an average temperature above 22\celsius.\footnote{We used information on temperature and population for a total of 172 countries, a complete list can be found in \ref{tab:geo_vis_countries}.} At the extremes, only 130 million people lived in regions with an average annual temperature below -5\celsius, while 1.5 billion people were exposed to temperatures above 30\celsius. This means that more than ten times as many people were exposed to extreme heat than to extreme cold. Rising global average temperatures would therefore significantly increase the number of people living in regions of extreme heat. Specifically, a temperature increase of 1\celsius{} would lead to a 20\% increase in the number of people living in areas with an average temperature above 30\celsius. A 2\celsius{} increase would lead to a 43\% increase, while a 3\celsius{} increase would lead to a 68\% increase, to a total of almost 2.4 billion people exposed to such extreme temperatures. This shows that the cooling effect we found in our analysis is likely to affect many more people than the heating effect.
\paragraph{}
Contrary to this expectation, our model predicts that a uniform global warming of 1\celsius{} with respect to the temperatures in 2020 will lead to an increase in electricity demand for some countries and a reduction for others. For example, demand for Saudi Arabia is predicted to increase by 0.63\% (75 ktoe), while demand for the USA and Canada is predicted to decrease by 0.21\% (268 ktoe) and 0.29\% (43 ktoe), respectively. For our sample, these changes in demand almost cancel out, leading to a total reduction of residential electricity demand by 0.05\% (269 ktoe). The predicted reduction in electricity demand may stem from two factors: (1) increased cooling needs are offset by reduced heating needs, and (2) the model does not account for future AC diffusion. As a result, potential growth in residential electricity demand in developing countries, driven by increasing AC adoption, remains unaccounted for.
\paragraph{}
For natural gas, we see a much stronger reduction due to global warming; for example, for Germany, a decrease of 24.1\% (5238 ktoe) and for the USA, a decrease of 5.2\% (5712 ktoe) are predicted. In total, the model predicts that a 1\celsius{} uniform global warming leads to a decrease of residential natural gas demand by 22993 ktoe, or 5\% of total demand in our sample.
\paragraph{}
The residential light fuel oil demand in Greece is predicted to decrease by 1.8\% (22 ktoe). In total, the model predicts, for a 1\celsius{} warming, a decrease of 168 ktoe, which is 0.47\% of the total demand accounted for in our sample.
\paragraph{}
These predictions should be viewed as tendencies rather than exact predictions since uncertainty is still substantial, especially for light fuel oil demand. As climate change makes temperature extremes more likely, the implications for global energy systems will be profound, requiring coordinated international efforts to mitigate the socio-economic impacts of changing residential energy demand.
\paragraph{}
This study uses a broad panel with highly aggregated data. This allows long run macro-level effects to be studied in depth. However, further research with more granular data is needed to better understand short-term dynamics. In addition, it remains to be explored how these new temperature response estimates can be fed into climate-energy models and how they can adequately incorporate distributional information about these parameters. Given the sensitivity of the results to the specification details of the temperature effect, we recommend future research to explore different temperature effect specifications or to explore different modeling strategies such as splines.
\newpage
\nocite{nasav21}
\bibliographystyle{apalike} % Defines the bibliography style
\bibliography{export} 
\newpage

\appendix
\setcounter{page}{1}
\renewcommand{\thepage}{A\arabic{page}}

\section{Summary Statistics}
\label{appendixA}
\setcounter{figure}{0}
\setcounter{table}{0}
\numberwithin{table}{section}
\numberwithin{figure}{section}

\begin{figure}[!htpb]
    \centering
    \textbf{Electricity Demand}
    \includegraphics[width=\linewidth]{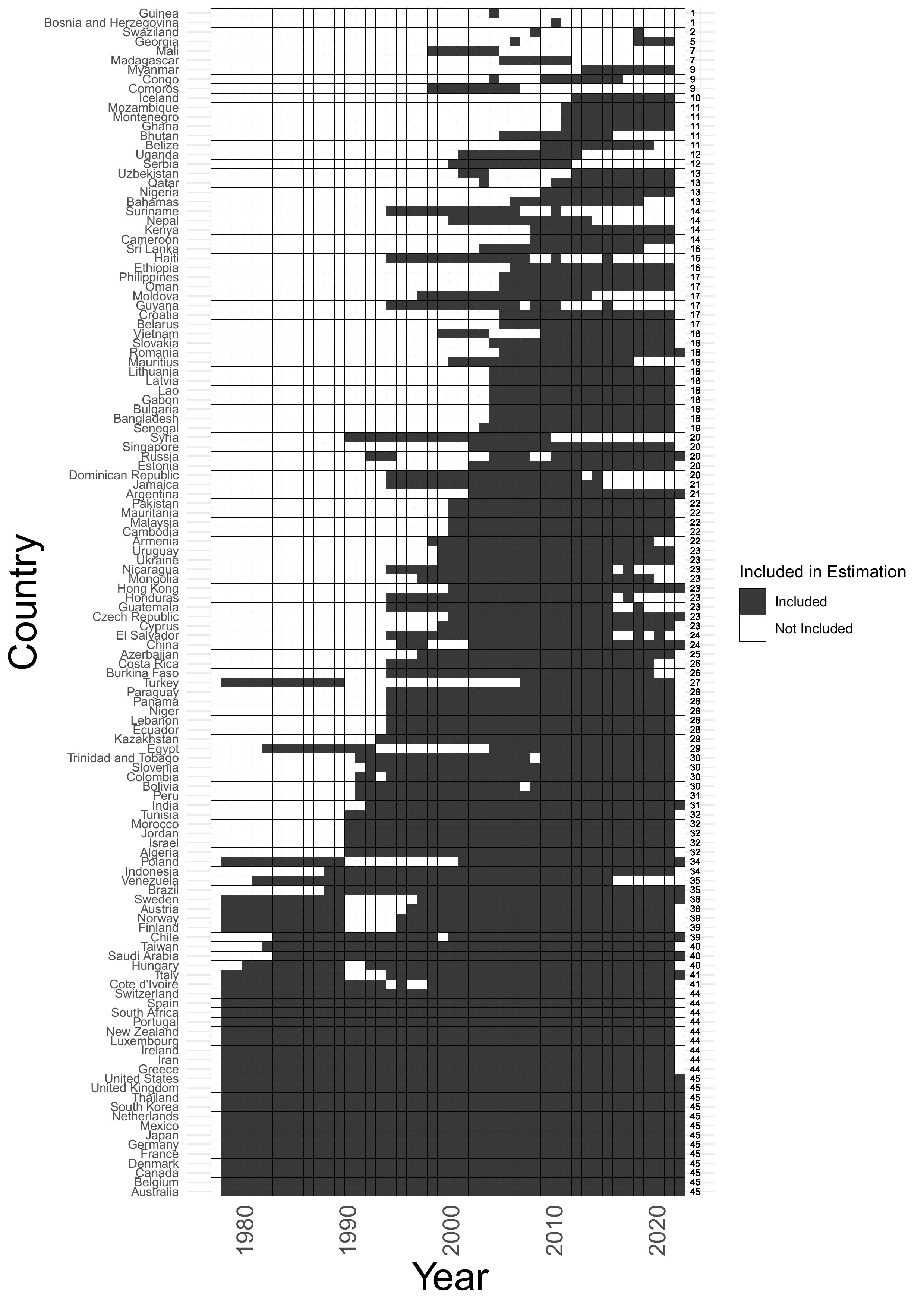}
    \caption{Observations which are used in the final estimation. Total number of periods for each country depicted on the right axis.
    \scriptsize{Data from ENERDATA.}}
    \label{fig:country_year}
\end{figure}
\newpage
\begin{figure}[!htpb]
    \centering
    \textbf{Natural Gas Demand}
    \includegraphics[width=\linewidth]{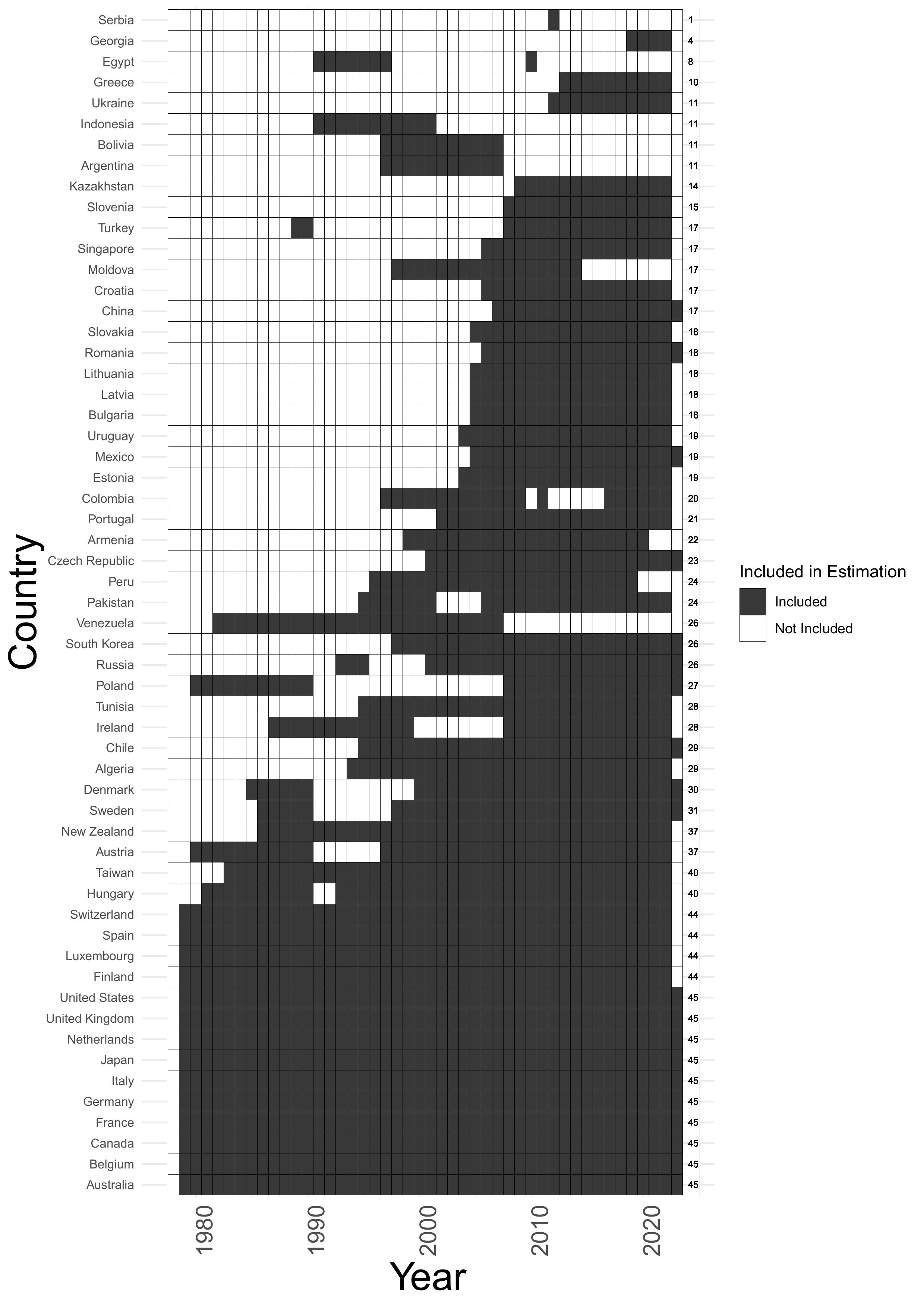}
    \caption{Observations which are used in the final estimation. Total number of periods for each country depicted on the right axis.
    \scriptsize{Data from ENERDATA.}}
    \label{fig:country_year_gas}
\end{figure}
\begin{figure}[!htpb]
    \centering
    \textbf{Light Fuel Oil Demand}
    \includegraphics[width=\linewidth]{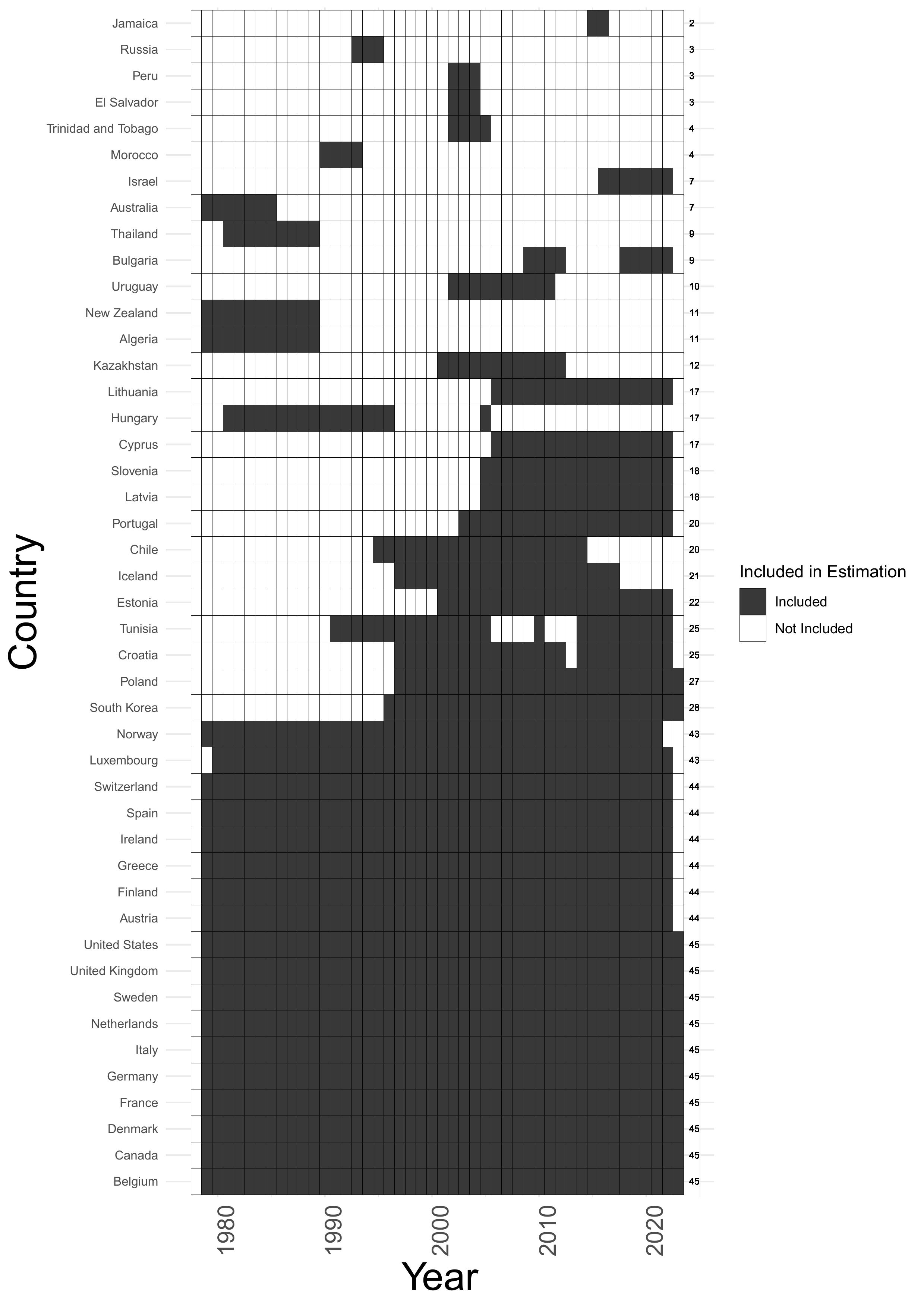}
    \caption{Observations which are used in the final estimation. Total number of periods for each country depicted on the right axis.
    \scriptsize{Data from ENERDATA.}}
    \label{fig:country_year_oil}
\end{figure}
\newpage
\section{Prior Sensitivity and Predictive Checks}
\label{AppendixB}
\setcounter{figure}{0}
\setcounter{table}{0}
\renewcommand{\thetable}{B\arabic{table}}
\renewcommand{\thefigure}{B\arabic{figure}}

\begin{figure}[!htpb]
    \textbf{Electricity}
    \centering
    \includegraphics[width=\linewidth]{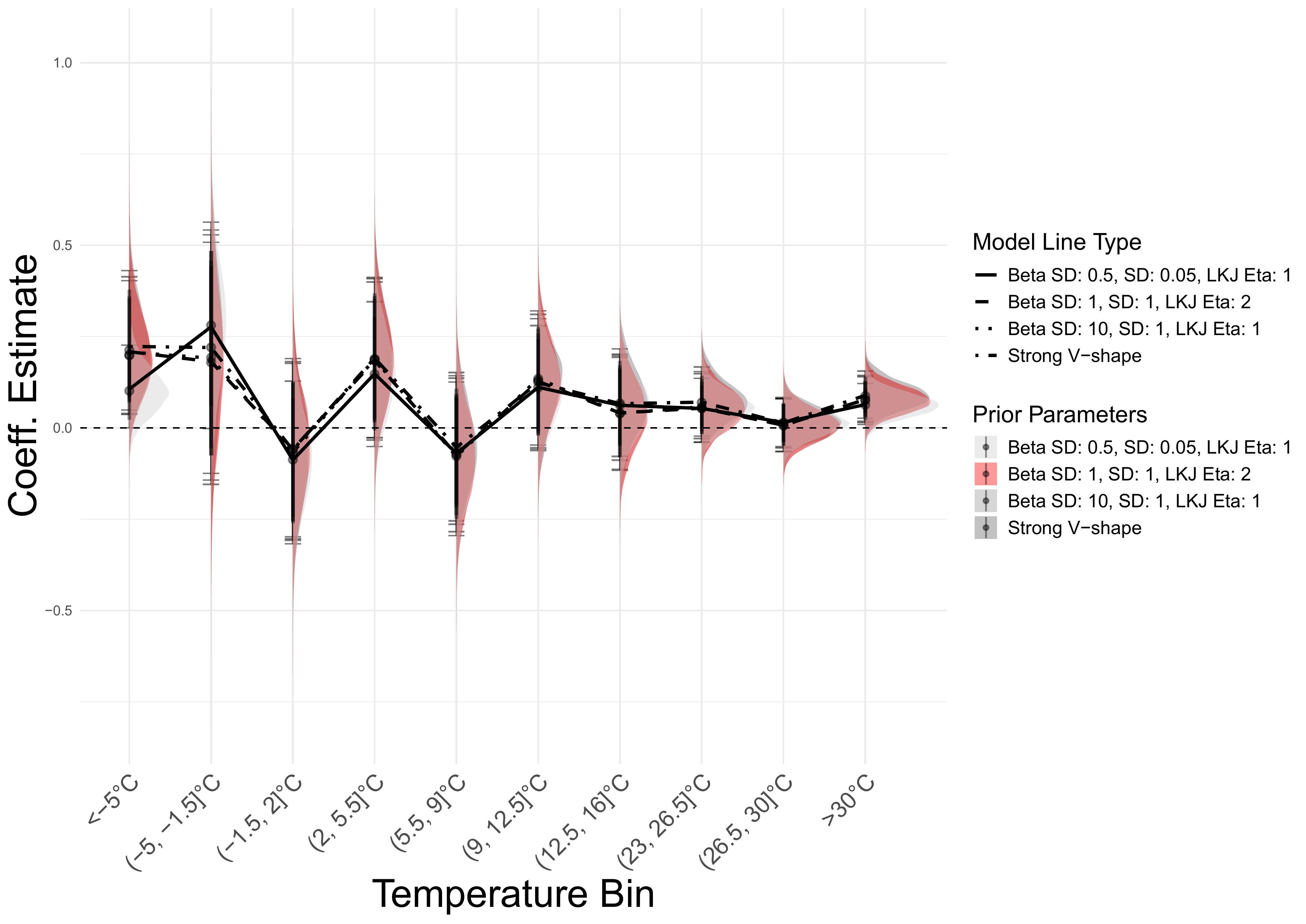}
    \caption{Estimates obtained through selected prior specifications including a specification, implying a strong V-shape for the temperature response. The main specification is highlighted in red. Including 50\% and 90\% credible intervals.
    \scriptsize{Estimated using Bayesian Partial Pooling Model (NUTS-sampler). Data from ENERDATA; \citetalias{nasav2}, \citetalias{population_data}, \citetalias{population1}, \citetalias{population2}.}}
    \label{fig:prior_sensitivity}
\end{figure}
\begin{figure}[!htpb]
    \textbf{Natural Gas}
    \centering
    \includegraphics[width=\linewidth]{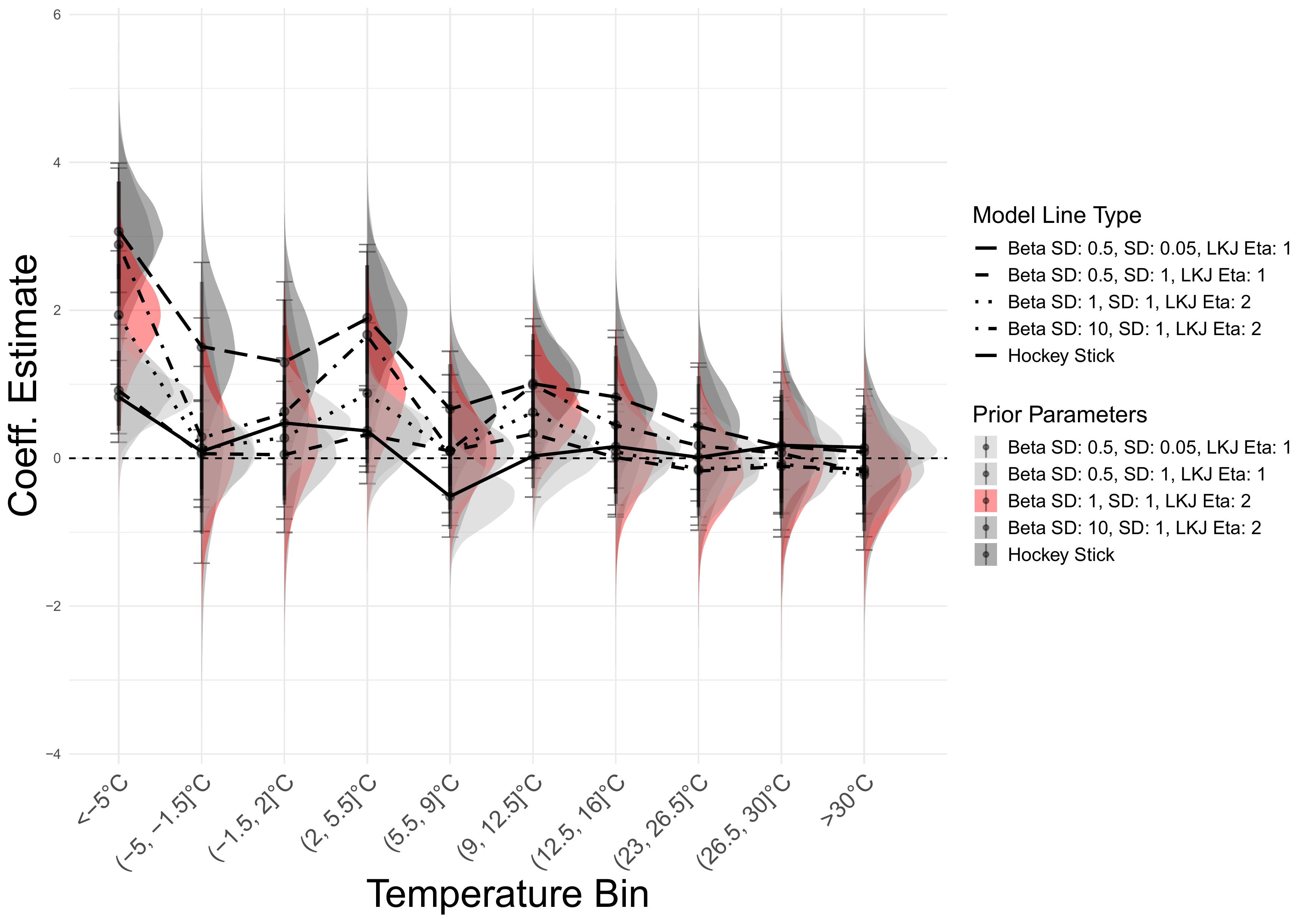}
    \caption{Estimates obtained through selected prior specifications including a specification, implying a Hockey-Stick-shape for the temperature response. The main specification is highlighted in red. Including 50\% and 90\% credible intervals.
    \scriptsize{Estimated using Bayesian Partial Pooling Model (NUTS-sampler). Data from ENERDATA; \citetalias{nasav2}, \citetalias{population_data}, \citetalias{population1}, \citetalias{population2}.}}
    \label{fig:prior_sensitivity_gas}
\end{figure}
\begin{figure}[!htpb]
    \textbf{Light Fuel Oil}
    \centering
    \includegraphics[width=\linewidth]{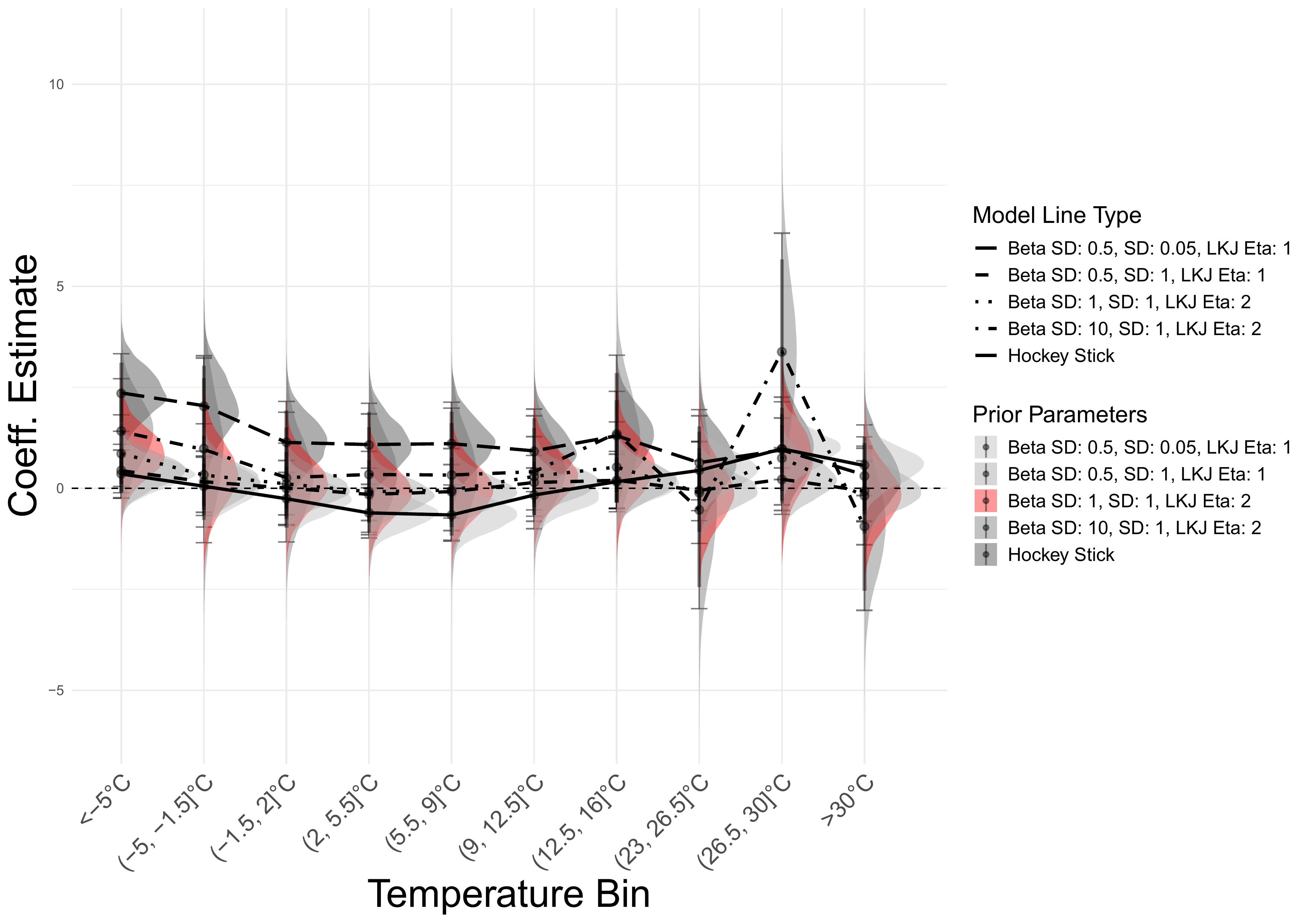}
    \caption{Estimates obtained through selected Prior specifications including a specification, implying a Hockey-Stick-shape for the temperature response. The main specification is highlighted in red. Including 50\% and 90\% credible intervals.
    \scriptsize{Summary statistics of posterior distribution for temperature variables. Estimated using Bayesian Partial Pooling Model (NUTS-sampler). Data from ENERDATA; \citetalias{nasav2}, \citetalias{population_data}, \citetalias{population1}, \citetalias{population2}.}}
    \label{fig:prior_sensitivity_oil}
\end{figure}
\begin{figure}[h!]
    \centering
    % First row: Gas
    \begin{minipage}{0.48\textwidth}
        \centering
        \includegraphics[width=\textwidth]{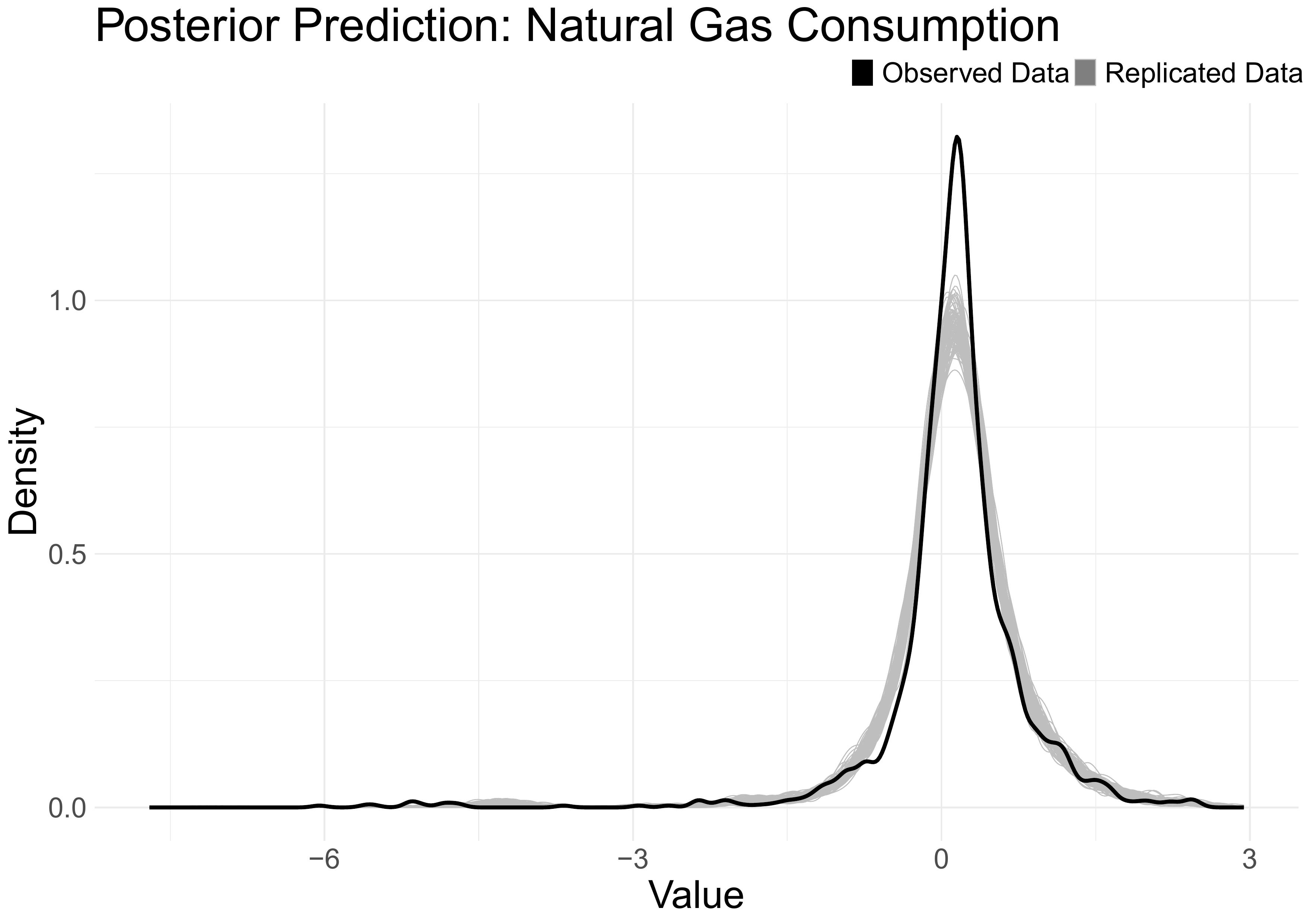}
        \label{fig:postpcheck_gas}
    \end{minipage}
    \hfill
    \begin{minipage}{0.48\textwidth}
        \centering
        \includegraphics[width=\textwidth]{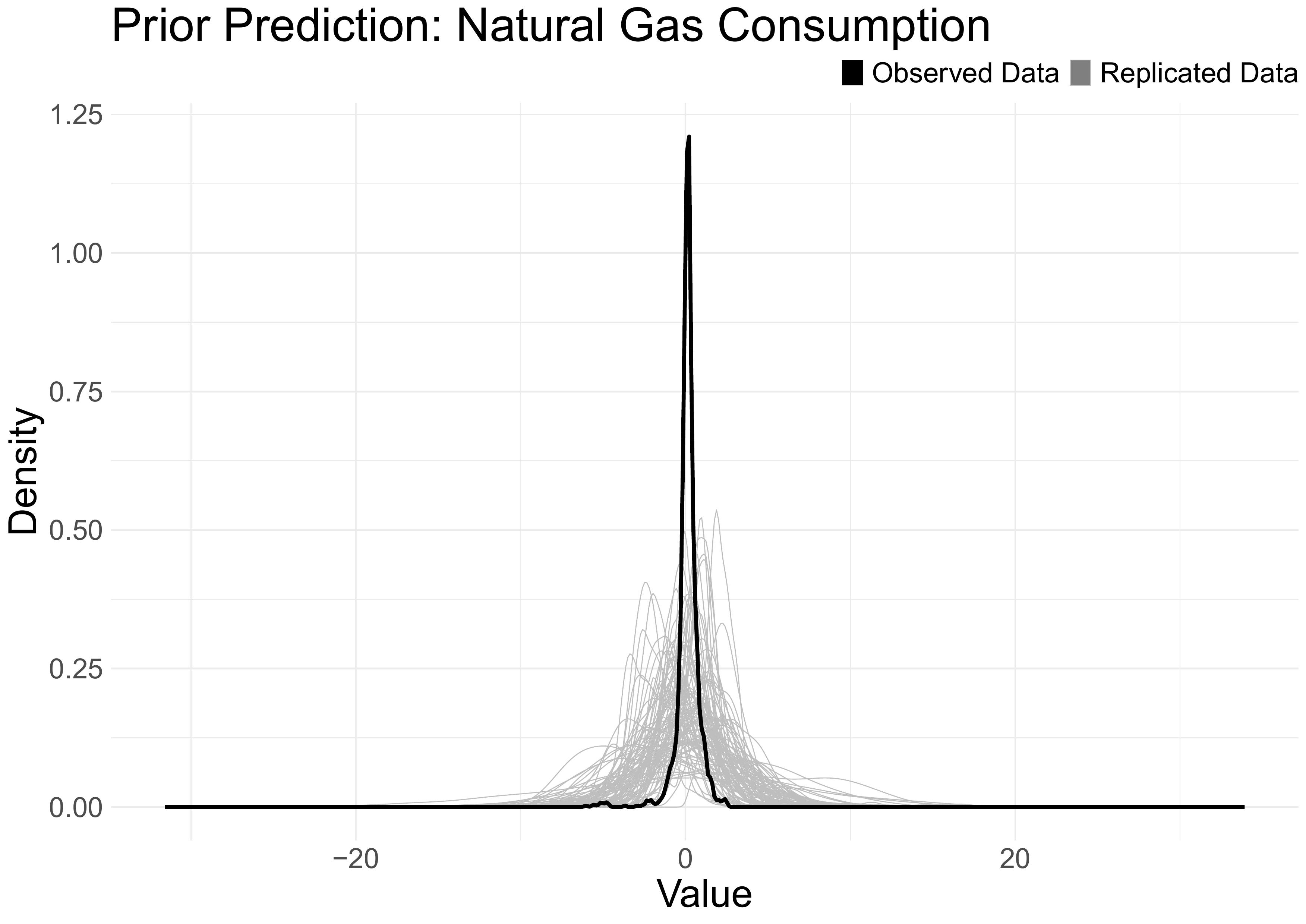}
        \label{fig:priorpcheck_gas}
    \end{minipage}
    
    \vspace{0.5cm}% Add vertical space between rows
    
    % Second row: Oil
    \begin{minipage}{0.48\textwidth}
        \centering
        \includegraphics[width=\textwidth]{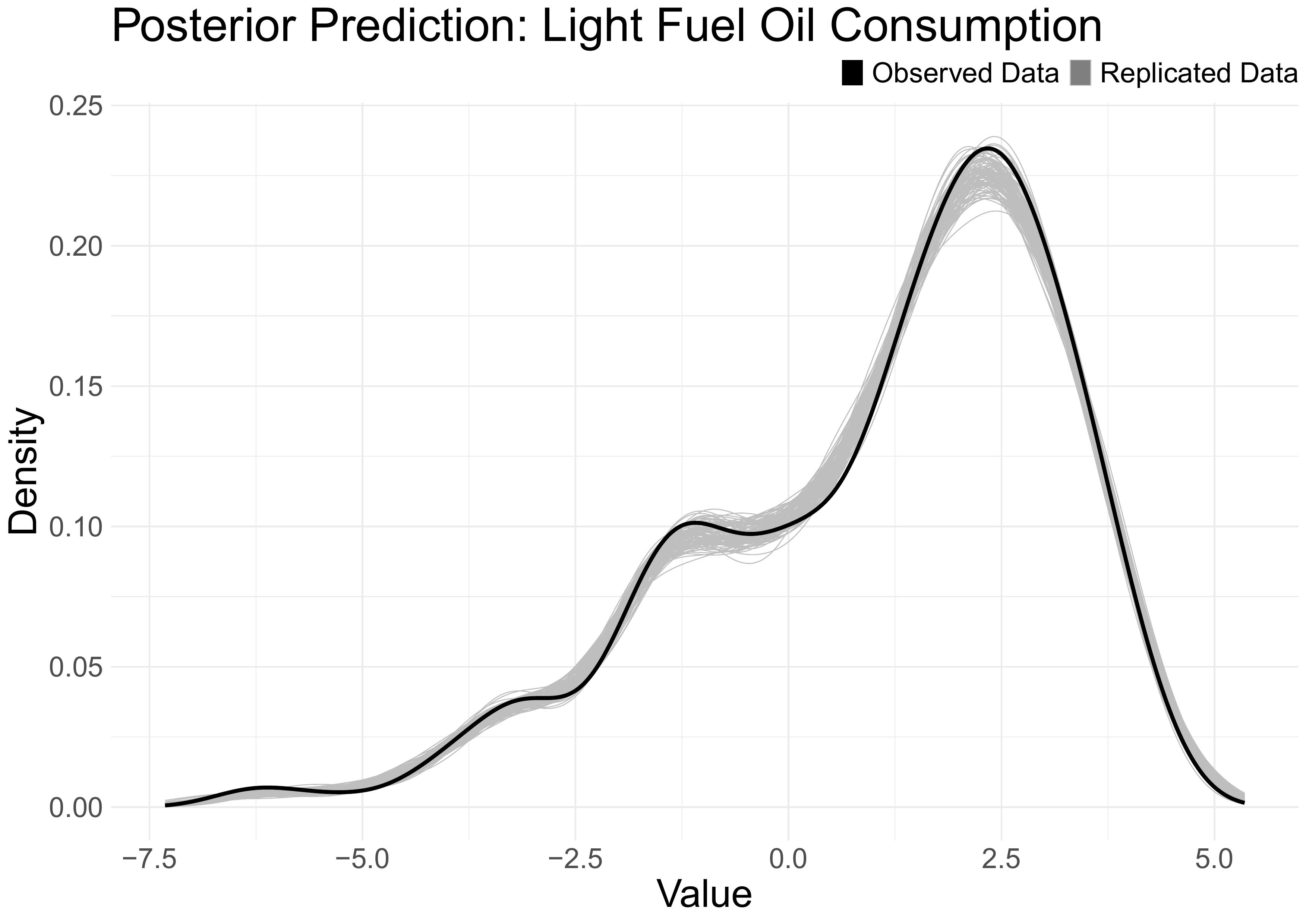}
        \label{fig:postpcheck_oil}
    \end{minipage}
    \hfill
    \begin{minipage}{0.48\textwidth}
        \centering
        \includegraphics[width=\textwidth]{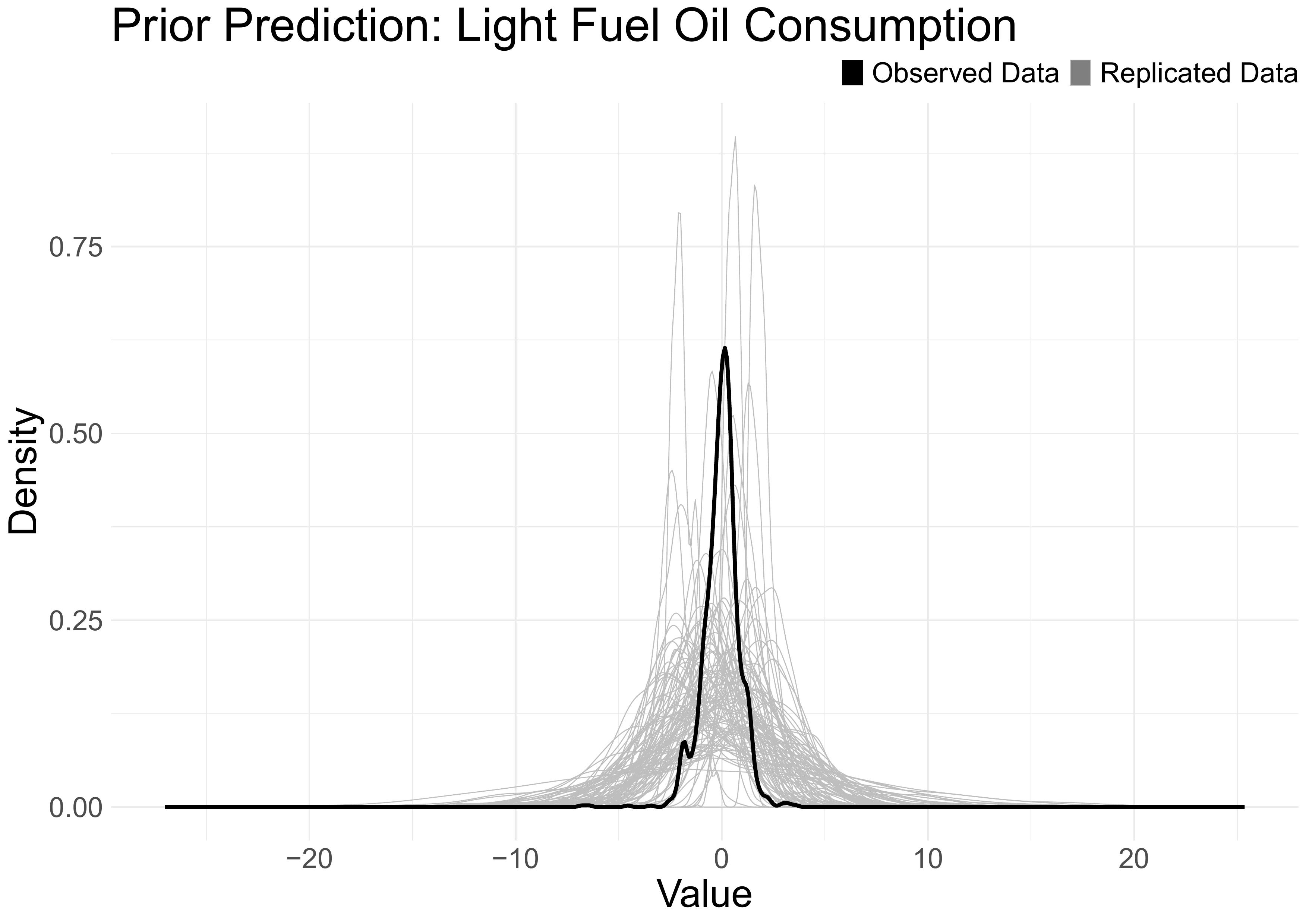}
        \label{fig:priorpcheck_oil}
    \end{minipage}
    \caption{Comparison of prior and posterior predictions for natural gas and light fuel oil demand.
    \scriptsize{Estimated using Bayesian Partial Pooling Model (NUTS-sampler). Data from ENERDATA; \citetalias{nasav2}, \citetalias{population_data}, \citetalias{population1}, \citetalias{population2}.}}
    \label{fig:predictive_checks}
\end{figure}

\newpage
\section{Additional Figures}
\label{appendixC}
\setcounter{figure}{0}
\setcounter{table}{0}
\renewcommand{\thetable}{C\arabic{table}}
\renewcommand{\thefigure}{C\arabic{figure}}

\begin{landscape}
\begin{figure}[!htbp]
\centering
\textbf{Electricity Demand}
\includegraphics[width=\linewidth]{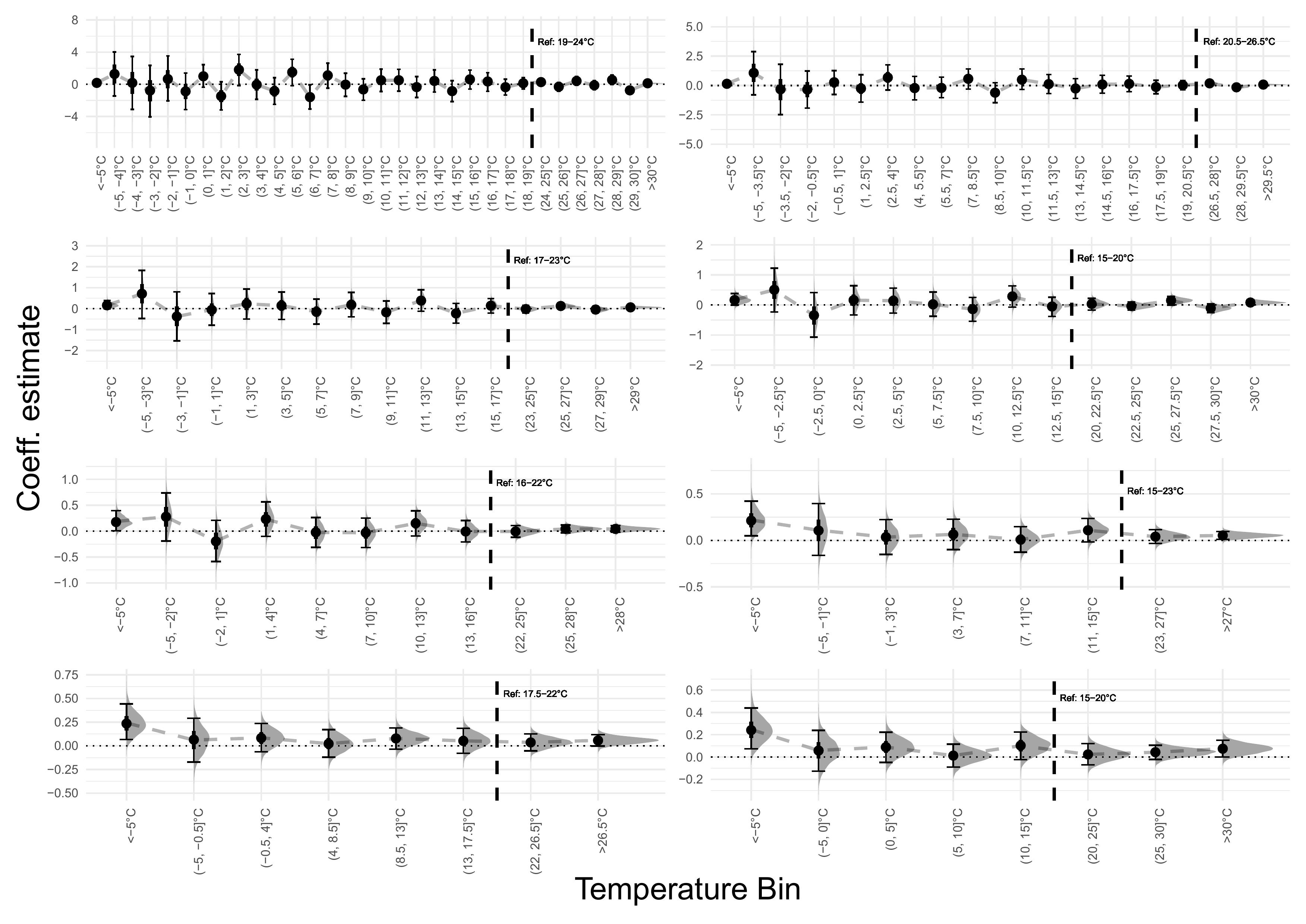}
\caption{Estimated impact of a shift of temperature exposure of the population to different temperature bins (\celsius{}) on log residential electricity demand, relative to the reference bin including 50\% and 90\% credible intervals.
\\
\scriptsize{Estimated using Bayesian Partial Pooling Model (NUTS-sampler). Data from ENERDATA; \citetalias{nasav2}, \citetalias{population_data}, \citetalias{population1}, \citetalias{population2}.}}
\label{fig:alternative_coeff_bayes_ele}
\end{figure}

\begin{figure}[!htbp]
\centering
\textbf{Natural Gas Demand}
\includegraphics[width=\linewidth]{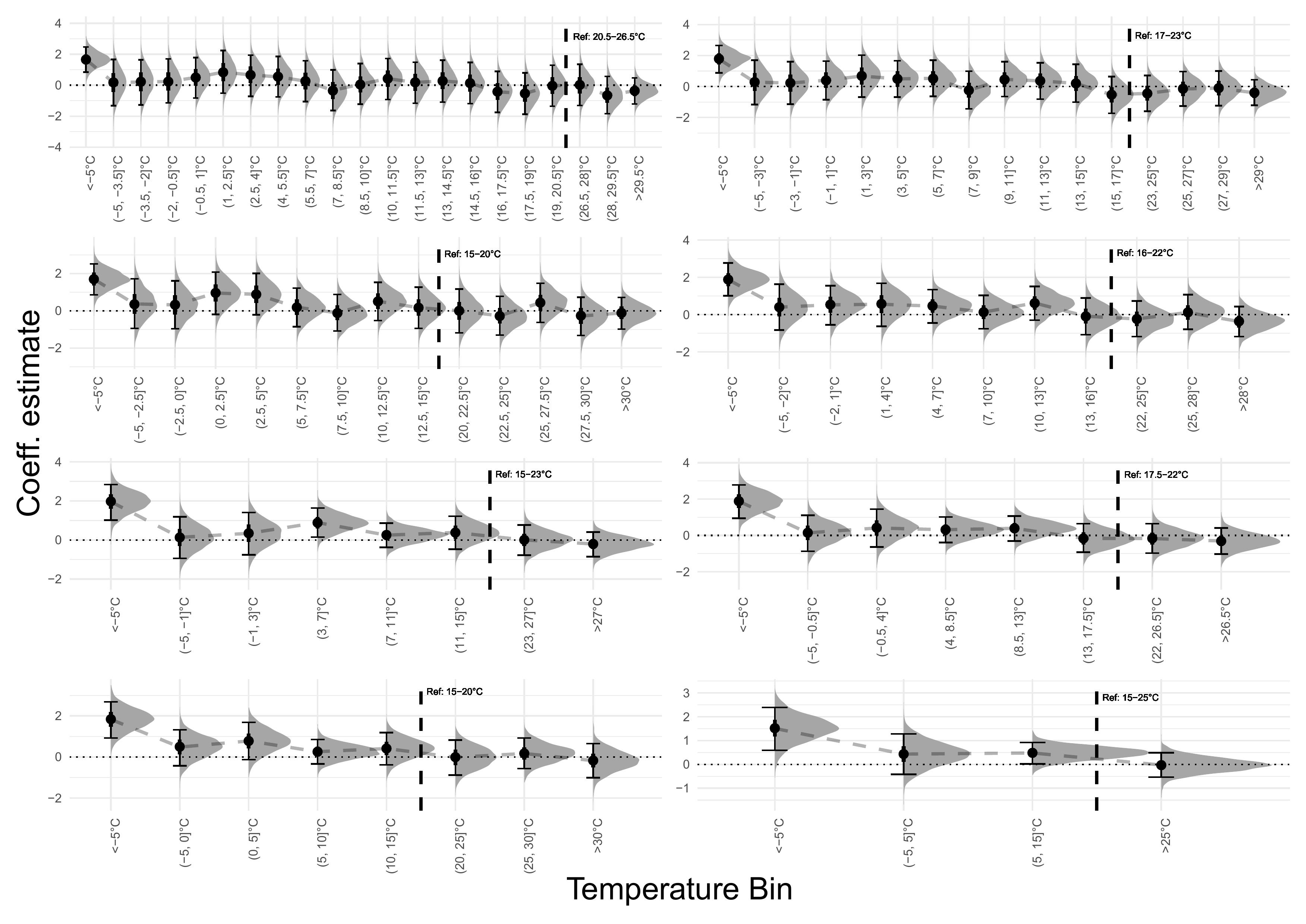}
\caption{Estimated impact of a shift of temperature exposure of the population to different temperature bins (\celsius{}) on log residential natural gas demand, relative to the reference bin including 50\% and 90\% credible intervals.
\\
\scriptsize{Estimated using Bayesian Partial Pooling Model (NUTS-sampler). Data from ENERDATA; \citetalias{nasav2}, \citetalias{population_data}, \citetalias{population1}, \citetalias{population2}.}}
\label{fig:alternative_coeff_bayes_gas}
\end{figure}

\begin{figure}[!htbp]
\centering
\textbf{Light Fuel Oil Demand}
\includegraphics[width=\linewidth]{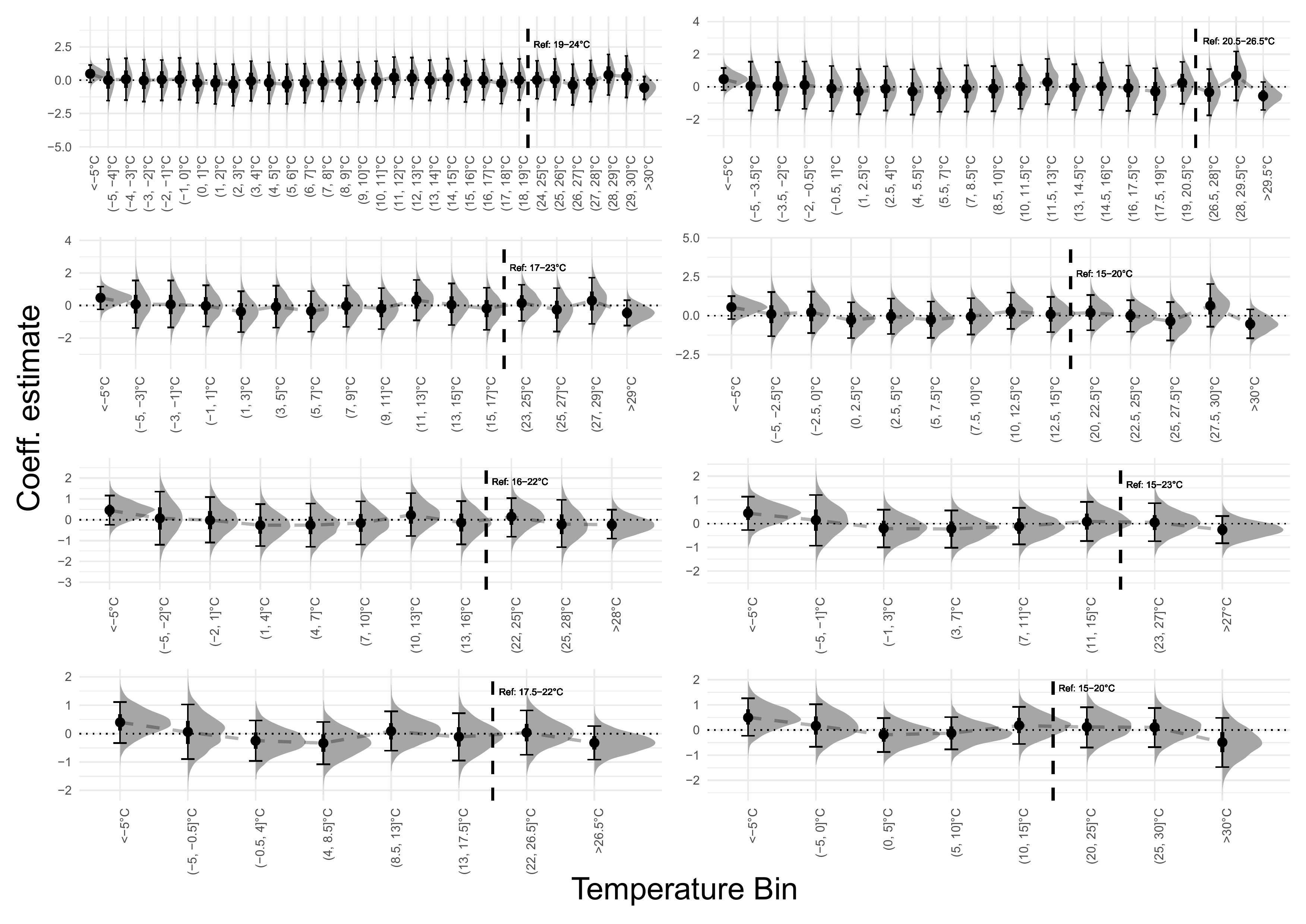}
\caption{Estimated impact of a shift of temperature exposure of the population to different temperature bins (\celsius{}) on log residential light fuel oil demand, relative to the reference bin including 50\% and 90\% credible intervals.
\\
\scriptsize{Estimated using Bayesian Partial Pooling Model (NUTS-sampler). Data from ENERDATA; \citetalias{nasav2}, \citetalias{population_data}, \citetalias{population1}, \citetalias{population2}.}}
\label{fig:alternative_coeff_bayes_oil}
\end{figure}
\end{landscape}
\newpage
\begin{landscape}
\begin{figure}
    \centering
    \textbf{Electricity Demand}
    \includegraphics[width=\linewidth]{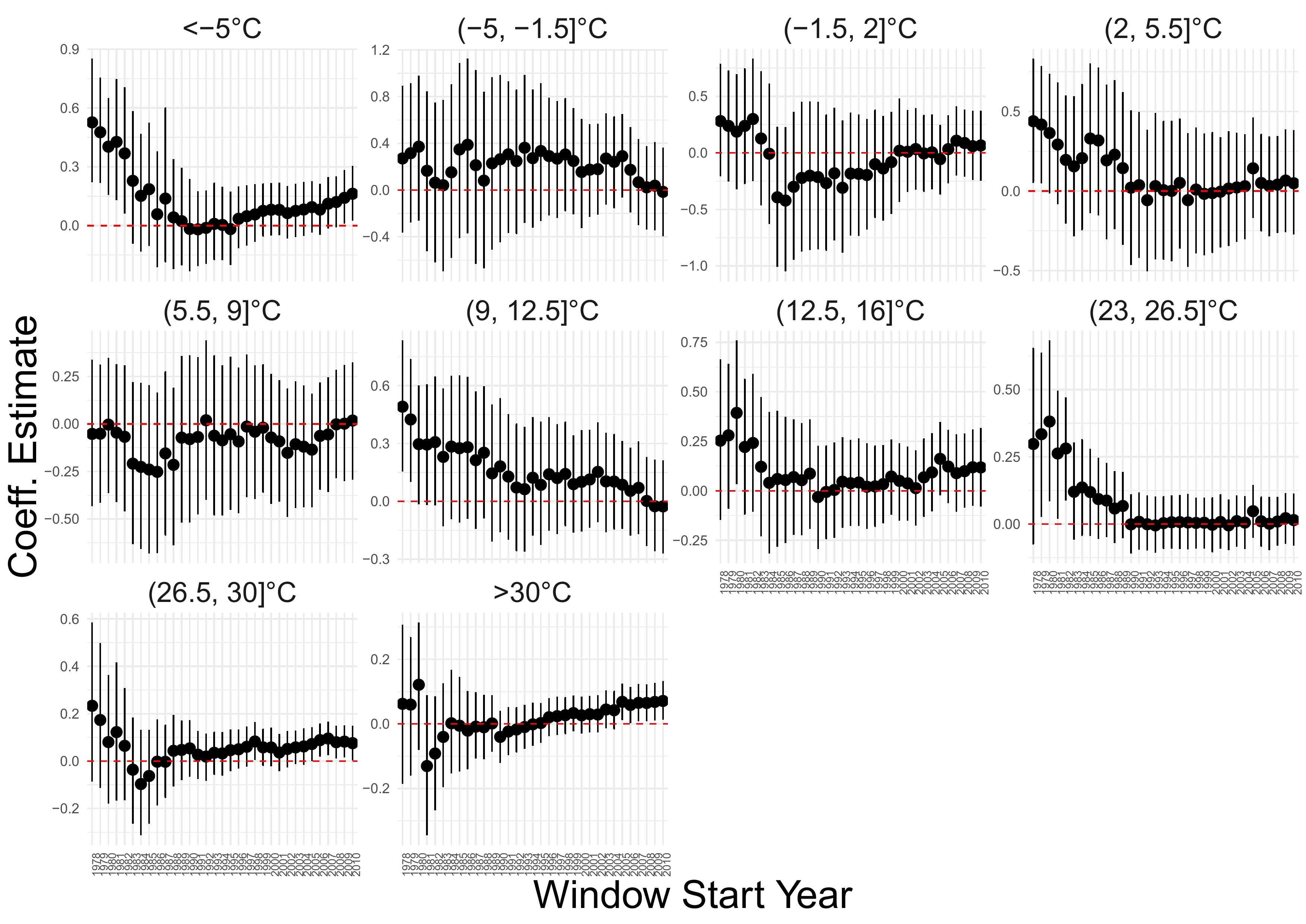}
    \caption{Estimated impact of a shift of temperature exposure of the population to ten different temperature bins (\celsius{}), relative to the 16\celsius{} to 23\celsius{} bin. On log residential electricity demand Using a 3.5\celsius{} bin width and including 90\% credible intervals. Based on a 15 year rolling window.
\\
\scriptsize{Estimated using Bayesian Partial Pooling Model (NUTS-sampler). Data from ENERDATA; \citetalias{nasav2}, \citetalias{population_data}, \citetalias{population1}, \citetalias{population2}.}}
    \label{fig:window_electricity}
\end{figure}
\end{landscape}
\newpage
\begin{landscape}
\begin{figure}
    \centering
    \textbf{Natural Gas Demand}
    \includegraphics[width=\linewidth]{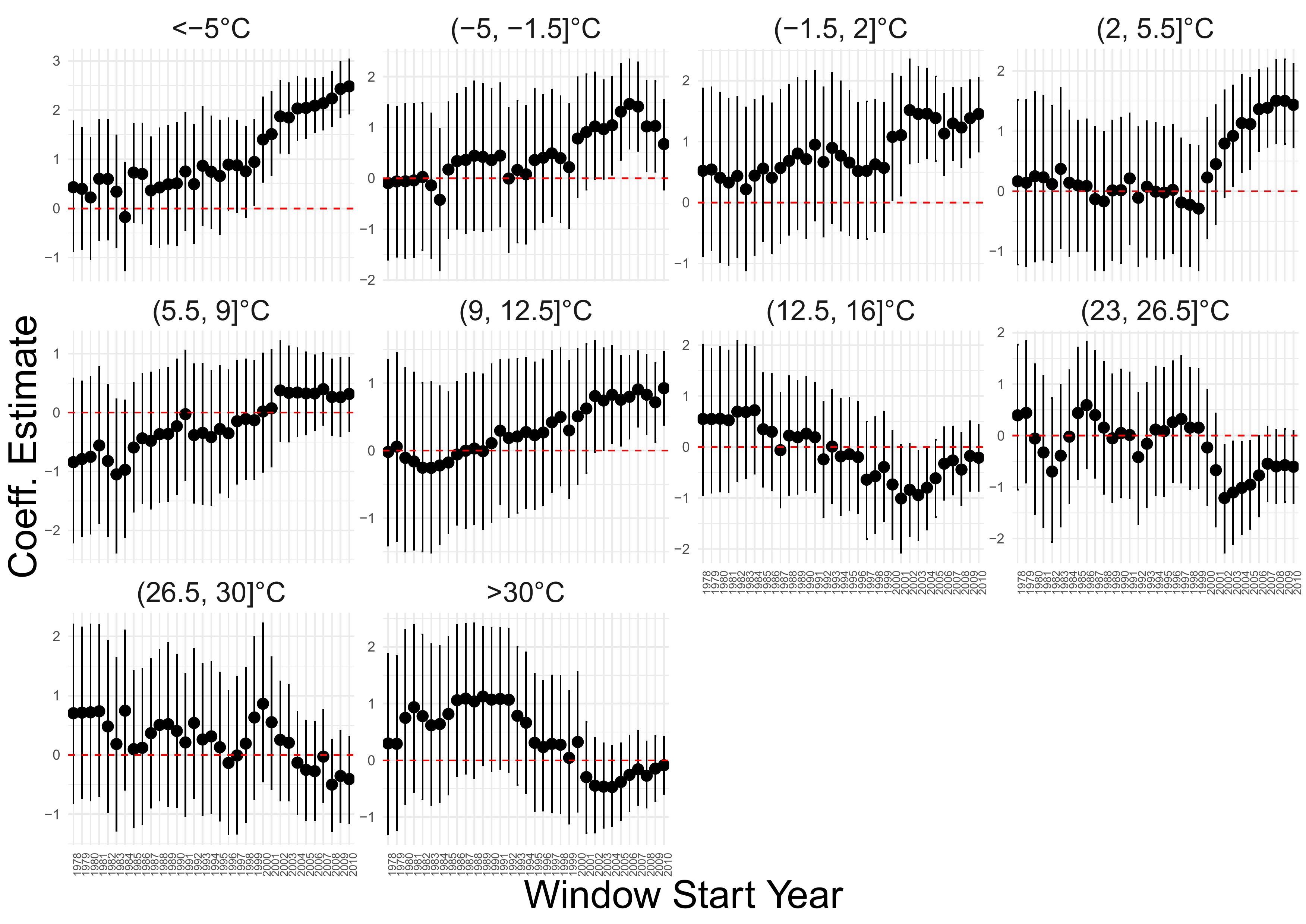}
    \caption{Estimated impact of a shift of temperature exposure of the population to ten different temperature bins (\celsius{}), relative to the 16\celsius{} to 23\celsius{} bin. On log residential natural gas demand Using a 3.5\celsius{} bin width and including 90\% credible intervals. Based on a 15 year rolling window.
\\
\scriptsize{Estimated using Bayesian Partial Pooling Model (NUTS-sampler). Data from ENERDATA; \citetalias{nasav2}, \citetalias{population_data}, \citetalias{population1}, \citetalias{population2}.}}
    \label{fig:window_gas}
\end{figure}
\end{landscape}
\newpage
\begin{landscape}
\begin{figure}
    \centering
    \textbf{Light Fuel Oil Demand}
    \includegraphics[width=\linewidth]{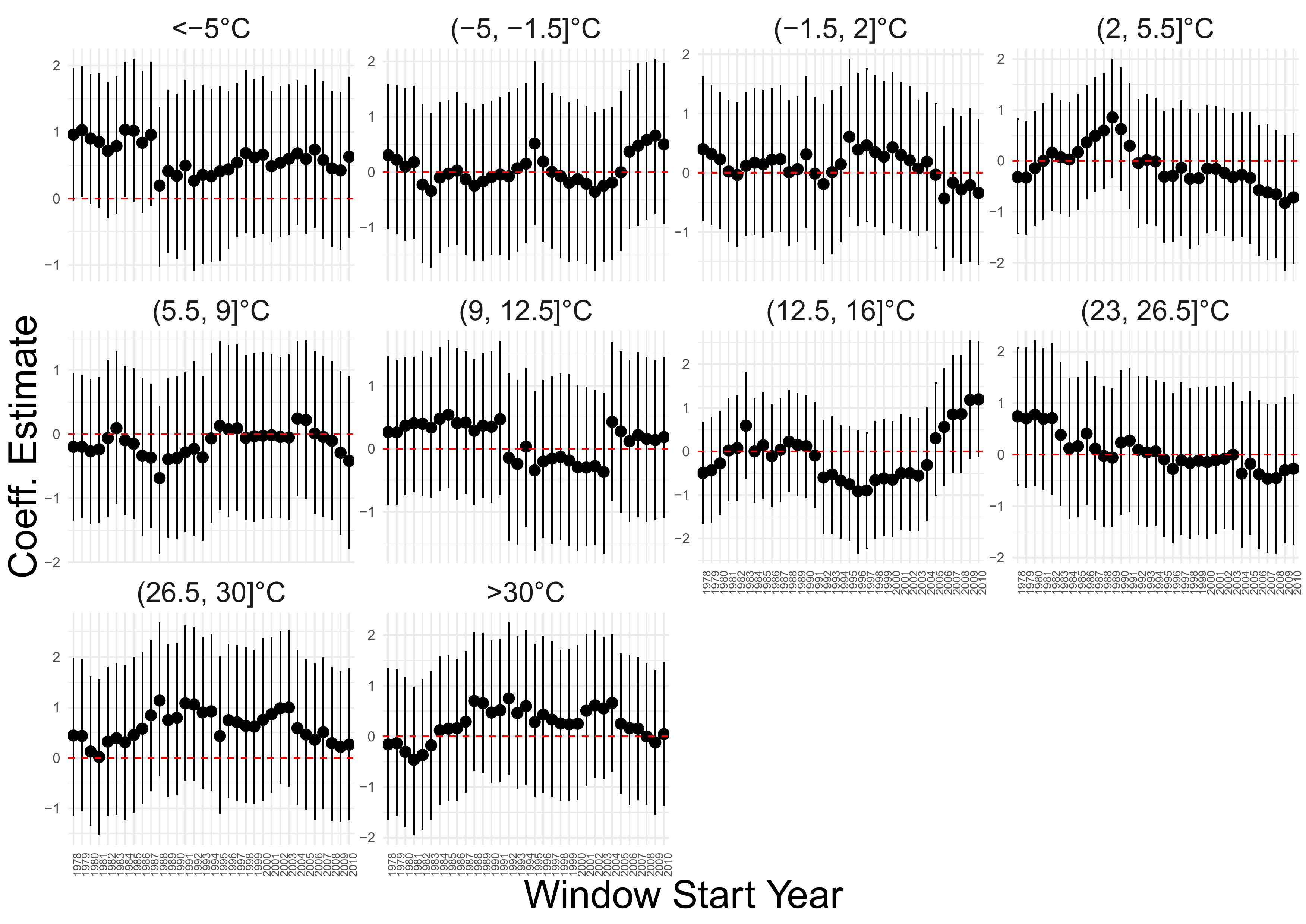}
    \caption{Estimated impact of a shift of temperature exposure of the population to ten different temperature bins (\celsius{}), relative to the 16\celsius{} to 23\celsius{} bin. On log residential lifgt fuel oik demand Using a 3.5\celsius{} bin width and including 90\% credible intervals. Based on a 15 year rolling window.
\\
\scriptsize{Estimated using Bayesian Partial Pooling Model (NUTS-sampler). Data from ENERDATA; \citetalias{nasav2}, \citetalias{population_data}, \citetalias{population1}, \citetalias{population2}.}}
    \label{fig:window_oil}
\end{figure}
\end{landscape}
\newpage
\begin{landscape}
\begin{figure}[!htbp]
\centering
\textbf{Electricity Demand}
\\
\includegraphics[width=0.8\linewidth]{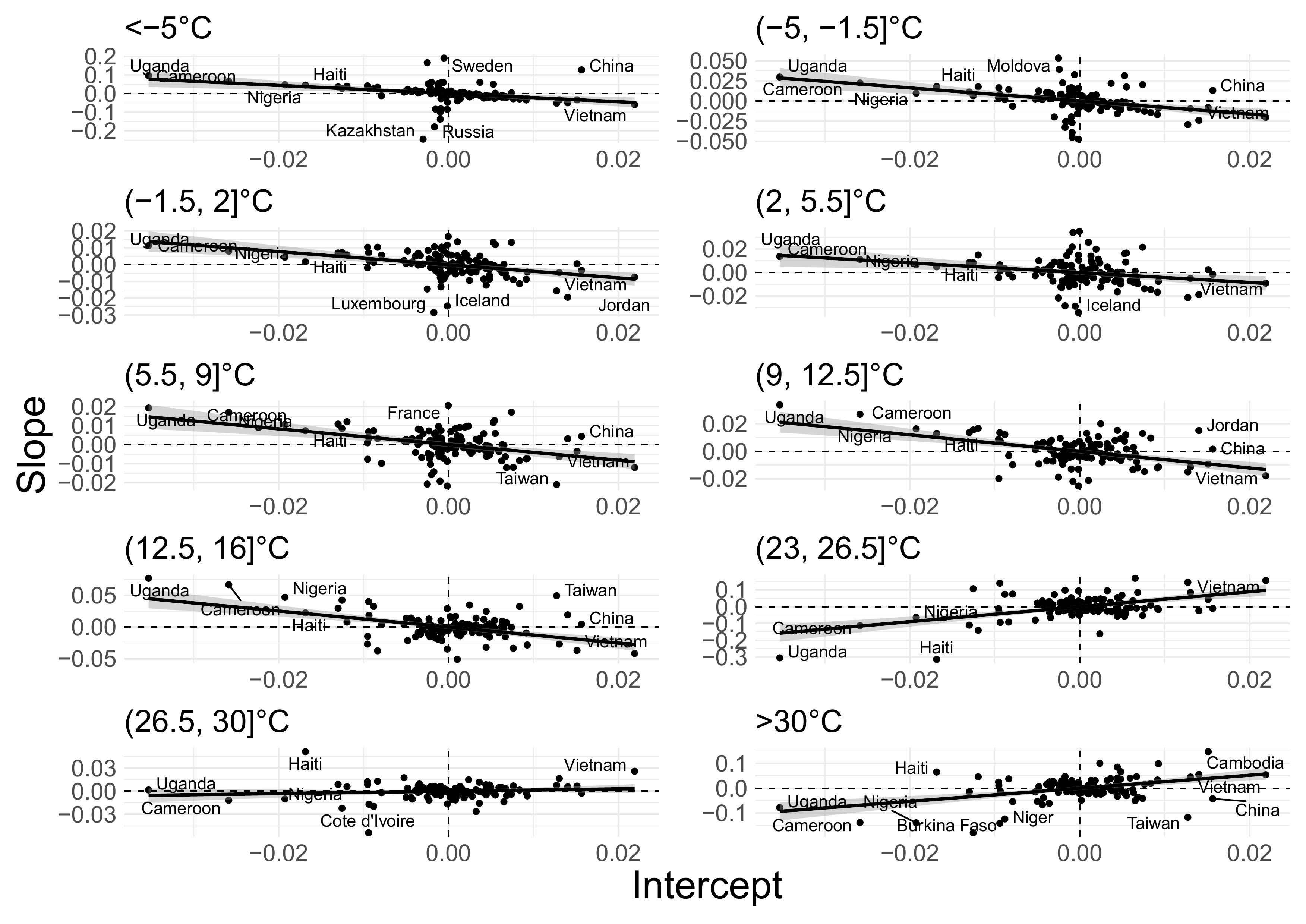}
\caption{Estimates of posterior means for individual intercepts and temperature slope parameters of log residential electricity demand. Using 3.5\celsius{} binwidth. Including linear trend with 95\% confidence bands.
\\
\scriptsize{Estimated using Bayesian Partial Pooling Model (NUTS-sampler). Data from ENERDATA;\citetalias{nasav2}, \citetalias{population_data}, \citetalias{population1}, \citetalias{population2}.}}

\label{fig:intercept_slope_ele}
\end{figure}
\newpage
\begin{figure}[!htbp]
\centering
\textbf{Natural Gas Demand}
\\
\includegraphics[width=0.8\linewidth]{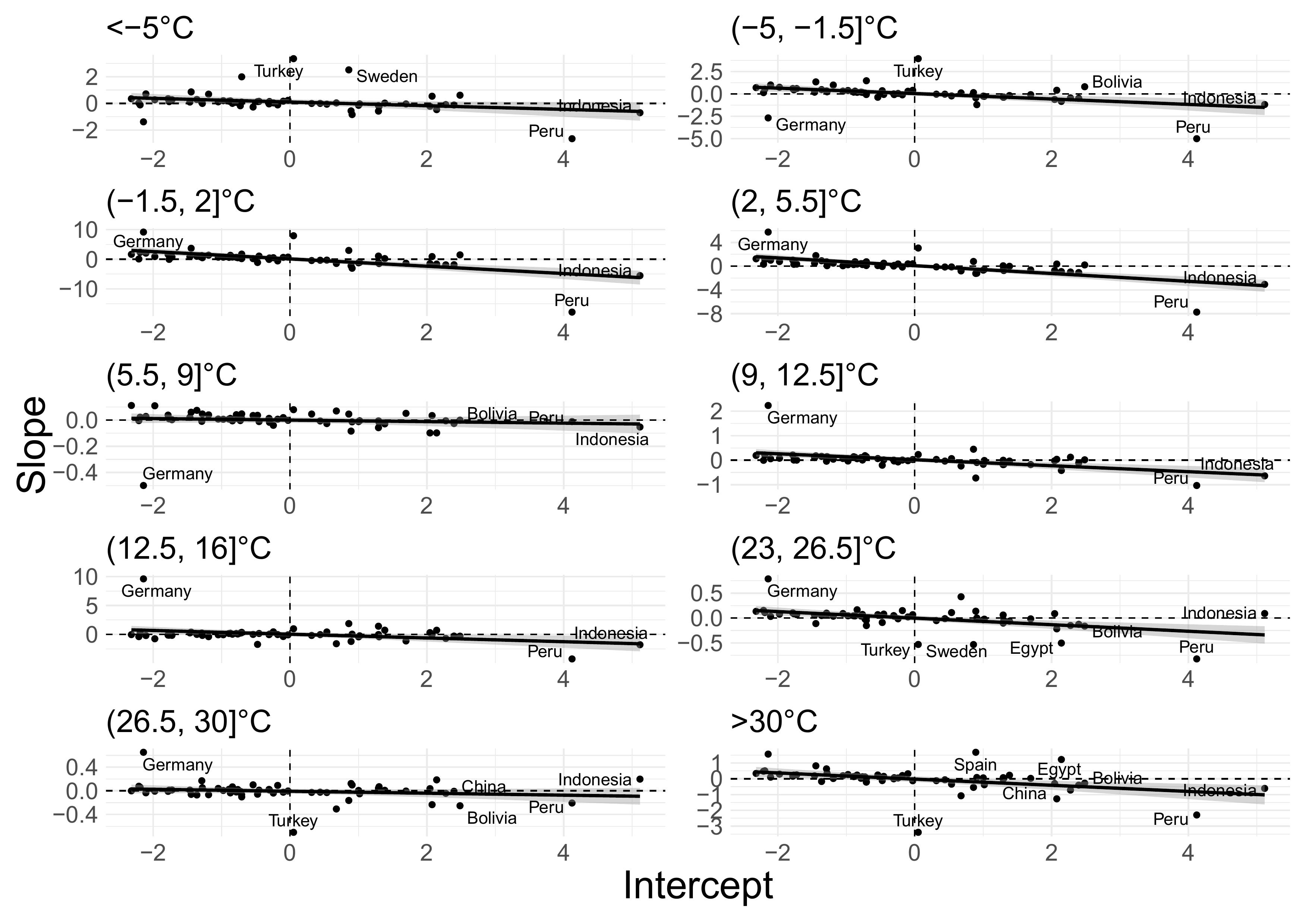}
\caption{Estimates of posterior means for individual intercepts and temperature slope parameters of log natural gas electricity demand. Using 3.5\celsius{} binwidth. Including linear trend with 95\% confidence bands.
\\
\scriptsize{Estimated using Bayesian Partial Pooling Model (NUTS-sampler). Including linear trend. Data from ENERDATA;\citetalias{nasav2}, \citetalias{population_data}, \citetalias{population1}, \citetalias{population2}.}}

\label{fig:intercept_slope_gas}
\end{figure}
\newpage
\begin{figure}[!htbp]
\centering
\textbf{Light Fuel Oil Demand}
\includegraphics[width=0.8\linewidth]{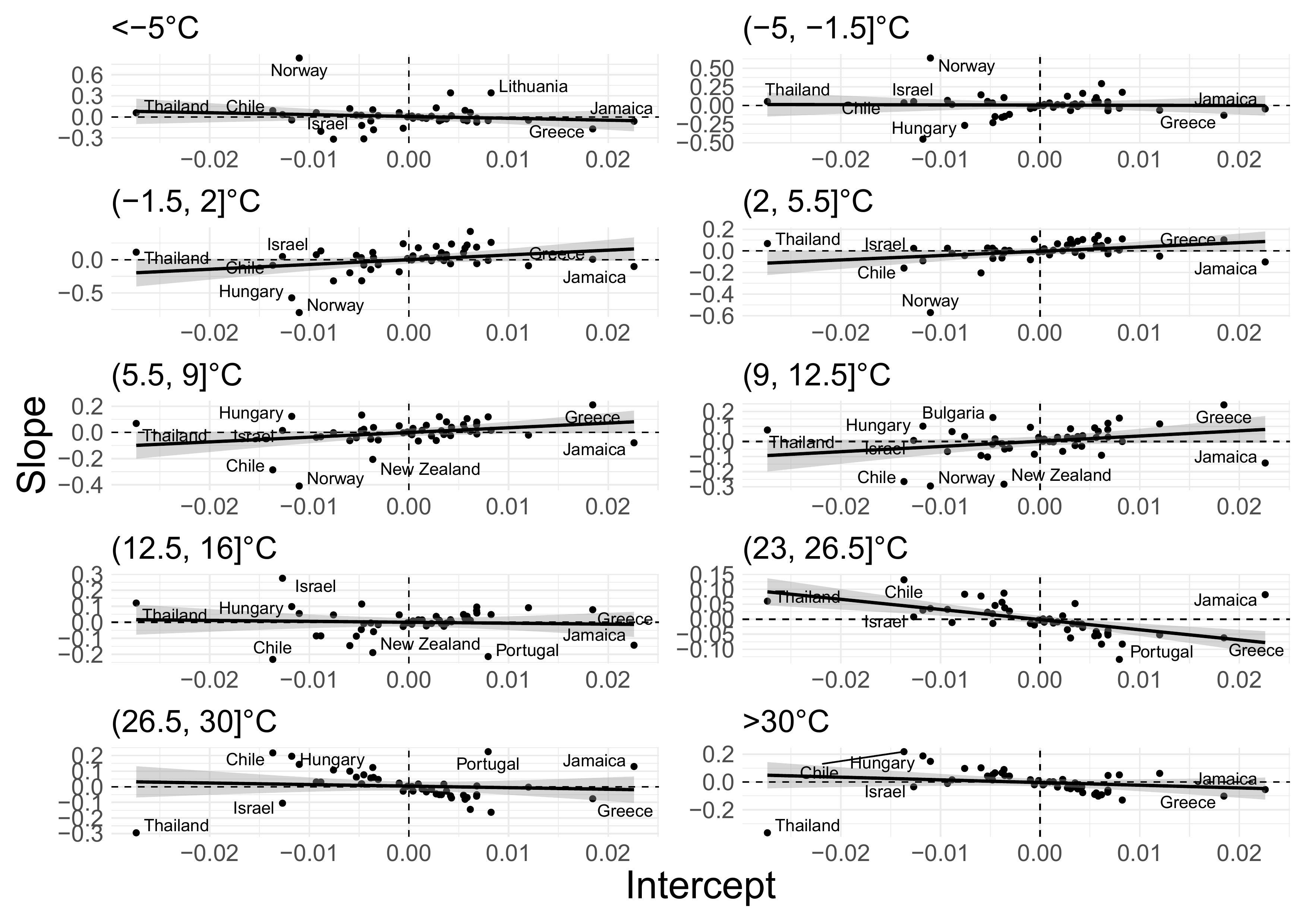}
\caption{Estimates of posterior means for individual intercepts and temperature slope parameters of log light fuel oil demand. Using 3.5\celsius{} binwidth. Including linear trend with 95\% confidence bands.
\scriptsize{Estimated using Bayesian Partial Pooling Model (NUTS-sampler).Data from ENERDATA;\citetalias{nasav2}, \citetalias{population_data}, \citetalias{population1}, \citetalias{population2}.}}
\label{fig:intercept_slope_oil}
\end{figure}
\end{landscape}
\begin{figure}[!htbp]
    \centering
    \includegraphics[width=\textwidth]{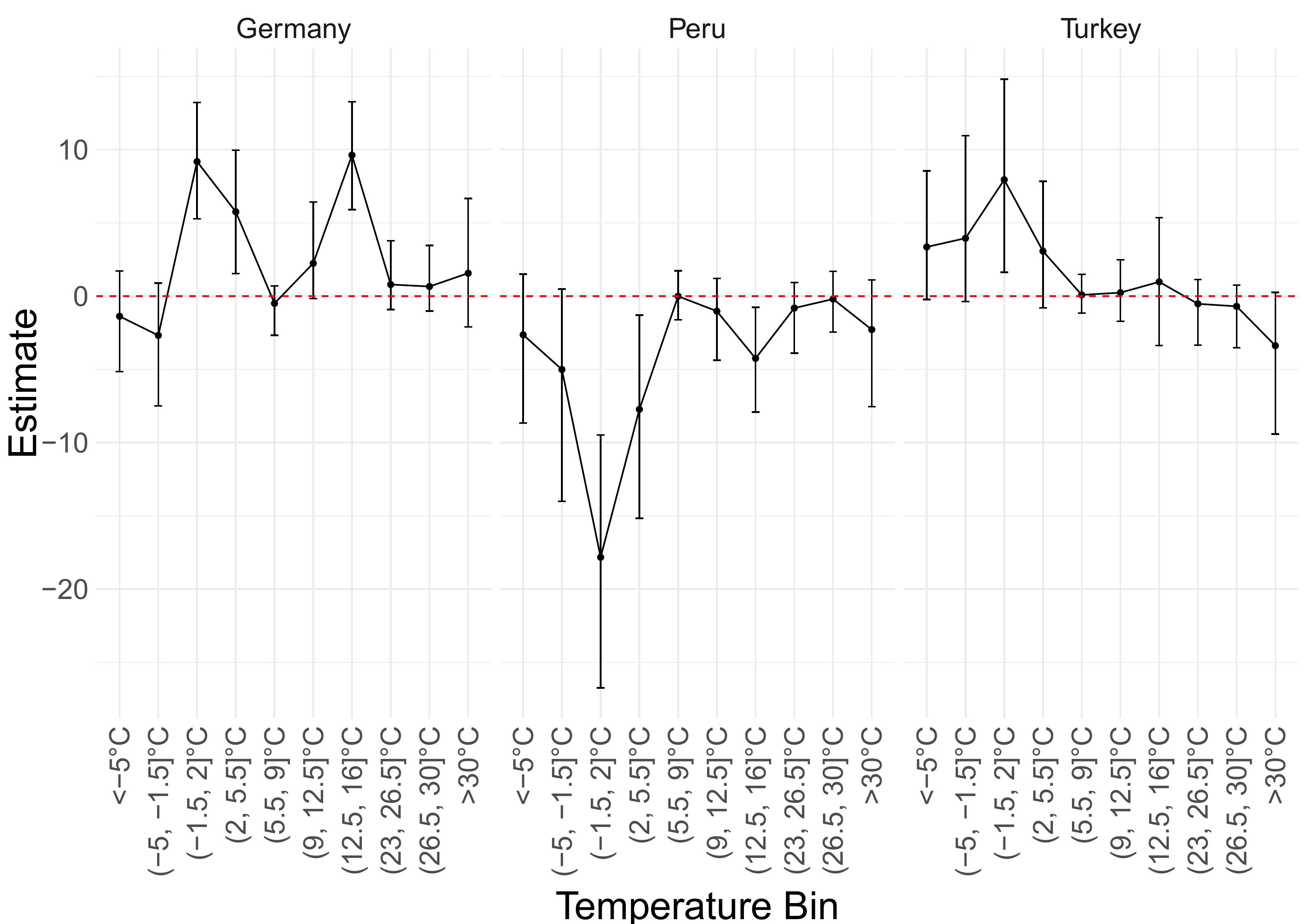}
    \caption{Temperature effects from -5\celsius{} to 30\celsius{} for selected Countries natural gas.
    \\
\scriptsize{Estimated using Bayesian Partial Pooling Model (NUTS-sampler). Data from ENERDATA; \citetalias{nasav2}, \citetalias{population_data}, \citetalias{population1}, \citetalias{population2}.}}
    \label{fig:country_slopes_gas}
\end{figure}
\begin{landscape}
    \begin{figure}[!htbp]
    \centering
    \textbf{Electricity}
    \\
    \includegraphics[width=\linewidth]{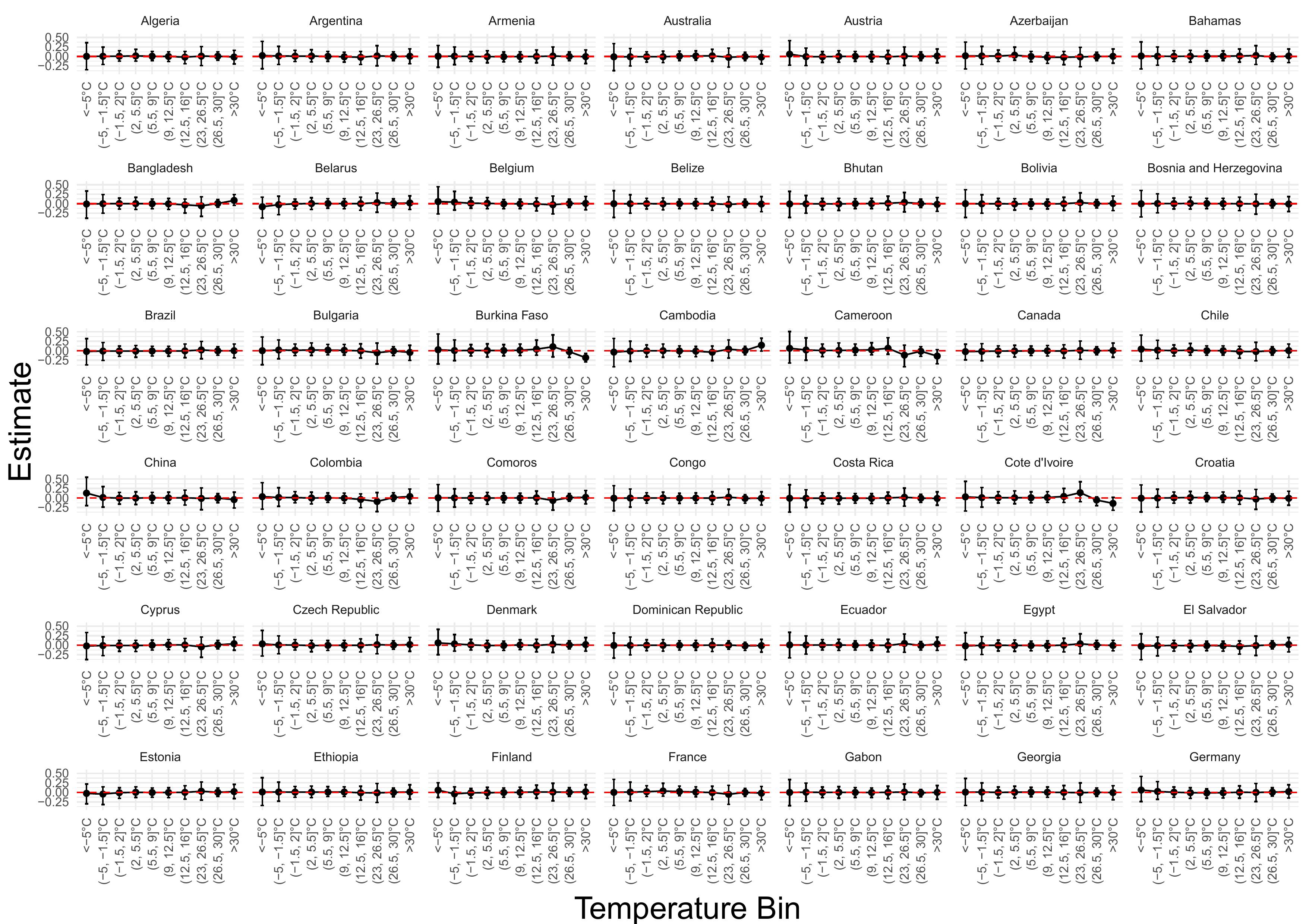}
    \caption{Estimated country specific deviation from the population level estimate for the effect of a shift of temperature exposure of the population to ten different temperature bins (\celsius{}) on log residential electricity demand, relative to the 16\celsius{} to 23\celsius{} bin. Using a 3.5\celsius{} bin width and including 90\% credible intervals.
    \\
\scriptsize{Estimated using Bayesian Partial Pooling Model (NUTS-sampler). Data from ENERDATA; \citetalias{nasav2}, \citetalias{population_data}, \citetalias{population1}, \citetalias{population2}.}}
    \label{fig:country_slopes_electricity1}
\end{figure}
\end{landscape}
\begin{landscape}
    \begin{figure}[!htbp]
    \centering
    \textbf{Electricity}
    \\
    \includegraphics[width=\linewidth]{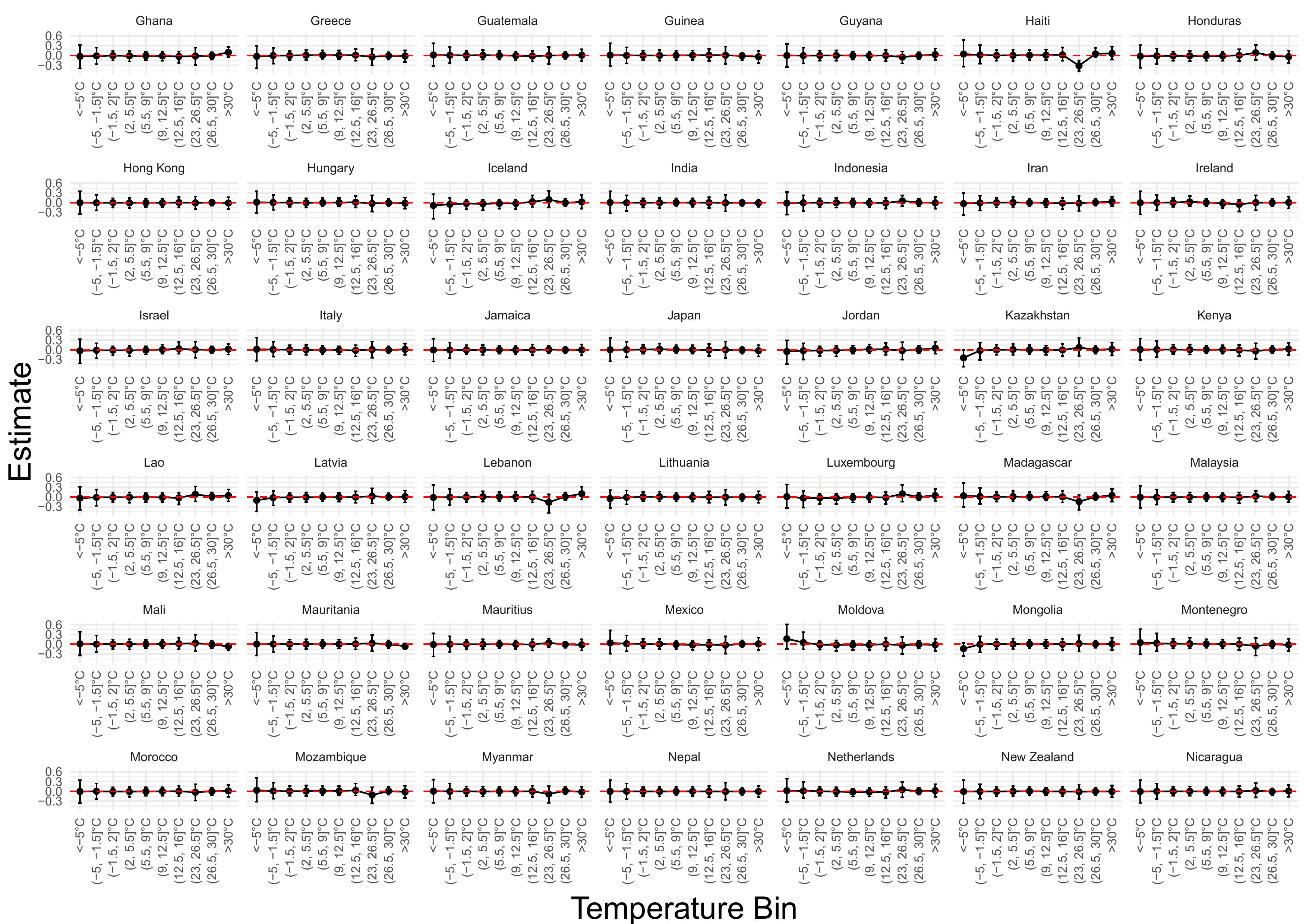}
    \caption{Estimated country specific deviation from the population level estimate for the effect of a shift of temperature exposure of the population to ten different temperature bins (\celsius{}) on log residential electricity demand, relative to the 16\celsius{} to 23\celsius{} bin. Using a 3.5\celsius{} bin width and including 90\% credible intervals.
    \\
\scriptsize{Estimated using Bayesian Partial Pooling Model (NUTS-sampler). Data from ENERDATA; \citetalias{nasav2}, \citetalias{population_data}, \citetalias{population1}, \citetalias{population2}.}}
    \label{fig:country_slopes_electricity2}
\end{figure}
\end{landscape}
\begin{landscape}
    \begin{figure}[!htbp]
    \centering
    \textbf{Electricity}
    \\
    \includegraphics[width=\linewidth]{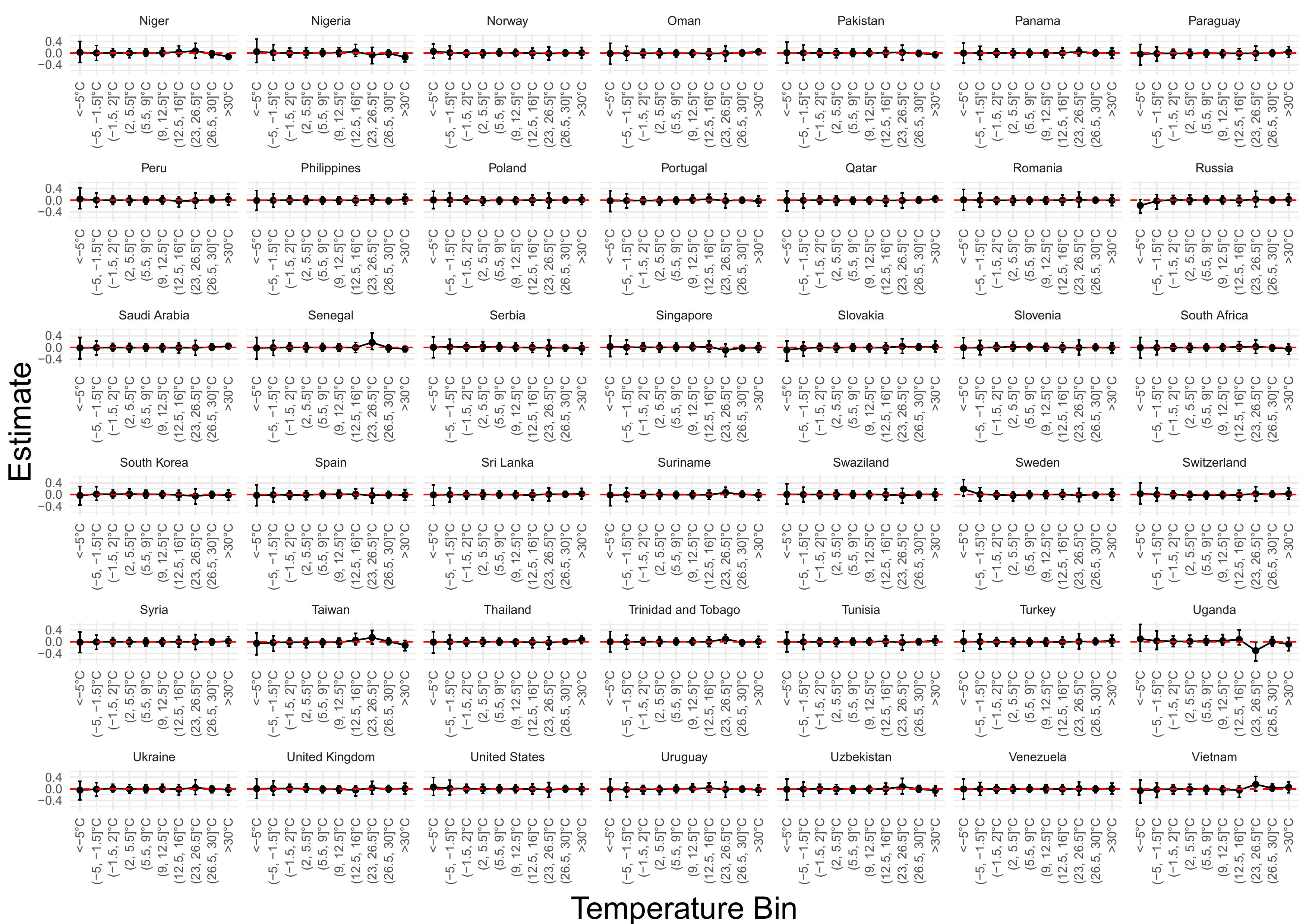}
    \caption{Estimated country specific deviation from the population level estimate for the effect of a shift of temperature exposure of the population to ten different temperature bins (\celsius{}) on log residential electricity demand, relative to the 16\celsius{} to 23\celsius{} bin. Using a 3.5\celsius{} bin width and including 90\% credible intervals.
    \\
\scriptsize{Estimated using Bayesian Partial Pooling Model (NUTS-sampler). Data from ENERDATA; \citetalias{nasav2}, \citetalias{population_data}, \citetalias{population1}, \citetalias{population2}.}}
    \label{fig:country_slopes_electricity_3}
\end{figure}
\end{landscape}
\begin{landscape}
    \begin{figure}[!htbp]
    \centering
    \textbf{Natural Gas}
    \\
    \includegraphics[width=\linewidth]{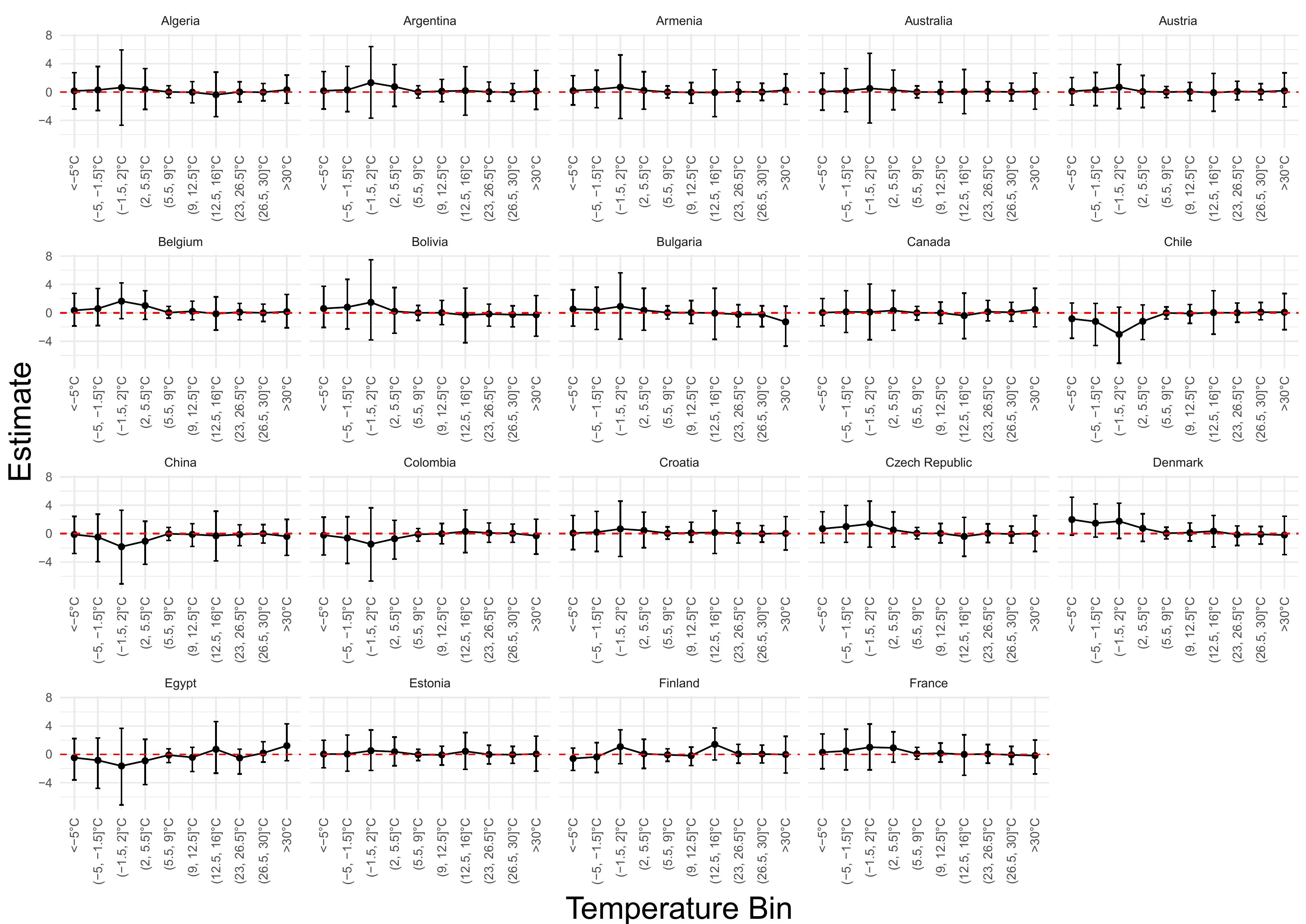}
    \caption{Estimated country specific deviation from the population level estimate for the effect of a shift of temperature exposure of the population to ten different temperature bins (\celsius{}) on log residential natural gas demand, relative to the 16\celsius{} to 23\celsius{} bin. Using a 3.5\celsius{} bin width and including 90\% credible intervals.
    \\
\scriptsize{Estimated using Bayesian Partial Pooling Model (NUTS-sampler). Data from ENERDATA; \citetalias{nasav2}, \citetalias{population_data}, \citetalias{population1}, \citetalias{population2}.}}
    \label{fig:country_slopes_gas1}
\end{figure}
\end{landscape}
\begin{landscape}
    \begin{figure}[!htbp]
    \centering
    \textbf{Natural Gas}
    \\
    \includegraphics[width=\linewidth]{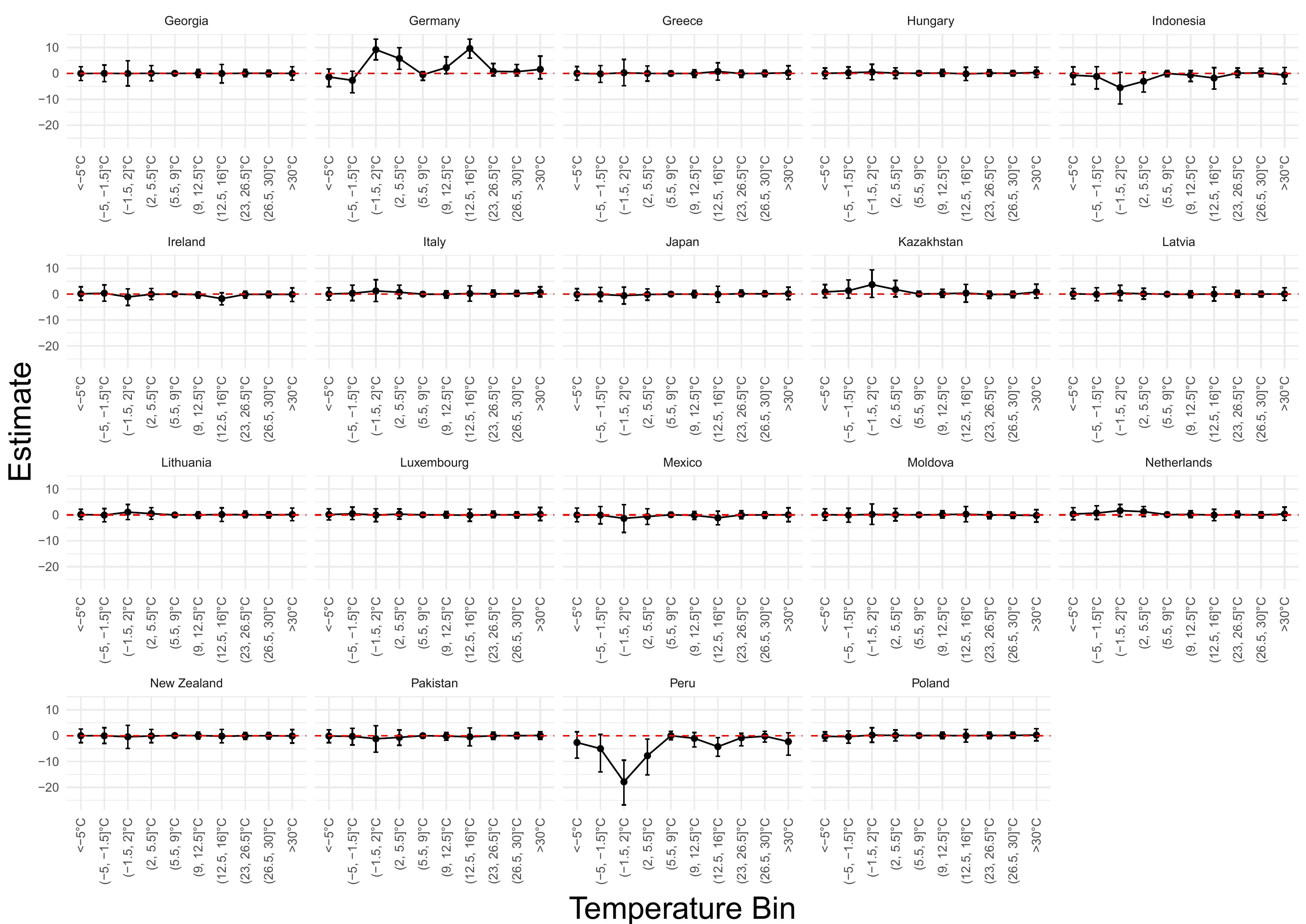}
    \caption{Estimated country specific deviation from the population level estimate for the effect of a shift of temperature exposure of the population to ten different temperature bins (\celsius{}) on log residential natural gas demand, relative to the 16\celsius{} to 23\celsius{} bin. Using a 3.5\celsius{} bin width and including 90\% credible intervals.
    \\
\scriptsize{Estimated using Bayesian Partial Pooling Model (NUTS-sampler). Data from ENERDATA; \citetalias{nasav2}, \citetalias{population_data}, \citetalias{population1}, \citetalias{population2}.}}
    \label{fig:country_slopes_gas2}
\end{figure}
\end{landscape}
\begin{landscape}
    \begin{figure}[!htbp]
    \centering
    \textbf{Natural Gas}
    \\
    \includegraphics[width=\linewidth]{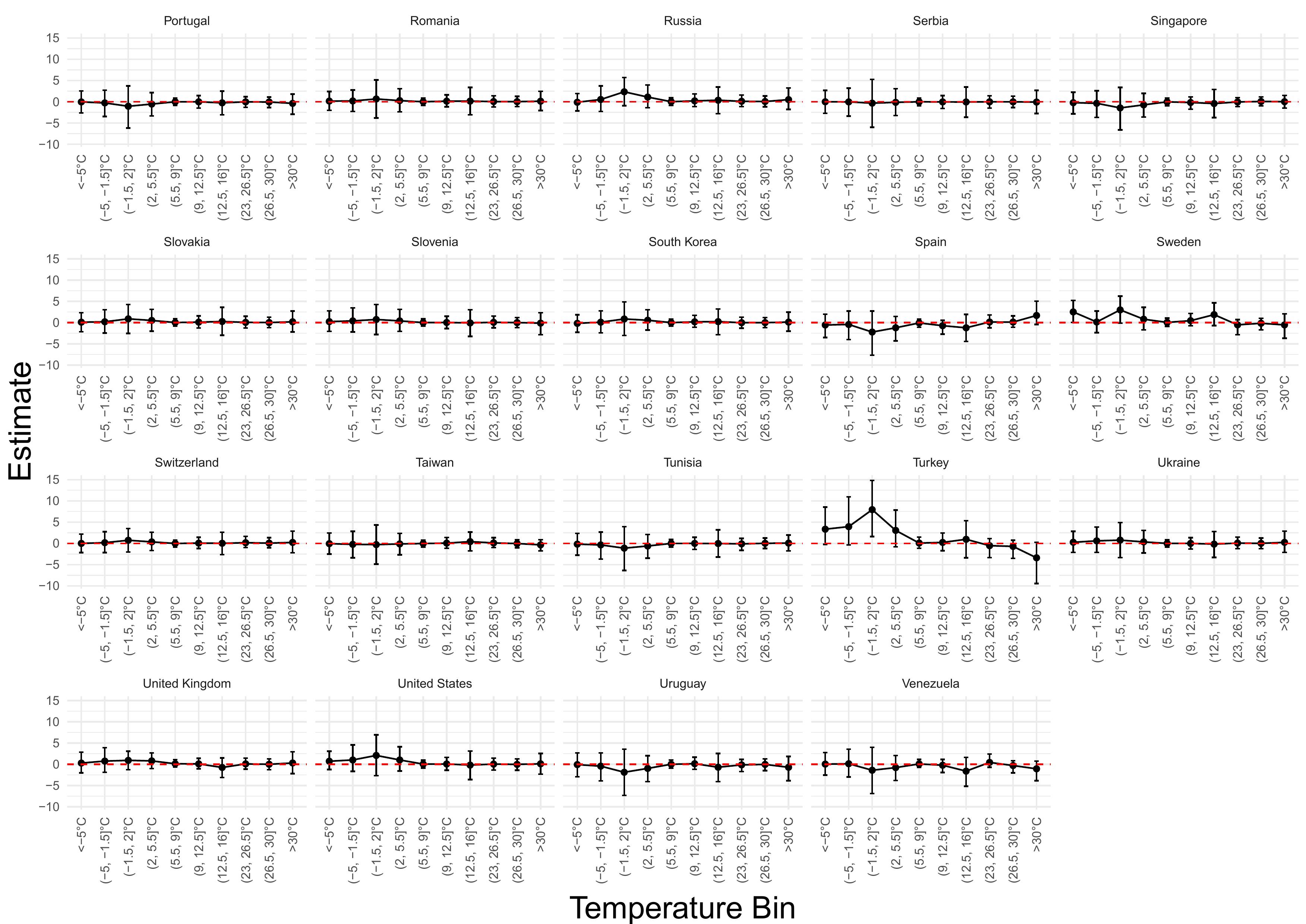}
    \caption{Estimated country specific deviation from the population level estimate for the effect of a shift of temperature exposure of the population to ten different temperature bins (\celsius{}) on log residential natural gas demand, relative to the 16\celsius{} to 23\celsius{} bin. Using a 3.5\celsius{} bin width and including 90\% credible intervals.
    \\
\scriptsize{Estimated using Bayesian Partial Pooling Model (NUTS-sampler). Data from ENERDATA; \citetalias{nasav2}, \citetalias{population_data}, \citetalias{population1}, \citetalias{population2}.}}
    \label{fig:country_slopes_gas3}
\end{figure}
\end{landscape}
\begin{landscape}
    \begin{figure}[!htbp]
    \centering
    \textbf{Light Fuel Oil}
    \\
    \includegraphics[width=\linewidth]{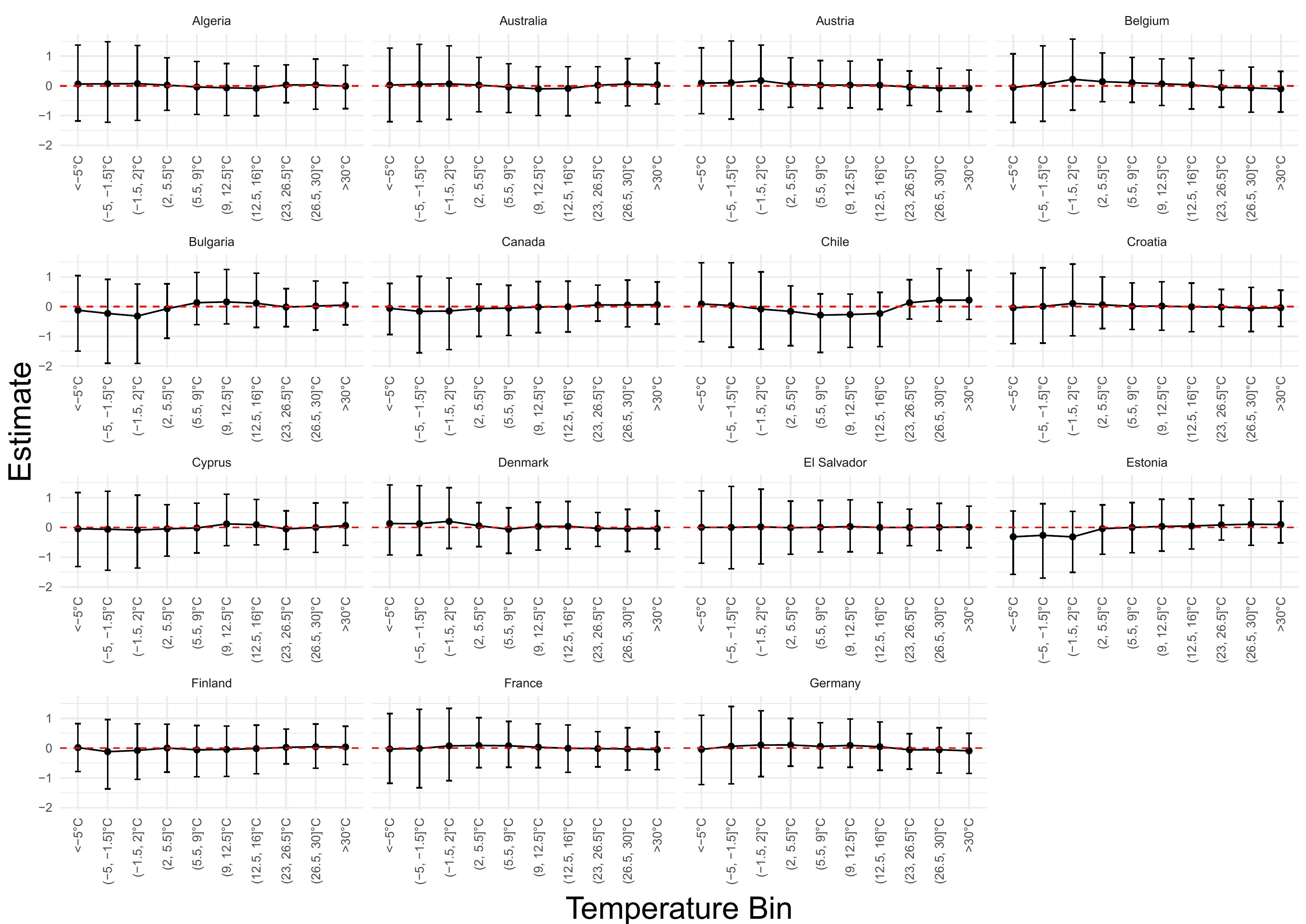}
    \caption{Estimated country specific deviation from the population level estimate for the effect of a shift of temperature exposure of the population to ten different temperature bins (\celsius{}) on log light fuel oil demand, relative to the 16\celsius{} to 23\celsius{} bin. Using a 3.5\celsius{} bin width and including 90\% credible intervals.
    \\
\scriptsize{Estimated using Bayesian Partial Pooling Model (NUTS-sampler). Data from ENERDATA; \citetalias{nasav2}, \citetalias{population_data}, \citetalias{population1}, \citetalias{population2}.}}
    \label{fig:country_slopes_oil1}
\end{figure}
\end{landscape}
\begin{landscape}
    \begin{figure}[!htbp]
    \centering
    \textbf{Light Fuel Oil}
    \\
    \includegraphics[width=\linewidth]{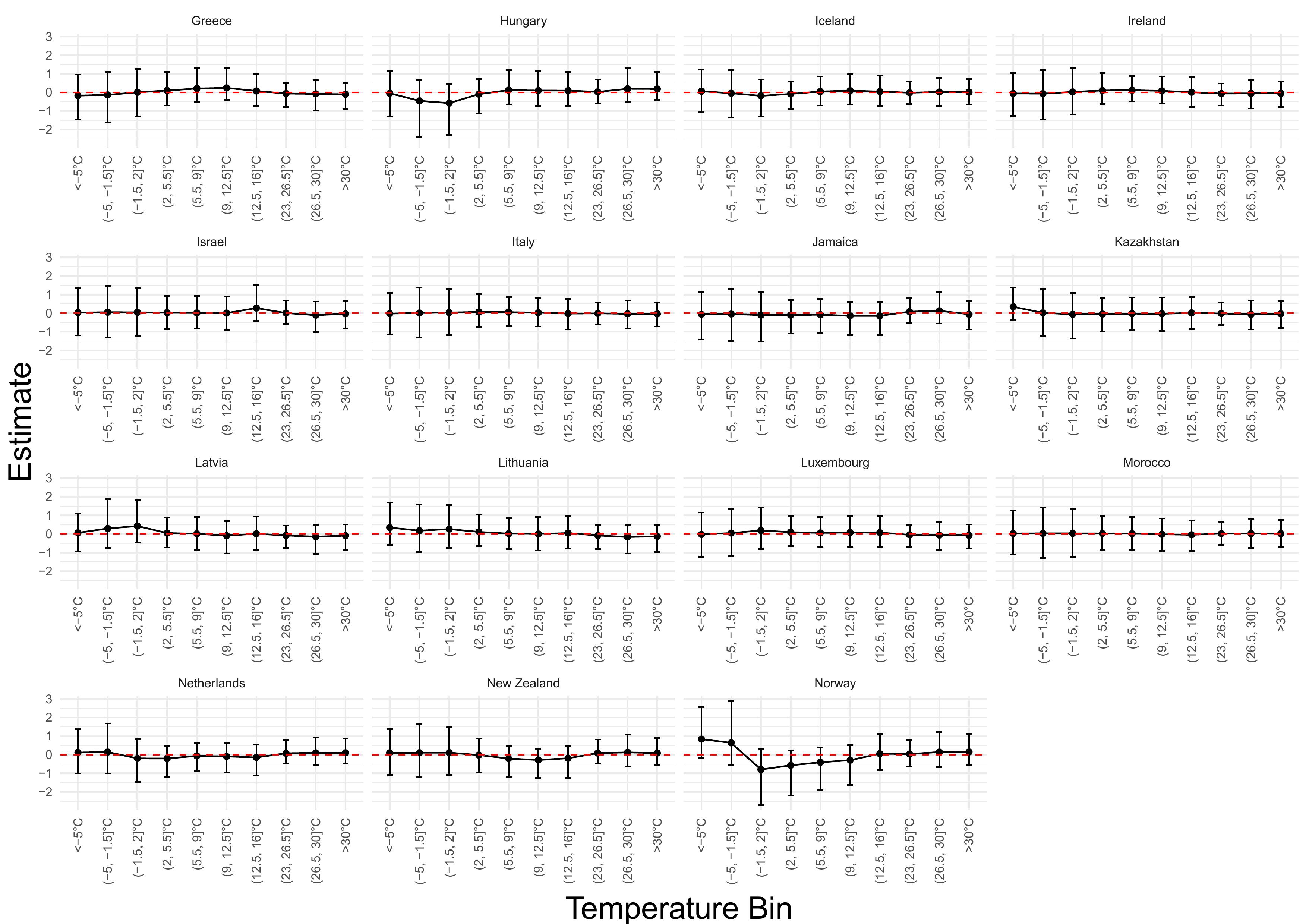}
    \caption{Estimated country specific deviation from the population level estimate for the effect of a shift of temperature exposure of the population to ten different temperature bins (\celsius{}) on log light fuel oil demand, relative to the 16\celsius{} to 23\celsius{} bin. Using a 3.5\celsius{} bin width and including 90\% credible intervals.
    \\
\scriptsize{Estimated using Bayesian Partial Pooling Model (NUTS-sampler). Data from ENERDATA; \citetalias{nasav2}, \citetalias{population_data}, \citetalias{population1}, \citetalias{population2}.}}
    \label{fig:country_slopes_oil2}
\end{figure}
\end{landscape}
\begin{landscape}
    \begin{figure}[!htbp]
    \centering
    \textbf{Light Fuel Oil}
    \\
    \includegraphics[width=\linewidth]{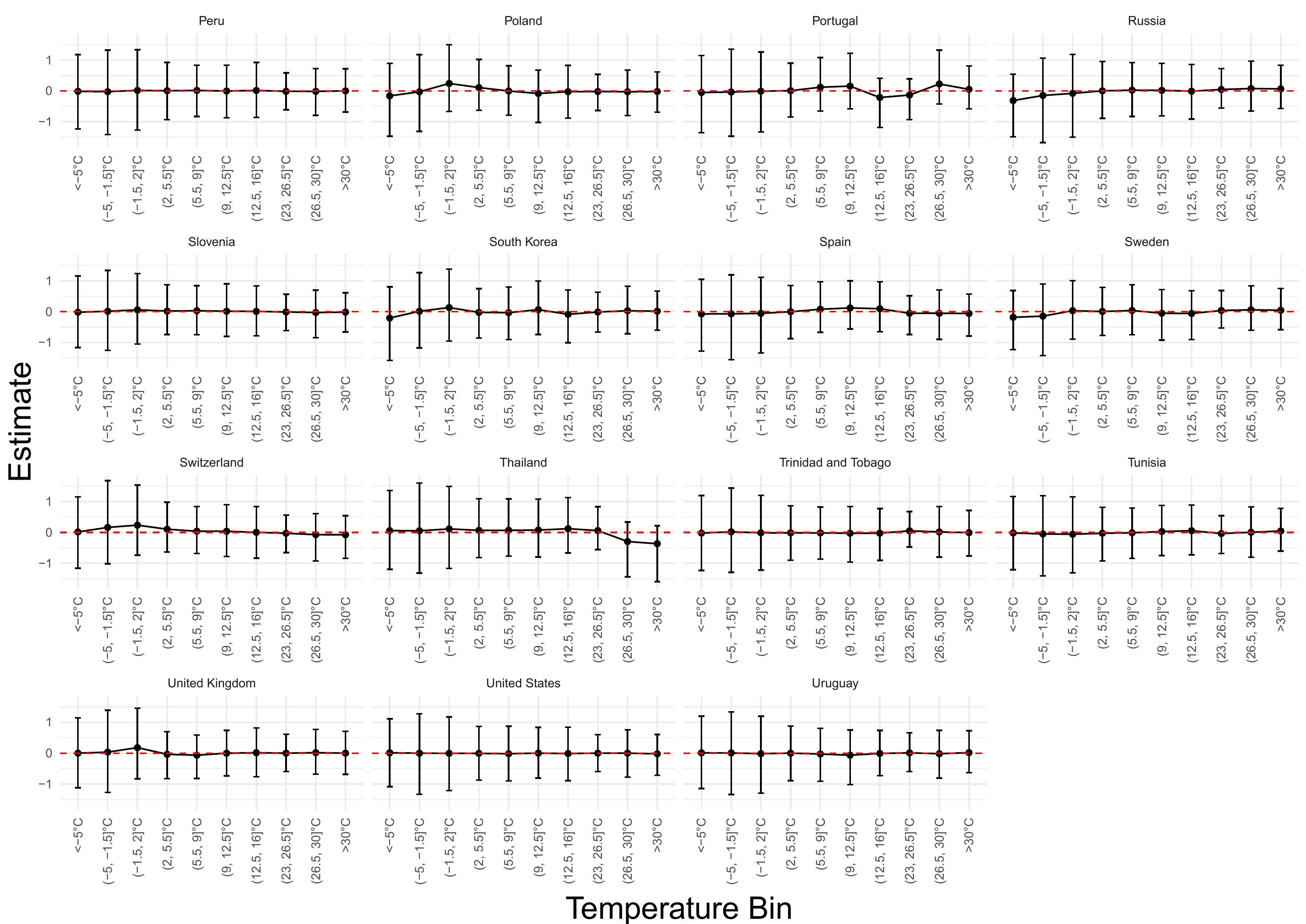}
    \caption{Estimated country specific deviation from the population level estimate for the effect of a shift of temperature exposure of the population to ten different temperature bins (\celsius{}) on log light fuel oil demand, relative to the 16\celsius{} to 23\celsius{} bin. Using a 3.5\celsius{} bin width and including 90\% credible intervals.
    \\
\scriptsize{Estimated using Bayesian Partial Pooling Model (NUTS-sampler). Data from ENERDATA; \citetalias{nasav2}, \citetalias{population_data}, \citetalias{population1}, \citetalias{population2}.}}
    \label{fig:country_slopes_oil3}
\end{figure}
\end{landscape}

\begin{landscape}
\begin{figure}
    \centering
    \includegraphics[width=0.8\linewidth]{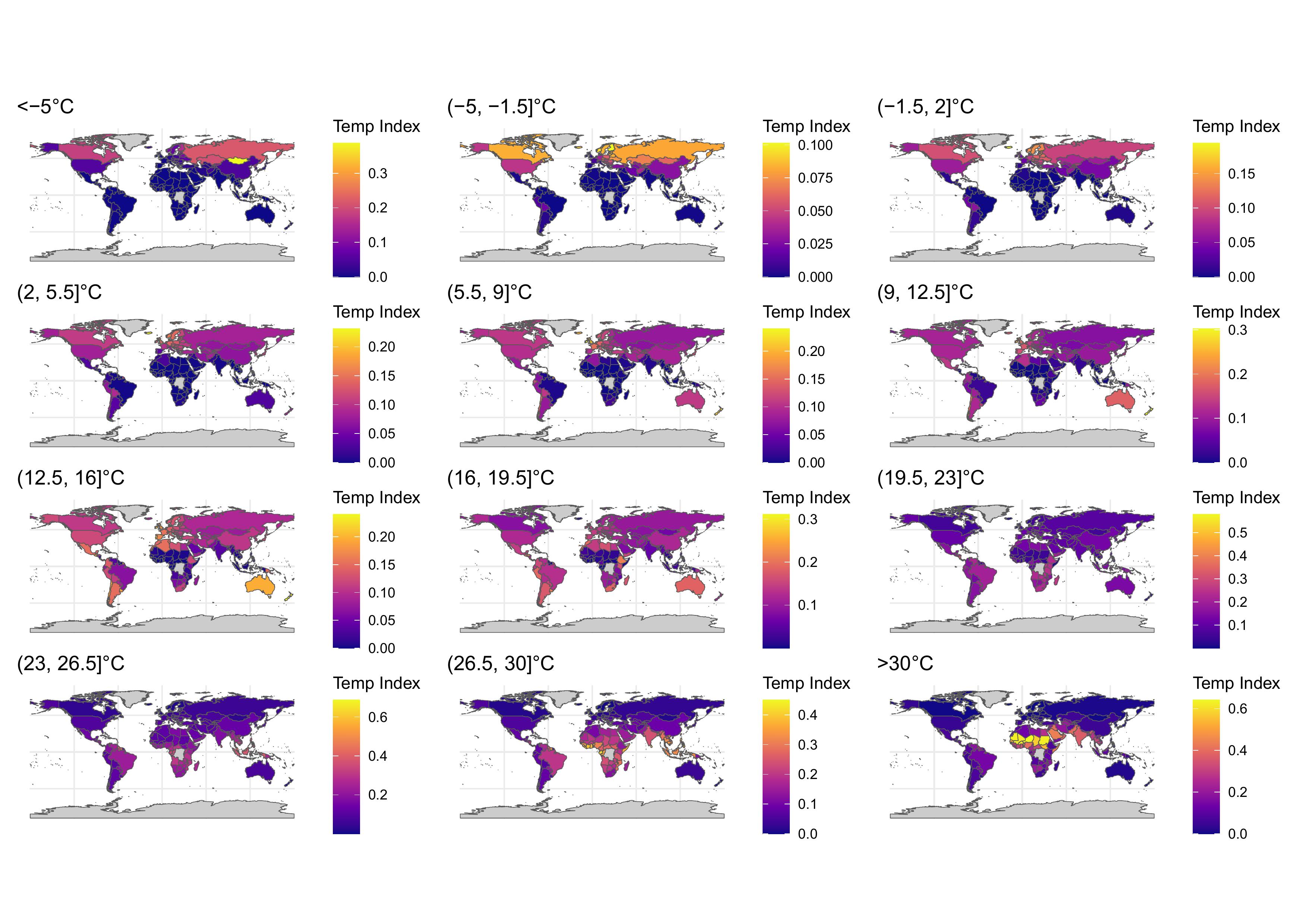}
    \caption{Regionalized temperature index for 12 different temperature intervals, averaged over 1978 to 2023.
    \scriptsize{data from \citetalias{nasav2,population_data,population1,population2}}}
    \label{fig:geo_vis}
\end{figure}
\end{landscape}
\newpage
\section{Replication Study}
\label{appendixD}
\setcounter{figure}{0}
\setcounter{table}{0}
\renewcommand{\thetable}{D\arabic{table}}
\renewcommand{\thefigure}{D\arabic{figure}}

\begin{figure}[!htbp]
\centering
\includegraphics[width=\textwidth]{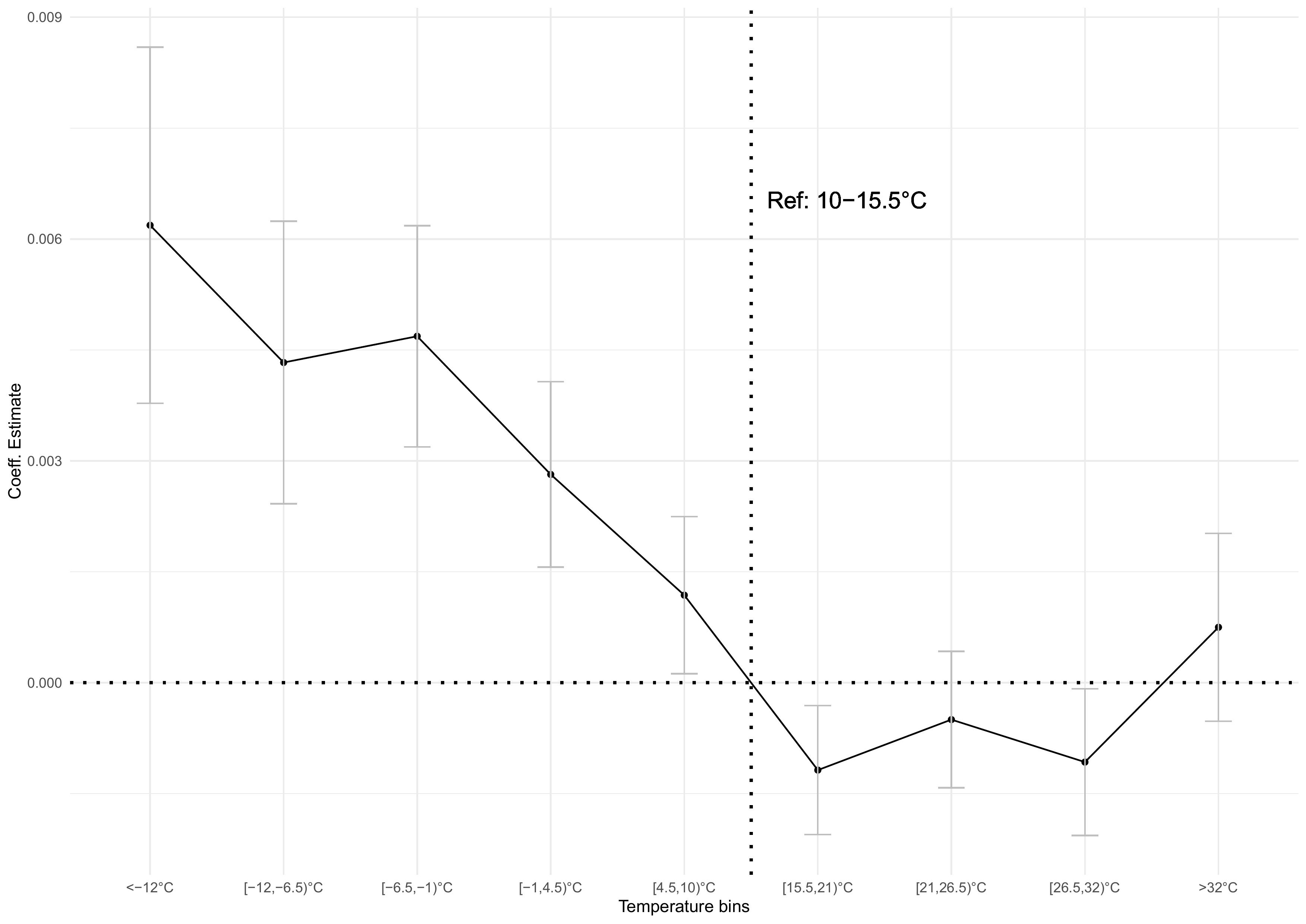}
\caption{Estimated impact of a day in nine daily mean temperature (\celsius{}) bins on log annual residential electricity demand, relative to a day in the 10° C-15.5\celsius{} bin.
\scriptsize{Slope parameter and confidence interval for temperature variables. Estimated using TWFE Estimator. Data from ENERDATA; \citetalias{nasav2}.}}
\label{fig:replication_study1}
\end{figure}

\begin{figure}[!htbp]
\centering
\includegraphics[width=\textwidth]{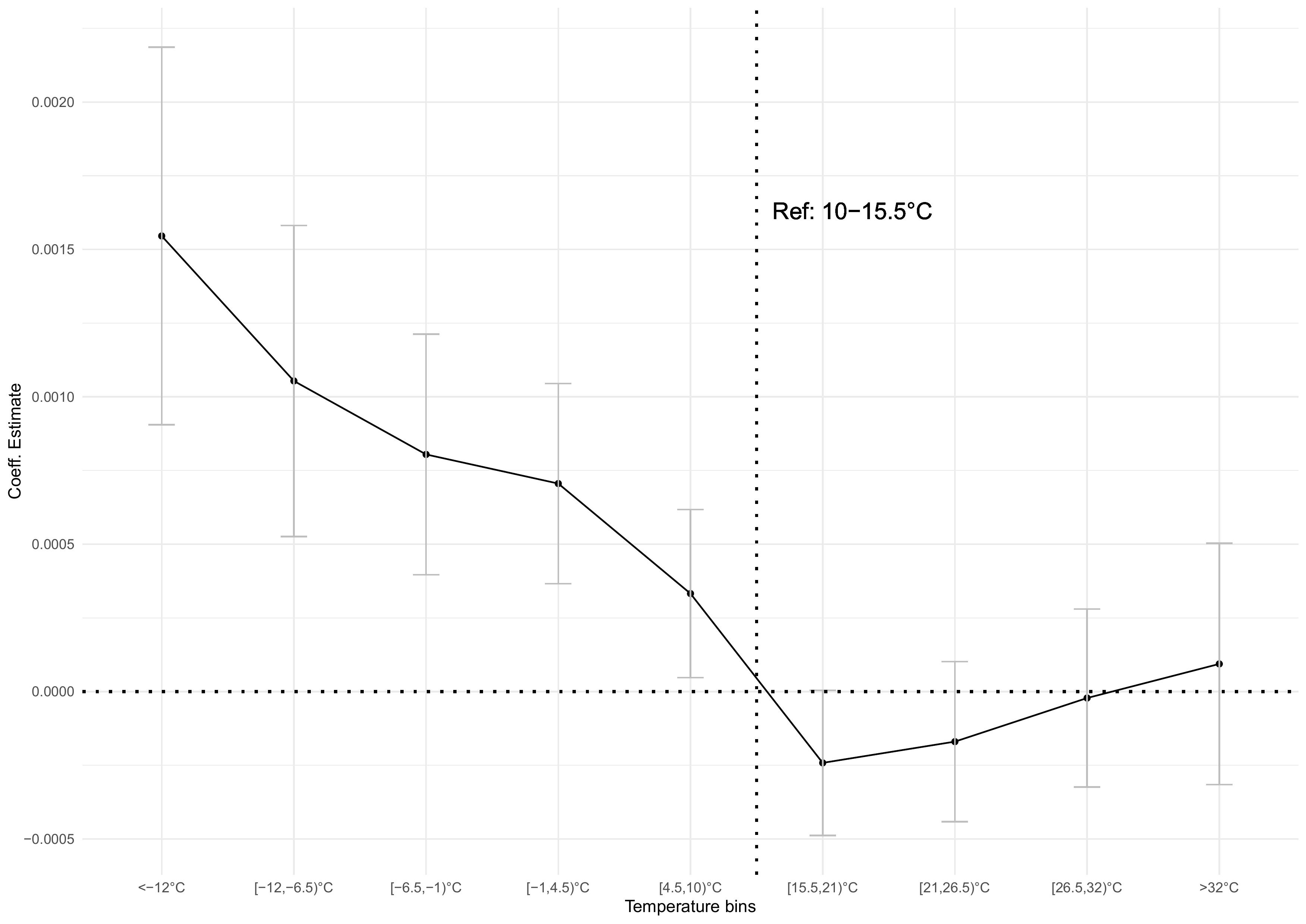}
\caption{Estimated impact of a day in nine daily mean temperature (\celsius{}) bins on log annual residential electricity demand, relative to a day in the 10° C-15.5\celsius{} bin.
\scriptsize{Slope parameter and confidence interval for temperature variables. Estimated using TWFE Estimator.Data from ENERDATA; \citetalias{nasav2}.}}
\label{fig:replication_study2}
\end{figure}

\begin{figure}[!htbp]
\centering
\includegraphics[width=0.8\linewidth]{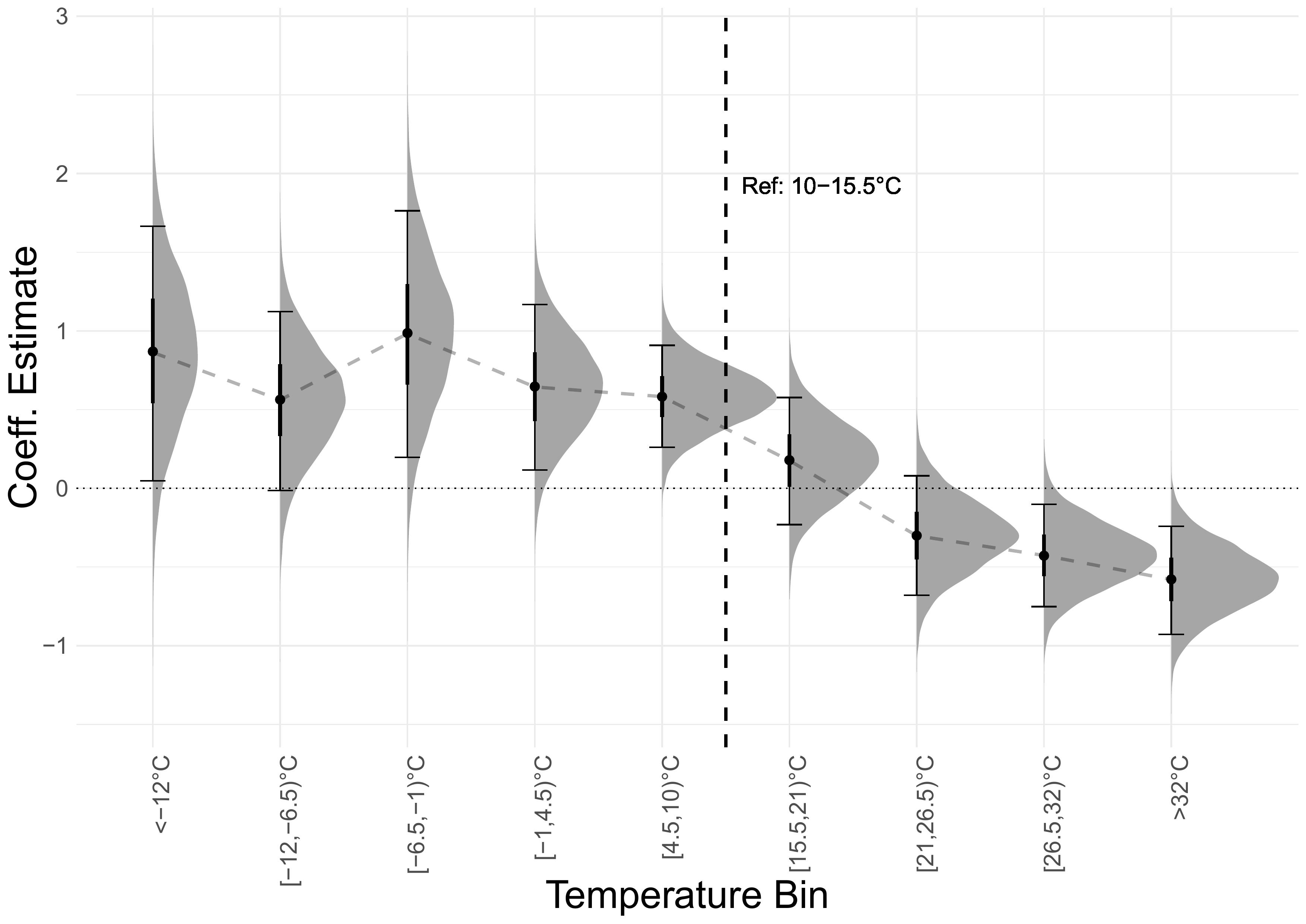}
\caption{Estimated impact of a shift of temperature exposure of the population to nine different temperature bins (\celsius{}), relative to the 10\celsius{} -15.5\celsius{} bin including 50\% and 90\% credible intervals.
\scriptsize{Estimated using Bayesian Partial Pooling Model (NUTS-sampler).Data from ENERDATA;\citetalias{nasav2,population_data,population1,population2}.}}
\label{fig:replication_study3}
\end{figure}

In this section we replicate the study of \cite{replicated_study} using our own data but leaving out the controls for precipitation since these are not included in our data set and are unlikely to affect the estimation results in a meaningful way.
\begin{align*}
    \log(y_{i,t}) &= \sum_{k=1}^K \Tilde{\beta}_k \Tilde{F}_{i,t}^k+\gamma \Tilde{\boldsymbol{X}}_{i,t} +\mu_i+\delta_t+\epsilon_{i,t} \\
\end{align*}
In this model, our dependent variable $\log(y_{i,t})$ is the natural logarithm of the per capita residential electricity demand. On the right-hand side, we then include the temperature exposure variables $\Tilde{F}_{i,t}^k$, which are coded as the number of days in year $t$ where the daily mean temperature of  country $i$ falls into the $k$th bin, see for comparison \citep{replicated_study}. $\mu_i$ and $\delta_t$ capture country and year fixed effects, respectively. $\Tilde{\boldsymbol{X}}_{i,t}$ contains the country-level natural logarithm of population and GDP and the squares thereof. $\epsilon_{i,t}$ denotes the stochastic error term.   
\paragraph{}
Note that temperature values were transformed from degree Fahrenheit to degree Celsius. Thus, bin widths slightly deviate from the original specification. Also, we use per-capita electricity demand and GDP instead of absolute values. Compared with the results of \cite{replicated_study} who focused solely on the USA, we find remarkable similarities, especially for the cooling effect see Figure \ref{fig:replication_study1}. When temperature drops below 0\celsius{}, we first see a relatively linear increase in the log-residential electricity demand. At temperatures below -6\celsius{} the effect seems to flatten out first and then increase linearly again. In contrast to \cite{replicated_study} we do not observe any cooling effect with this specification\footnote{When including other variables such as the one year lag of the dependent variable and electricity prices the shape of the response function largely stays the same but effect sizes are smaller, see \ref{fig:replication_study2}.}.
In the next step, we compare the results from the replicated study with the results using our own model formulation with a Bayesian estimation procedure. For better comparability, we adopt the temperature binning structure and the reference bin of \cite{replicated_study}. Figure \ref{fig:replication_study3} shows that the estimate of the temperature response differs substantially. The Bayesian model draws a more complete picture of the uncertainty of our estimates, highlighting the overconfidence of the fixed-effects approach of \cite{replicated_study}. Furthermore, the heating effect here is only present for relatively colder temperatures, starting from temperatures below 4.5\celsius{} and increases less strongly with lower temperatures. Furthermore, the model detects a reduced residential electricity demand for temperatures between 21\celsius{} and 32\celsius{} and for temperatures above 32\celsius{} the model assigns a high probability to an increased residential electricity demand compared to the reference temperature of 10\celsius{} to 15.5\celsius.
Table \ref{tab:bayes_replication} reveals that this binning structure substantially changes the estimates for the income and price elasticities, as well as for the autoregressive parameter. This is a stark indicator of the sensitivity of estimation results to specification of the binning structure, and we strongly advise future research to explicitly report a variety of binning structures and explore other flexible modeling options like e.g. Splines.
\begin{table}[!htbp]
  \centering
  \caption{Parameter Estimates with 95\% Credible Intervals}
\begin{tabular}{lrrrrrr}
  \toprule
    \multicolumn{7}{c}{Population Level Estimates}\\
    \midrule
Parameter & $\hat{R}$ & Mean & SD & 2.5\% & Median & 97.5\% \\
  \midrule
$\alpha$ & 1.00 & 2.60 & 0.10 & 2.40 & 2.60 & 2.70 \\
$\nu$ & 1.00 & 0.60 & 0.00 & 0.60 & 0.60 & 0.70 \\
$\beta_{below\,-12^{\circ}C}$ & 1.00 & 0.90 & 0.50 & -0.10 & 0.90 & 1.80 \\
$\beta_{-12^{\circ}C\, to \, -6.5^{\circ}C}$ & 1.00 & 1.00 & 0.50 & 0.00 & 1.00 & 1.90 \\
$\beta_{-6.5^{\circ}C\, to\, -1^{\circ}C}$ & 1.00 & 0.60 & 0.30 & 0.00 & 0.60 & 1.30 \\
$\beta_{-1^{\circ}C\,to\,4.5^{\circ}C}$ & 1.00 & 0.60 & 0.20 & 0.20 & 0.60 & 1.00 \\
$\beta_{4.5^{\circ}C \, to \, 10^{\circ}C}$ & 1.00 & 0.20 & 0.20 & -0.30 & 0.20 & 0.70 \\
$\beta_{15.5^{\circ}C\, to \,21^{\circ}C}$ & 1.00 & -0.30 & 0.20 & -0.80 & -0.30 & 0.20 \\
$\beta_{21^{\circ}C\, to\, 26.5^{\circ}C}$ & 1.00 & -0.40 & 0.20 & -0.80 & -0.40 & 0.00 \\
$\beta_{26.5^{\circ}C \, to \,32^{\circ}C}$ & 1.00 & -0.60 & 0.20 & -1.00 & -0.60 & -0.20 \\
$\beta_{above \,32^{\circ}C}$ & 1.00 & 0.60 & 0.30 & -0.10 & 0.60 & 1.20 \\
$\gamma_{\log(GDP)}$ & 1.00 & 0.40 & 0.00 & 0.40 & 0.40 & 0.40 \\
$\gamma_{\log(Price_{t-1})}$ & 1.00 & -0.10 & 0.00 & -0.10 & -0.10 & -0.10 \\
  \midrule
    \multicolumn{7}{c}{Group Level Standard Deviation Estimates}\\
    \midrule
Parameter & $\hat{R}$ & Mean & SD & 2.5\% & Median & 97.5\% \\
    \midrule
$sd_{Intercept}$ & 1.00 & 0.40 & 0.10 & 0.30 & 0.40 & 0.50 \\
$sd_{below\,-12^{\circ}C}$ & 1.00 & 0.50 & 0.40 & 0.00 & 0.40 & 1.60 \\
$sd_{-12^{\circ}C\, to \, -6.5^{\circ}C}$ & 1.00 & 0.40 & 0.30 & 0.00 & 0.30 & 1.20 \\
$sd_{-6.5^{\circ}C\, to\, -1^{\circ}C}$ & 1.00 & 0.30 & 0.30 & 0.00 & 0.30 & 1.00 \\
$sd_{-1^{\circ}C\,to\,4.5^{\circ}C}$ & 1.00 & 0.30 & 0.20 & 0.00 & 0.20 & 0.70 \\
$sd_{4.5^{\circ}C \, to \, 10^{\circ}C}$ & 1.00 & 0.50 & 0.30 & 0.00 & 0.50 & 1.10 \\
$sd_{15.5^{\circ}C\, to \,21^{\circ}C}$ & 1.00 & 0.90 & 0.40 & 0.20 & 0.90 & 1.70 \\
$sd_{21^{\circ}C\, to\, 26.5^{\circ}C}$ & 1.00 & 1.10 & 0.30 & 0.60 & 1.10 & 1.70 \\
$sd_{26.5^{\circ}C \, to \,32^{\circ}C}$ & 1.00 & 1.10 & 0.20 & 0.60 & 1.10 & 1.50 \\
$sd_{above \,32^{\circ}C}$ & 1.00 & 2.70 & 0.40 & 2.00 & 2.70 & 3.40 \\
   \bottomrule
\end{tabular}
\caption{Selected statistics for the estimated posterior densities for population level parameter and group level standard deviations. Alternative binning structure mimicking \cite{replicated_study}. 
\scriptsize{Estimated using Bayesian Partial Pooling Model (NUTS-sampler). Data from ENERDATA; \citetalias{nasav2}, \citetalias{population_data}, \citetalias{population1}, \citetalias{population2}.}}
\label{tab:bayes_replication}
\end{table}

\newpage

\section{Supplementary Tables}
\label{appendixE}  
\setcounter{figure}{0}
\setcounter{table}{0}
\renewcommand{\thetable}{E\arabic{table}}
\renewcommand{\thefigure}{E\arabic{figure}}

\begin{table}[]
\centering
\caption{List of countries used for visualization}
\label{tab:geo_vis_countries}
\begin{tabular}{llll} \hline
Afghanistan & Albania & Algeria & Angola \\
Argentina & Armenia & Australia & Austria \\
Azerbaijan & Bahamas & Bangladesh & Belarus \\
Belgium & Belize & Benin & Bhutan \\
Bolivia & Bosnia and Herzegovina & Botswana & Brazil \\
Brunei Darussalam & Bulgaria & Burkina Faso & Burundi \\
Myanmar & Cambodia & Cameroon & Canada \\
Cape Verde & Central African Republic & Chad & Chile \\
China & Hong Kong & Colombia & Comoros \\
Congo & Costa Rica & Cote d'Ivoire & Croatia \\
Cuba & Cyprus & Czech Republic & Denmark \\
Djibouti & Dominica & Dominican Republic & Ecuador \\
Egypt & El Salvador & Equatorial Guinea & Eritrea \\
Estonia & Ethiopia & Fiji & Finland \\
France & Gabon & Gambia & Georgia \\
Germany & Ghana & Greece & Guatemala \\
Guinea & Guinea-Bissau & Guyana & Haiti \\
Honduras & Hungary & Iceland & India \\
Indonesia & Iran & Iraq & Ireland \\
Israel & Italy & Jamaica & Japan \\
Jordan & Kazakhstan & Kenya & Kuwait \\
Kyrgyzstan & Lao & Latvia & Lebanon \\
Lesotho & Liberia & Libya & Lithuania \\
Luxembourg & North Macedonia & Madagascar & Malawi \\
Malaysia & Mali & Mauritania & Mauritius \\
Mexico & Moldova & Mongolia & Montenegro \\
Morocco & Mozambique & Namibia & Nepal \\
Netherlands & New Zealand & Nicaragua & Niger \\
Nigeria & North Korea & Norway & Oman \\
Pakistan & Panama & Papua New Guinea & Paraguay \\
Peru & Philippines & Poland & Portugal \\
Qatar & Romania & Russia & Rwanda \\
Samoa & Sao Tome and Principe & Saudi Arabia & Senegal \\
Serbia & Sierra Leone & Singapore & Slovakia \\
Slovenia & Solomon Islands & Somalia & South Africa \\
South Korea & Spain & Sri Lanka & Sudan \\
South Sudan & Suriname & Swaziland & Sweden \\
Switzerland & Syria & Taiwan & Tajikistan \\
Tanzania & Thailand & Togo & Trinidad and Tobago \\
Tunisia & Turkey & Turkmenistan & Uganda \\
Ukraine & United Arab Emirates & United Kingdom & United States \\
Uruguay & Uzbekistan & Vanuatu & Venezuela \\
Vietnam & Yemen & Zambia & Zimbabwe \\ \hline
\end{tabular}
\end{table}

\end{document}